\documentclass[12pt,a4paper]{article}            
 \usepackage[skins,theorems]{tcolorbox}
\tcbset{highlight math style={enhanced,
  colframe=red,colback=white,arc=0pt,boxrule=1pt}}
  \usepackage[bookmarksopen, bookmarksnumbered, bookmarksopenlevel=2]{hyperref}
  \usepackage{tikz}
  \usepackage{tikz-3dplot}
 \usetikzlibrary{calc}
 \usetikzlibrary{decorations} %
 \usepackage[UKenglish]{babel}
 \usepackage[toc,page]{appendix}
 \usepackage{amsmath}
 \usepackage{amssymb}
 \usepackage{graphicx}
 \usepackage{hhline}
 \usepackage[bf]{caption}
\usepackage{cite}
\usepackage[vcentermath]{youngtab}
\usepackage{geometry}
\usepackage{slashed}
\usepackage{color}
\usepackage{stackrel}
\usepackage{tikz-cd} 
\usepackage{cancel} 
\usepackage{booktabs} 
\usepackage{diagbox}

\usepackage{mathtools}

\DeclarePairedDelimiter\floor{\lfloor}{\rfloor}

\usepackage{empheq}
\usepackage{arydshln}

\newcommand{\bqa}{\begin{eqnarray}}
\newcommand{\eqa}{\end{eqnarray}}

\newcommand{\nn}{\nonumber}

\def\IC{\mathbb{C}}

\def\proof{\vspace{.7\baselineskip}\noindent\bf{Proof}}

\usepackage{enumitem}

\newtheorem{PROOF}{Proof}

\hypersetup{
    pdftitle={},
    pdfauthor={},
    pdfsubject={}
}
\numberwithin{equation}{section}
\numberwithin{table}{section}

\usepackage{tikz}
\usetikzlibrary{shadows} 
\usepackage[framemethod=tikz]{mdframed}

\global\mdfdefinestyle{myboxstyle}{%
  shadow=true,
  linecolor=black,
  shadowcolor=black,
  shadowsize=6pt,
  nobreak=false,
  innertopmargin=10pt,
  innerbottommargin=10pt,
  leftmargin=5pt,
  rightmargin=5pt,
  needspace=1cm,
  skipabove=10pt,
  skipbelow=15pt,
  middlelinewidth=1pt,
  afterlastframe={\vspace{5pt}},
  aftersingleframe={\vspace{5pt}},
  tikzsetting={%
draw=black,
very thick} }

\newmdenv[style=myboxstyle]{whitebox} \newmdenv[style=myboxstyle,backgroundcolor=black!20]{graybox}

\newmdenv[style=myboxstyle,nobreak=true]{blockwhitebox}
\newmdenv[style=myboxstyle,backgroundcolor=black!20,nobreak=true]{blockgraybox}
\newmdenv[nobreak=true,hidealllines=true]{blockbox}

\makeatletter

\newtheorem{theorem}{Theorem}
\newtheorem{corollary}[theorem]{Corollary}

\newtheorem{Lemma}{Lemma}

\newtheorem{proposition}{Proposition}

\newcommand{\be}{\begin{equation}}
\newcommand{\ee}{\end{equation}}
\newcommand{\beq}{\begin{equation}}
\newcommand{\eeq}{\end{equation}}
\newcommand{\ba}{\begin{aligned}}
\newcommand{\ea}{\end{aligned}}

\newcommand{\bea}{\begin{eqnarray}}
\newcommand{\eea}{\end{eqnarray}}

\newcommand{\cO}{\mathcal{O}}

\newcommand{\cP}{\mathcal{P}}

\newcommand{\cN}{\mathcal{N}}

\newcommand{\cB}{\mathcal{B}}

\newcommand{\cI}{\mathcal{I}}
\newcommand{\cJ}{\mathcal{J}}
\newcommand{\cR}{\mathcal{R}}

\newcommand{\cV}{\mathcal{V}}

\newcommand{\cQ}{\mathcal Q}
\newcommand{\cY}{\mathcal Y}

\newcommand\bi{\begin{itemize}}
\newcommand\ei{\end{itemize}}

\def\unit{{1\kern-.65ex {\rm l}}}
\def\1{{1\kern-.65ex {\rm l}}}

\def\IZ{\mathbb{Z}}
\def\IP{\mathbb{P}}

\def\IR{{\mathbb{R}}}

\newcount\hour \newcount\minute
\hour=\time \divide \hour by 60
\minute=\time
\def\now{%
\ifnum \hour<13
  \ifnum \hour=0 \advance \hour by 12 \number\hour:\else \number\hour:\fi%
     \ifnum \minute<10 0\fi%
     \number\minute%
\ A.M.%
\else \advance \hour by -12 \number\hour:%
  \ifnum \minute<10 0\fi%
  \number\minute%
  \ P.M.%
\fi%
}

\def\fnote#1#2{\begingroup\def\thefootnote{#1}\footnote{#2}
     \addtocounter{footnote}{-1}\endgroup}

\makeatother

\begin{document}

\begin{flushright}
{\tt\normalsize CTPU-PTC-21-40}\\
{\tt\normalsize  ZMP-HH/21-23}
\end{flushright}

\vskip 40 pt
\begin{center}
{\large \bf 
Elliptic K3 Surfaces at Infinite Complex Structure\\   \vspace{2mm}
 and their Refined Kulikov models
} 

\vskip 11 mm

Seung-Joo Lee${}^{1}$
and Timo Weigand${}^{2}$

\vskip 11 mm
\small ${}^{1}${\it Center for Theoretical Physics of the Universe, \\ Institute for Basic Science, Daejeon 34126, South Korea} \\[3 mm]
\small ${}^{2}${\it II. Institut f\"ur Theoretische Physik, Universit\"at Hamburg, \\  Luruper Chaussee 149, 22607 Hamburg, Germany } \\[3 mm]
\phantom{\small ${}^{2}$}{\it Zentrum f\"ur Mathematische Physik, Universit\"at Hamburg, \\ Bundesstrasse 55, 20146 Hamburg, Germany  }   \\[3 mm]

\fnote{}{Email: seungjoolee at ibs.re.kr, 
timo.weigand at desy.de}

\end{center}

\vskip 7mm

\begin{abstract}

Motivated by the Swampland Distance and the Emergent String Conjecture of Quantum Gravity, we analyse the infinite distance degenerations in the complex structure moduli space of elliptic K3 surfaces.
 All complex degenerations of K3 surfaces are known to be classified according to their associated Kulikov models of Type I (finite distance), Type II or Type III (infinite distance).
 For elliptic K3 surfaces, we characterise the underlying Weierstrass models in detail. Similarly to the known two classes of Type II Kulikov models for elliptic K3 surfaces
 we find that the Weierstrass models of the more elusive Type III Kulikov models can be brought into two canonical forms. 
We furthermore show that all infinite distance limits are related
 to degenerations of Weierstrass models with non-minimal singularities in codimension one or to models with degenerating generic fibers as in the Sen limit.
 We explicitly work out the general structure of blowups and base changes required to remove the non-minimal singularities.
 These results form the basis for a classification of the infinite distance limits of elliptic K3 surfaces as probed by F-theory in the companion paper \cite{HetFpaper}.
 The Type III limits, in particular, are (partial) decompactification limits as signalled by an emergent affine enhancement of the symmetry algebra.

\end{abstract}

\vfill

\thispagestyle{empty}
\setcounter{page}{0}
\newpage

\tableofcontents

\setcounter{page}{1}
\newpage

\section{Introduction}

Degenerations of complex varieties are of significant interest in geometry and physics alike.
In string theory, if a variety serves as the compactification space to a lower-dimensional theory,
degenerations of its complex structure typically result in new massless degrees of freedom, oftentimes in combination with extra symmetries.
Degenerations at finite distance in the complex structure moduli space include, among others, configurations in which a finite number of light degrees of freedom occur.
This recurring theme is realised, {\it par excellence}, 
in F-theory \cite{Vafa:1996xn,Morrison:1996na,Morrison:1996pp} compactified on an elliptic fibration. The finite distance degenerations of the fiber over codimension-one loci of the base follow the classification by Kodaira and N\'eron \cite{Kodaira2,Kodaira3,Neron}, obtained originally for elliptic K3 surfaces. The Kodaira fibers on a K3 surface realise singularities of ADE type, viewed as the finite gauge algebra on a stack of 7-branes located at the position of  the singular fiber. For a review see for instance \cite{Weigand:2018rez}. Interpreting the degenerations either purely geometrically or from the point of view of the associated gauge theory has inspired continuous progress towards
a deeper understanding of both sides. 
Other types of finite distance degenerations can give rise to infinitely many light degrees of freedom, but in a genuinely strongly coupled theory \cite{Witten:1996qb}. Examples of this type are related to fibers of non-minimal Kodaira type over loci of codimension two on the base of an elliptic Calabi-Yau threefold.\footnote{We refer to the review \cite{Heckman:2018jxk} for the extensive literature on such non-minimal singularities in codimension two and their interpretation in physics.}

The present work is dedicated to the geometry of complex structure deformations at infinite distance. Our focus will be on such degenerations for elliptically fibered K3 surfaces. 
As we will see these are related, in part, to non-minimal fiber types appearing in codimension one, and the symmetry enhancements associated with such degenerations 
are of affine or, more generally, loop algebra type. 

The physics motivation for analysing infinite distance degenerations stems from the desire to understand the boundaries of moduli space for theories of quantum gravity, such as M- or F-theory compactified 
on the K3 surface \cite{HetFpaper}. 
The Swampland approach to quantum gravity \cite{Vafa:2005ui}, reviewed in \cite{Brennan:2017rbf,Palti:2019pca,vanBeest:2021lhn,Grana:2021zvf}, makes specific predictions for the behaviour of a consistent gravity theory at infinite distance in its moduli space. Among them is the appearance of infinitely many
massless degrees of freedom \cite{Ooguri:2006in}. The latter are conjectured to admit a universal interpretation as Kaluza-Klein or weakly coupled string excitation towers \cite{Lee:2019wij}.
To test these ideas, detailed knowledge of the geometry of the degeneration
and the way how it is probed by string or M-theory is required, and it is the goal of the present work to provide this information
for the complex structure degenerations at infinite distance of an elliptic K3 surface.
Our viewpoint takes a complementary, perhaps more geometric angle compared to the primarily Hodge theoretic approach \cite{Blumenhagen:2018nts,Grimm:2018ohb,Grimm:2018cpv,Joshi:2019nzi,Grimm:2019ixq,Gendler:2020dfp,Grimm:2020cda,Grimm:2021ikg,Bastian:2021eom,Palti:2021ubp,Bastian:2021hpc,Grimm:2021ckh} to infinite distance limits in the complex structure moduli space of Calabi-Yau threefolds and fourfolds. As one of the new aspects, we also
analyse the role of brane moduli within the swampland conjectures of \cite{Ooguri:2006in,Lee:2019wij}, by interpreting the complex structure deformations of the elliptic K3 surface as open string moduli of 7-branes in F-theory.
Infinite distance limits in K\"ahler moduli space, on the other hand, have been investigated in various contexts in \cite{Lee:2018urn,Lee:2018spm,Corvilain:2018lgw,Lee:2019tst,Baume:2019sry,Lee:2019xtm,Xu:2020nlh,Klaewer:2020lfg,Klawer:2021ltm}.

With this motivation in mind, this article characterises the Weierstrass models for elliptic K3 surfaces at infinite distance in complex structure moduli space, in a form
suitable for a physics interpretation in the companion paper \cite{HetFpaper}. 
Our results have been obtained independently of the recent analysis of \cite{alexeev2021compactifications,Brunyantethesis}, but to the best of our understanding are compatible with them wherever the two approaches overlap.

According to the classic theory of semi-stable degenerations \cite{MumfordToroidal}, complex structure degenerations of K3 surfaces can be brought into the form of a Kulikov model \cite{Kulikov1,Kulikov2,PerssonPink,FriedmanMorrison} of Type I, Type II or Type III.
Models of Type I occur at finite distance, those of Type II and III at infinite distance.
For elliptic K3 surfaces, the Type I degenerations correspond to the minimal Kodaira degenerations of the elliptic fiber.
Models of Type II have been analysed early on in the physics literature \cite{Morrison:1996pp,Aspinwall:1997ye}. Up to birational transformations and base change, they enjoy a refined classification as models of Type II.a or II.b \cite{Clingher:2003ui}, in which the K3 surface splits either into two rational elliptic surfaces intersecting along a common elliptic fiber or into two rational fibrations intersecting over a bi-section. 
The first degeneration points to a decompactification of F-theory to ten dimensions in a dual heterotic frame \cite{Morrison:1996pp}, while the second is a weak coupling limit in a perturbative Type IIB frame known as a Sen limit \cite{Aspinwall:1997ye,Clingher:2012rg,Sen:1996vd}.

By contrast, the structure of elliptic Type III Kulikov models is much less understood and forms the primary subject of this article (and of the independent analysis \cite{alexeev2021compactifications,Brunyantethesis}). We will see that the Weierstrass model associated with a Kulikov Type III model
can likewise be brought into one of two canonical forms, which we call models of Type III.a and III.b. 
The degenerating K3 surface $Y_0$ is a chain of Weierstrass models,
\be
Y_0 = \cup_{i=0}^P Y^i \, .
\ee
All components $Y^i$ have fibers of Kodaira type I$_{n_i}$ for $n_i \geq 0$ over generic points of their base, such that adjacent components $Y^i$ and $Y^{i+1}$ intersect over elliptic fibers of Kodaira Type I$_{k>0}$.
Of special importance for the classification are the end components $Y^0$ and $Y^P$. They can either be rational elliptic surfaces (in which case the value of $n_i=0$) or degenerate Weierstrass models with a generic I$_{n_i>0}$ fiber. In this case, the singularity type of the fiber enhances to a
 $D$-type singularity over two special points of the base component, possibly together with $A$-type singularities over additional special points.
 Here we are characterising the singularities from the viewpoint of the degenerate K3 surface, in a sense made precise in the main text.
If both end components are of this latter type with $n_i>0$, the model is called of Type III.b, otherwise it is of Type III.a. These two types of elliptic Type III models are illustrated in Figures \ref{fig:TypeIIIageneral} and \ref{fig:TypeIIIbgeneral}.

All Type III Weierstrass models (and also the Type II.a models reviewed above) are the result of engineering a suitable non-minimal singularity over one or several points on an elliptic K3 surface.
In terms of the standard Weierstrass form 
\bea
y^2 = x^3 + f x z^4 + g  z^6  
\eea
this means that the vanishing orders of $f$, $g$ and their discriminant $\Delta = 4 f^3 + 27 g^2$ simultaneously  reach or exceed
the values of $4$, $6$ and $12$. When this happens the singularity in the fiber does not allow for a crepant resolution.
However, one can perform a sequence of blowups in the base \cite{Aspinwall:1997ye}. We will show in detail how the blowups lead to the Type II.a, Type III.a or III.b Kulikov models, depending on the specifics of the non-minimal degeneration. Type II.b degenerations, on the other hand, occur by degenerating the fiber over generic points of the Weierstrass model, but without any non-minimal fiber types.

We begin in Section \ref{sec_KulikovTypeII} with a brief review of the concept of Kulikov models for semi-stable degenerations of general K3 surfaces and recall the two canonical Type II models
for elliptic K3 surfaces \cite{Clingher:2003ui}. 

In Section \ref{subsec_TypeIIImath} we classify the Weierstrass models for Type III Kulikov degenerations. In Section \ref{sec_Possiblecomp} we characterise the individual surface components of a Type III Kulikov model, arguing for the appearance of 
generic fibers of Kodaira Type no worse than  I$_n$ mentioned already above and analysing the possible enhancements over special points.
Among other things, we will see that on components with generic I$_{n>0}$ fibers, the aforementioned special enhancements to  
 $D_k$-type singularities include the values $0\leq k \leq 3$, which are absent on K3 surfaces or more generally non-degenerate elliptic surfaces. From the physics point of view, this reflects a local weak coupling structure along the I$_n$ component with $n>0$. If $n=0$, the surface is a rational elliptic surface, and all types of (minimal) Kodaira fibers can occur.

The individual surface components are then combined into the admissible Kulikov Type III Weierstrass models in Section \ref{sec_TypeIIIellipticclass}, leading to the Type III.a and III.b configurations
described above.

In Section \ref{sec_BlowupWeier} we summarise how these Type III models originate in non-minimal singularities of an underlying Weierstrass model. 
The precise statements are formulated in Theorems \ref{ref-Theorem1} and \ref{ref-Theorem2}.  
A chain of blowups can always be found, possibly up to base change, to remove the non-minimal singularity at the cost of degenerating the surface to one with several components, which turn out to be exactly
of the form characterised in Sections  \ref{sec_Possiblecomp} and \ref{sec_TypeIIIellipticclass}. 
For the interpretation of the physics in \cite{HetFpaper} it will be very important that the intersection loci of these surfaces can be assumed to be disjoint from the special fibers 
of the Weierstrass model. This can be achieved, if necessary, by additional blowups allowed in turn by an appropriate base change. Importantly, they do not change the structure of the special fibers in the interior of the components, as guaranteed by Theorem \ref{nointersectiontheorem}.

For better readability we have relegated all the technical details of this analysis and in particular the proofs of Theorems  \ref{ref-Theorem1},  \ref{ref-Theorem2} and \ref{nointersectiontheorem} to the appendices.

In Section \ref{sec_Examples}, we provide various examples for the engineering of non-minimal singularities, as well as the resulting blowups and base changes required to
bring the geometry into a Kulikov Weierstrass model. 

The interpretation of the physics as probed by F-theory in the infinite distance limit and the relation to affine algebras as a hallmark of (partial) decompactification limits is the subject of the companion paper \cite{HetFpaper}. Its main results are
briefly summarised in Section \ref{sec_Ftheoryinterpret} and in Section \ref{sec_Conclusions}.

\section{Review: Kulikov Models} \label{sec_KulikovTypeII}

Our goal is to understand infinite distance limits in the complex structure moduli space of F-theory \cite{Vafa:1996xn,Morrison:1996na,Morrison:1996pp,Weigand:2018rez} compactified on an elliptic K3 surface $X$ \cite{HetFpaper}.
The mathematical framework to study such limits systematically is the theory of degenerations in the complex structure moduli space of K3 surfaces \cite{FriedmanMorrison}.
We will begin by reviewing, in Section \ref{subsec_Semi}, the notion of Kulikov models of Type I, Type II and Type III for general K3 surfaces, which admit a systematic treatment of the geometry of complex structure degenerations.
Type I models lie at finite distance, while Type II and Type III models lie at infinite distance in complex structure moduli space.
For elliptic K3 surfaces, degenerations of Type II are well understood and come in two canonical forms, dubbed Type II.a and II.b in \cite{Clingher:2003ui}.
These results are reviewed in Section \ref{subsec_TypeIIamth}.

\subsection{Semi-stable degenerations and Kulikov models} \label{subsec_Semi}

\begin{figure}[t!]
\centering
\includegraphics[width=7cm]{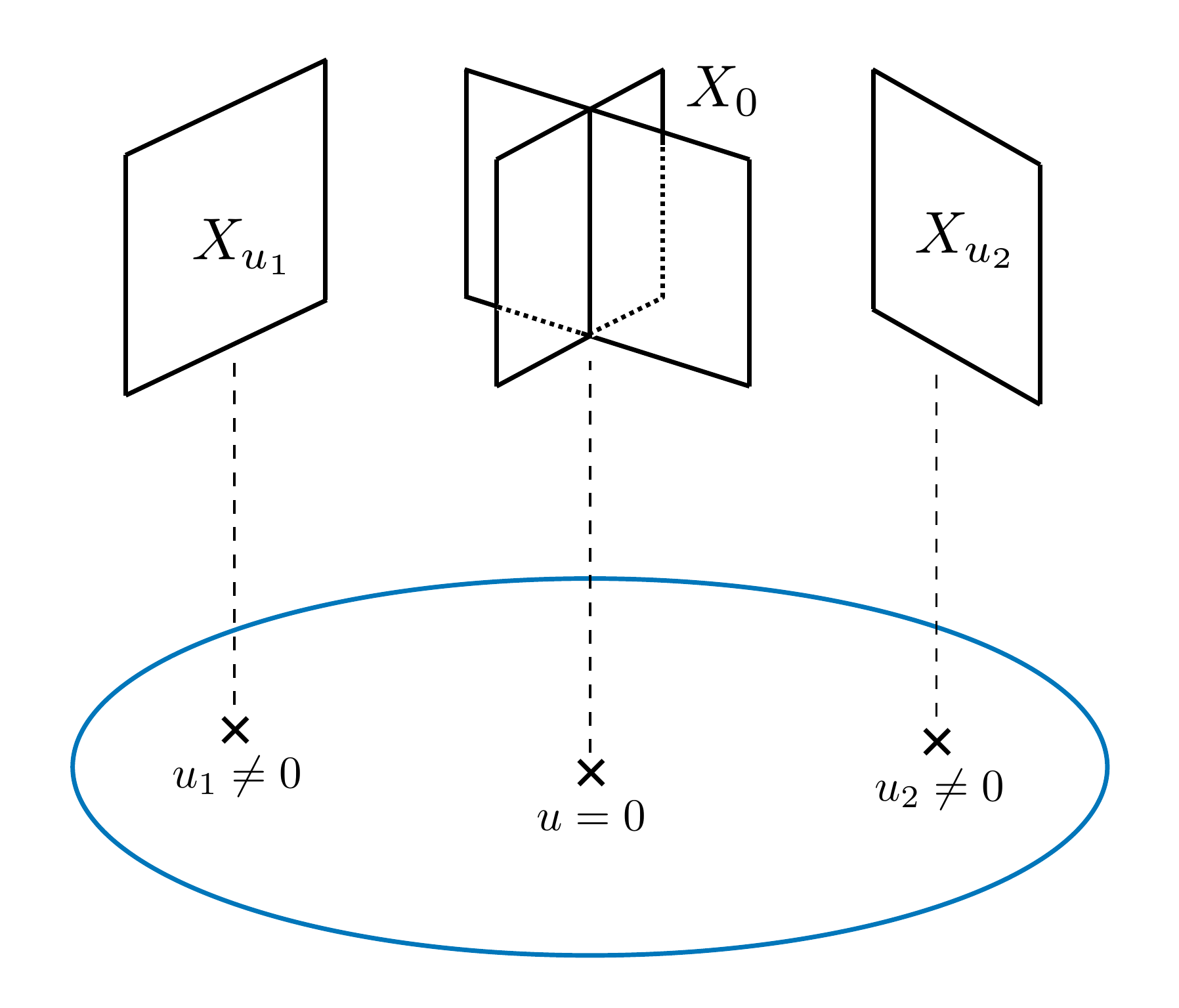}
\caption{
Semi-stable degeneration of K3 surfaces.}
\label{fig:stabledegen}
\end{figure}

By a K3 deformation one understands a one-parameter family $X_u$ of K3 surfaces whose complex parameter $u$ takes values in a disk, $u \in D = \{ u \in \mathbb C: |u| < 1\}$.\footnote{The restriction to a one-parameter, as opposed to a multi-parameter, family is merely for simplicity. For the purpose of classifying the physics at the end point of the (infinite distance) degeneration \cite{HetFpaper} it is sufficient to consider such one-parameter families.}
The generic member of the family $X_u$  for $u \neq 0$ is a smooth K3 surface, while $X_0$ denotes a degenerate K3.
If we think of this family as a fibration over $D$ with fiber $X_u$ at $u\in D$, the degeneration defines a threefold $\cal X$ together with a projection
\bea  \label{3fold-fam}
\rho: {\cal X}   \to D   \,.
\eea
This is illustrated in Figure \ref{fig:stabledegen}.
By the semi-stable reduction theorem such degenerations can always be brought into a semi-stable form \cite{MumfordToroidal}.
Semi-stability means that $\cal X$ is smooth as a threefold and
the central fiber $X_0$ is a reduced variety whose singularities are all of normal crossing type.
In other words, 
\be \label{X0deg}
X_0 = \cup_{i=1}^n X^i \,,
\ee 
where each component appears with multiplicity one and all singularities arise from local normal crossings.
The operations which may be required to achieve this form of the degeneration are birational transformations on ${\cal X}$ which act as isomorphisms on the generic family members $X_{u \neq 0}$, 
as well as so-called base changes which amount to a reparametrisation of the family by replacing 
\be  \label{basechange-def}
u \to u^k     \qquad (\text{base change})  \,.
\ee
For every semi-stable degeneration, it can in addition be arranged that the threefold ${\cal X}$ is Ricci flat \cite{Kulikov1,Kulikov2,PerssonPink}, again possibly up to birational transformations of ${\cal X}$ or base changes.
Such degenerations are called {\it Kulikov models}.

According to a rather crude classification, all Kulikov models can be characterised as being of Type I, Type II or Type III \cite{FriedmanMorrison,Kulikov2,Persson}. 
Models of Type I describe finite distance degenerations. In this case $X_0$ is a smooth irreducible variety.
Degenerations of complex structure at infinite distance, on the other hand, give rise to Type II or Type III Kulikov models. 
A detailed account can be found for instance in the the expositions in \cite{Harder_2015,Braun:2016sks}.
Some of the properties of these models of most importance to us are as follows:
\begin{itemize}
\item In a Type II Kulikov model, the central fiber $X_0$ degenerates as in  (\ref{X0deg}) such that the dual intersection graph of the irreducible components $X^i$ forms a chain of surfaces.
The end components $X^1$ and $X^n$ are rational surfaces and the remaining components $X^i$ for $i = 2, \ldots, n-1$ have a minimal model that is ruled over an elliptic curve. Furthermore the so-called double curves $C^{i,i+1}:=X^i \cap X^{i+1}$ are elliptic curves which share the same complex structure. 
\item 
In a Type III Kulikov model, the central fiber $X_0$ degenerates as in~\eqref{X0deg} such that the components $X^i$ are rational surfaces whose dual intersection graph forms a triangulation of $S^2$. If $C^{i,j} := X^i\cap X^j \neq \emptyset$, then
$C^{i,j}$ is a rational curve. The double curves satisfy additional properties which can be found e.g. in  \cite{Braun:2016sks} and references therein.

\end{itemize}

In  a Type II Kulikov model, the asymptotic form of the complex structure on the central fiber $X_0$ takes value in a 2-dimensional sublattice of $H^2(X_0,\mathbb Z)$ denoted by
\bea
 W_1 = \langle e_1, e_2 \rangle \,,
 \eea
where the generating classes $e_1$ and $e_2$ obey the relation 
\bea
e_i \cdot e_j =0 \,.
\eea
The dual 2-cycles $\gamma_i \in H_2(X_0,\mathbb Z)$ are transcendental 2-tori, whose calibrated volume vanishes
in the infinite distance limit,
\bea \label{calibratedvolumegammai}
{\rm Vol}(\gamma_i) = \int_{\gamma_i} \Omega_0   = 0 \,.
\eea
Here $\Omega_0$ refers to the $(2,0)$ form on the degenerate K3 surface $X_0$.
In Kulikov degenerations of Type III, by contrast, there exists only a single class, $\gamma$, of transcendental elliptic curves in $H_{2}(X_0, \mathbb Z)$ satisfying the relation 
(\ref{calibratedvolumegammai}).

As we will see in the companion paper \cite{HetFpaper}, in F-theory compactifications these transcendental 2-tori are responsible for  in general {\it part, but not all} of the towers of states which become massless at the fastest rate in the infinite distance limit.
In particular a correct identification of the physics of F-theory in a large complex structure limit of Type II and III will require 
a more detailed understanding of the geometry of the degenerate surface than encoded merely in the existence of the transcendental vanishing cycles.
This is precisely what we embark on in the sequel.

\subsection{Elliptic Kulikov models of Type II }  \label{subsec_TypeIIamth}

Of special interest for us are Kulikov models of Type II and III for K3 surfaces which are elliptically fibered, as these are the relevant K3 surfaces for F-theory compactifications to eight dimensions.

For elliptically fibered K3 surfaces, the infinite distance limits which give rise to Type II Kulikov models are in fact very well understood in the literature.
Up to birational transformations and base change (\ref{basechange-def}), elliptic K3 surfaces admit only two types of Type II Kulikov models.
More precisely, if we require that the Picard group of the generic element $X_u$ of the fibration contains  the two elements corresponding to a holomorphic section and the elliptic fiber\footnote{The more precise term is that the degeneration is polarised by a hyperbolic lattice $U$  \cite{Braun:2016sks}.}, the possible Type II Kulikov limits lead to two-component degenerations
\be
X_0 = X^1 \cup X^2  \,,
\ee
and fall into one of the following two classes as characterised in \cite{Clingher:2003ui}:
\begin{itemize}
\item {\bf Models of Type II.a: } \\
 $X^1$ and $X^2$ are both dP$_9$ surfaces intersecting over their common elliptic fiber as 
\be
X^1 \cap X^2 = E \,.
\ee
\item {\bf Models of Type II.b:} \\
 The surfaces $X^1$ and $X^2$ are both rational ruled surfaces, i.e. a $\mathbb P^1$-fibration over a base $\mathbb P^1_{[s:t]}$. Such $X_0$ can be obtained as the resolution of an elliptic fibration whose generic fiber degenerates to a {\it non-split} I$_2$ Kodaira fiber.\footnote{See the discussion at the end of Section \ref{sec_TypeIIIellipticclass} for the distinction between split and non-split fibers in this context.} The elliptic curve $E = X^1 \cap X^2$ is then a bisection of this degenerate fibration $X_0$ and forms a double cover of the base branched over four points.
 \end{itemize}

The physics of the associated infinite distance limits for F-theory compactified on $X_0$ is very different for both kinds of Type II limits \cite{HetFpaper}.
Type II.a limits are familiar from F-theory-heterotic duality \cite{Morrison:1996na,Morrison:1996pp}, where they describe the complete decompactification limit
of the torus $T_{\rm het}^2$ on which the dual heterotic string is compactified. The complex structure of $T_{\rm het}^2$ stays finite and is identified with the remaining
free complex structure of the double curve $E$. The limit is an effectively ten-dimensional decompactification limit in which the volume of $T_{\rm het}^2$ becomes infinite
such that the ratio of radii of both 1-cycles stays finite.

Limits of Type II.b, on the other hand, were first studied in the F-theory context in \cite{Aspinwall:1997ye}. They realise perturbative weak coupling limits known as Sen limits \cite{Sen:1996vd}, in which the theory reduces to an effectively 8d Type IIB orientifold compactified on a torus $T_{\rm IIB}^2$, which is
identified again with the elliptic double curve $E$. A systematic study of the Sen limit as a stable degeneration has been given in \cite{Clingher:2012rg}.

\section{Elliptic Kulikov Models of Type III} \label{subsec_TypeIIImath}

We now embark on a systematic exploration of Type III Kulikov models associated to infinite distance degenerations
of elliptic K3 surfaces. Very recently, explicit divisor models for such Type III degenerations have been presented in the mathematics literature in \cite{alexeev2021compactifications}, building on \cite{Brunyantethesis,alexeev2019stable}.
The analysis which we are now going to present has been obtained independently of these results but is, to the best of our understanding, generally consistent with the findings of \cite{alexeev2021compactifications}.\footnote{The analysis of~\cite{alexeev2021compactifications} does not assume the existence of a section to the fibration.}

Our object of study is a family ${\cal X}$ of K3 surfaces whose generic members $X_u$ 
admit an elliptic fibration over a base curve $B_u$. 
This implies that also the threefold ${\cal X}$ enjoys an elliptic fibration
 \bea \label{3foldell}
 \pi:   {\cal X}    \to {\cal B}   \,,
 \eea
 where $\cal B$ is the family of rational curves $B_u$ forming the base of $X_u$.
Since we are analysing F-theory on the degenerate K3 surface $X_0$, we are only interested in complex structure degenerations which are compatible with the fibration structure, even at infinite
distance.
This means that also the degenerate K3 surface $X_0$ admits an elliptic  fibration
\bea
\pi_0:   X_0 \to B_0 \,, 
\eea
whose base curve $B_0$ may split into a union of several rational curves.
Similarly each surface component $X^i$ in the decomposition~\eqref{X0deg} of $X_0$ must inherit the structure of a fibration over one of the components of $B_0$.

We can associate to the generic member $X_u$ for $u\neq 0$ of the degeneration ${\cal X}$  a Weierstrass model
 \bea \label{Weierstrassfam-def}
 Y_u: \qquad y^2 = x^3 + f_u(s,t) x z^4 + g_u(s,t) z^6 \
 \eea
 with discriminant
  \bea
  \Delta_u = 4 f_u^3 + 27 g_u^2 \,.
  \eea
Here $[s :t]$ denote homogenous coordinates on the base $\mathbb P^1_{[s:t]}$ of the K3 surface $X_u$, and $f_u$ and $g_u$ are homogeneous polynomials of degree $8$ and $12$ in $[s :t]$, respectively.

The resulting family of Weierstrass models will be denoted by  ${\cal Y}$. It is obtained 
by contracting the exceptional fibers of ${\cal X}$, viewed as an elliptic fibration over ${\cal B}$, with blowdown map $\varphi$:
\bea \label{XtoYdef}
\varphi : {\cal X}  \to {\cal Y}  \,.
\eea
The restriction, $\varphi_0$, of this blowdown map to the central fiber of ${\cal X}$ contracts the exceptional curves in the elliptic fiber of the degenerate K3 $X_0$:\footnote{This corresponds to blowing down some of the components $X^j$ of $X_0$.}  
\bea  \label{varphi_0-def}
\varphi_0:  X_0   \longrightarrow Y_0 = \cup_{i=0}^P Y^i  \,.
\eea

Note that since the total space ${\cal Y}$ of the K3 degenerations becomes singular by the contraction, we leave the framework of semi-stable degenerations
and hence that of Kulikov models in particular. We will thus refer to $\cal Y$ as a {\it Kulikov Weierstrass model}, to distinguish it from the associated Kulikov model $\cal X$.\footnote{We will further restrict, in Section \ref{sec_Possiblecomp}, the meaning of a Kulikov Weierstrass model to only denote configurations in which no special fibers lie at the intersection of two base components. This can in fact always be arranged by base change and birational transformations.}

Conversely, blowing up the singular elliptic fibers of $Y^i$ gives back the components $X^j$ of $X_0$. More precisely, 
to each $Y^i$ we can associate a set of surface components $X^{j_i}$ and a contraction $\varphi_{0,i}$ such that 
\bea \label{def-blowdown}
\varphi_{0,i}:  \cup_{j_i}   X^{j_i}   \longrightarrow Y^i     \,.
\eea

In Section \ref{sec_Possiblecomp} we characterise the possible surface components $Y^i$ which can arise in a Kulikov limit of either Type II or Type III.
These are then combined into full elliptic Type III Kulikov Weierstrass models in Section \ref{sec_TypeIIIellipticclass}, where we will distinguish between two canonical forms of Type III.a and Type III.b.
In Section \ref{sec_BlowupWeier} we explain how these occur as blowups along the base of suitably degenerate Weierstrass models. Technical details of this analysis are collected in the Appendices.

\subsection{Components of degenerate elliptic K3 surfaces} \label{sec_Possiblecomp}

Let us take a closer look at the surface $Y_0$ associated with the degenerate elliptic K3 $X_0$ by blowing down the exceptional fibers as in (\ref{varphi_0-def}).
$Y_0$ can be described as a singular Weierstrass model\footnote{Strictly speaking, we should be writing $f_0$ and $g_0$ to make clear that these are the central elements of the family (\ref{Weierstrassfam-def}), but we will drop this subscript for notational simplicity.}
\bea
y^2 = x^3 + f x z^4 + g  z^6  \,
\eea
with
\be \label{fgDeltadef1}
f \in H^0(B_0,{\cal O}_{B_0}(8)) \,, \qquad g \in H^0(B_0,{\cal O}_{B_0}(12)) \,, \qquad \Delta = 4 f^3 + 27 g^2 \in  H^0(B_0,{\cal O}_{B_0}(24)) \,.
\ee
Each of its components $Y^i$ is a possibly degenerate elliptic fibration over one of the  rational curves $B^i$ forming the components of the central base 
\be
B_0 = \cup_{i=0}^P B^i \,.
\ee
By degenerate we mean in particular that 
singular fibers can occur already over generic points of the base curves $B^i$. 
We refer to such singularities as the {\it codimension-zero singularities} of the surface component $Y^i$.\footnote{Equivalently, these are the codimension-one singularities of the 3-fold ${\cal Y}$ along the divisors $Y^i$.}
The type of singularity in the fiber is read off from the associated Weierstrass model by determining the vanishing orders of $f$, $g$ and
$\Delta$ over generic points of the base components $B^i$ following the Kodaira-N\'eron classification \cite{Kodaira2,Kodaira3,Neron} (see e.g. \cite{Weigand:2018rez} for background).
To this end,
we view the base component $B^i$ as the vanishing locus of a local coordinate $e^i$, 
\bea
B^i = \{e_i = 0\} \subset \cal B \,,
\eea
within the two-fold base $\cal B$ of~\eqref{3foldell}
and express the Weierstrass data in the form
\bea \label{fgDeltaK3van}
f = \prod_{i = 0}^P  e_i^{a_i}   f' \,,   \qquad 
g = \prod_{i = 0}^P e_i^{b_i}   g' \,,   \qquad 
\Delta =\prod_{i = 0}^P e_i^{c_i}   \Delta' \,.
\eea
Here $f'$, $g'$ and $\Delta'$ do not contain any overall factors of $e_i$.
Then the codimension-zero singularities on $Y^i$ are determined via Kodaira's table \ref{tab_Kodaira} from the vanishing orders $(a_i, b_i, c_i)$.

The first important observation is that only two qualitatively different types of surface components $Y^i$ can occur in the family of Weierstrass models associated with a semi-stable degeneration ${\cal X}$ of elliptic K3-surfaces:
\begin{enumerate}
\item 
{\bf I$_{n=0}$-components}, whose codimension-zero fibers are of Kodaira Type I$_{0}$ (i.e., smooth);  
\item {\bf I$_{n>0}$-components}, whose codimension-zero fibers are of Kodaira Type  I$_{n_i}$ with $n_i >0$.
\end{enumerate}

In terms of the Weierstrass data (\ref{fgDeltaK3van}), this amounts to the statement that
\bea   \label{aibiconstr}
(a_i \geq 0\,,~  b_i =0)     \quad \text{ or} \quad (a_i  =0\,,~  b_i  \geq 0)\,, \qquad \forall i=0, \ldots, P\,,
\eea
and furthermore
\bea
c_i =: n_i  \geq 0   \,.
\eea

\begin{table}
\centering
\begin{tabular}{|c|c|c|c|c|}
\hline
Algebra &Kodaira Type &${\rm ord}(f)$ & ${\rm ord}(g)$  & ${\rm ord}(\Delta)$ \\\hline
 $A_n$  & I$_{n+1}$   & 0 & 0 & $n+1$ \\
 $-$      & II & $\geq 1$ & $1$ & $2$ \\
 $A_1$      & III & $1$ & $\geq 2$ & $3$ \\
 $A_2$      & IV & $\geq 2$ & $2$ & $4$ \\
 $D_n$ &  I$^\ast_{n-4}$   &$\geq 2$ & $\geq 3$ & $n+2$ \\
 $E_6$   & IV$^\ast$ & $\geq 3 $& 4 & 8 \\
 $E_7$& III$^\ast$  & 3 &$ \geq 5$ & $9$ \\
$E_8$ & II$^\ast$ &   $\geq$ 4 & $ 5$ & $10$ \\\hline\hline
non-canonical   & non-minimal &   $\geq$ 4 & $\geq$ 6 & $\geq$ 12 \\\hline
\end{tabular}
\caption{Kodaira table of vanishing orders for a Weierstrass model of a K3 surface.
\label{tab_Kodaira}}
\end{table}

To see this, recall from Section \ref{subsec_Semi} that in the original Kulikov model with central fiber $X_0 = \cup_i X^i$, semi-stability requires that each component $X^i$ appear with multiplicity one
and that the singularities of $X_0$ be of locally normal crossing type. This implies that the singularities in the codimension-zero elliptic fibers of $Y^i$ can only be of Kodaira type ${\rm I}_{n_i}$ for $n_i \geq 0$:
After resolving the singularities in the fibers (corresponding to the inverse of the blowdown $\varphi_{0,i}$ defined in (\ref{def-blowdown})) we arrive at the corresponding (resolved) Kodaira fibers over the base component $B^i$.
Fibers of Kodaira Type II, III, or IV have singularities which are not of normal crossing type: Type II fibers have cuspidal singularities, in Kodaira Type III fibers two rational curves touch tangentially and in Type IV fibers three rational curves intersect in one point.  Such fibers are therefore excluded by the normal-crossing property of $X_0$. Fibers of Kodaira Type ${\rm I}_{m}^\ast$ or ${\rm II}^\ast$, ${\rm III}^\ast$, ${\rm IV}^\ast$, on the other hand, contain exceptional rational curves of multiplicity bigger than one. Since the fibration of the exceptional curves
give rise to some of the component $X^{j_i}$ in $X_0$, this implies that these components
appear with multiplicity bigger than one as well.
Altogether, we have excluded all fibers of Kodaira type different from Type ${\rm I}_{n_i}$, $n_i \geq 0$. For fibers of the latter type, the singularities over generic points of the base in the associated Kulikov model are indeed of locally normal crossing type, as required.

For later purposes it is useful to define the restriction of the Weierstrass sections $f$ and $g$ to the components $B^i$:
 \bea \label{Lidef1}
 f_i = f|_{B^i} \in  H^0(B^i, L_i^4)  \,, \qquad g_i = g|_{B^i} \in  H^0(B^i, L_i^6)   \,.
 \eea
Here, each $L_i$ is a line bundle on $B^i$, whose possible degree is restricted by the total degrees of $f$ and $g$ given in (\ref{fgDeltadef1}) to take the values
 \be   \label{Lide2}
{\rm deg}(L_i) \in \{0,1,2\} \,.
\ee
Note that as a result of (\ref{aibiconstr}), at most one of $f_i$ and $g_i$, if any, can vanish identically on a given component $B^i$.

The degree of $L_i$ restricts the types of surface components $Y^i$ with generic 
 I$_0$ fibers in the following way: An I$_0$ component is
\begin{itemize}
\item
a trivial elliptic fibration if ${\rm deg}(L_i) =0$, 
\item
a rational elliptic surface (often referred to as a dP$_9$ surface) if ${\rm deg}(L_i) =1$, or
\item
 a K3-surface is ${\rm deg}(L_i) =2$.
 \end{itemize}
Similarly, for positive I$_{n_i}$ components all values of  ${\rm deg}(L_i) \in \{0,1,2\}$ can occur.

 The singularity type of the fiber can enhance over special points of $B^i$. We will refer to such singularities as {\it codimension-one singularities on $Y^i$}.\footnote{These correspond to the codimension-two singularities from the point of view of the 3-fold ${\cal Y}$.}
 For I$_0$ components, the special fibers lie over the intersection points with the discriminant of the 
elliptic 3-fold ${\cal Y}$. Such points include the intersection loci with the positive I$_{n}$ surfaces as well as special points in the interior of $B^i$. 
For I$_{n>0}$ components, the enhancement loci are points where the discriminant of ${\cal Y}$ self-intersects (since the base of the I$_{n>0}$ component itself is part of the discriminant of ${\cal Y}$).
 
 There are two a priori different ways to characterise the singularities over such special points.
 From the perspective of the elliptic 3-fold ${\cal Y}$, the singularities are read off from the vanishing orders of the Weierstrass model in a straightforward manner via Kodaira's table.
For instance, in the simple case that the singularity enhancement occurs over a point ${\cal P} \in B^i$ given by a vanishing locus $h_{\cal P} =0$, one writes
\bea \label{orders6d}
(f, g, \Delta )|_{e_i=h_{\cal P}= \mu} =: (\mu^{\alpha_{i,\cal P}} \tilde f, \mu^{\beta_{i,\cal P}} \tilde g, \mu^{\gamma_{i,\cal P}} \tilde \Delta)  
\eea
such that $\tilde f$, $\tilde g$ and $\tilde \Delta$ have no overall factors of $\mu$ and reads off the
\bea   \label{orders6d-2}
\text{{\it 3-fold-vanishing orders} at} \, {\cal P}: \quad  {\rm ord}_{\cal Y}(f, g,\Delta)|_{\cal P}  :=  (\alpha_{i,\cal P},\beta_{i,\cal P},\gamma_{i,\cal P})   \,.
\eea
The 3-fold vanishing orders govern the singularities of the blowdown ${\cal Y}$ of the 3-fold family ${\cal X}$ with central fiber $Y_0$.

 On the other hand, we can also consider a surface component $Y^i$ by itself and define a notion of vanishing orders which is more directly related to the 7-brane content
 in F-theory compactified on $Y_0$. 
 As it turns out, the correct way to define these vanishing orders is to discard, in the expression (\ref{fgDeltaK3van}) for the discriminant $\Delta$, the overall powers of $e_i$
 as these merely account for the singularity over the generic points of $B^i$.
 Starting from (\ref{fgDeltaK3van}), we therefore define the restriction
\bea
\Delta'_i :=    \Delta' |_{e_i=0}   
\eea
as well as the 
\bea  \label{K3vanishingorderdef}
\text{{\it K3-vanishing orders} at} \, {\cal P} \in B^i : \quad     {\rm ord}_{Y_0}(f, g,\Delta)|_{\cal P}  := (v_{\cal P}(f_i),v_{\cal P}(g_i), v_{\cal P}( \Delta'_i) )   \,,
\eea
where $f_i$ and $g_i$ are the restrictions defined in (\ref{Lidef1}).
In (\ref{K3vanishingorderdef}), $v_{\cal P}(F)$ denotes the vanishing order of a function $F$ at point ${\cal P}$.
For example, if the point ${\cal P} \in B^i$ can be written as the equation $h_{\cal P} =0$, then the above vanishing orders are
 the powers of overall factors of $h_{\cal P}$ in $(f_i, g_i,  \Delta'_i)$.
 Note that if $f_i \equiv 0$, one sets $v_{\cal P}(f_i) = \infty$, and similarly for  $g_i$. 
  It is the K3-vanishing orders from which one reads off the physical 7-brane brane content localised on the component $B^i$ via Kodaira's table \ref{tab_Kodaira}.

In Appendix \ref{app_Non-minimal} we prove the following non-trivial properties of the codimension-one singularities:
\begin{itemize}
\item
{\bf Fiber minimality} (Proposition \ref{prop6}): \\
For the Weierstrass model associated with a Kulikov model, both the 3-fold and the K3 vanishing orders can be assumed to be minimal in the sense of the Kodaira classification in Table \ref{tab_Kodaira}.
Indeed if 
the 3-fold singularity were non-minimal, no blowup in the fiber (defined via (\ref{def-blowdown})) to a smooth Ricci flat 3-fold ${\cal X}$, the actual Kulikov model, could be found. 
Minimality of the 
K3-vanishing orders, on the other hand, can be achieved explicitly by performing  suitable blowups of the base. The nature of these blowups is detailed in full generality in Appendix \ref{nomen}
and illustrated in an example in Section \ref{sec_K3nonmin}.
\item
{\bf I$_k$ fibers at component intersections} (Proposition  \ref{prop1}): \\
In the Weierstrass model associated with a Kulikov model, the 3-fold vanishing orders at the intersection loci $B^i \cap B^j$ are always of Kodaira Type I$_k$.

\item 
{\bf No special fibers at component intersections} (Propositions \ref{prop8} and \ref{prop9}, see Theorem \ref{nointersectiontheorem}): \\
By base change and further blowups it can always be arranged that the the position of the special fibers is disjoint from the intersection loci $B^i \cap B^j$ of the base curves.
These operations do furthermore not change Kodaira type of the special fibers in the interior of the base components.
\end{itemize}

The proofs in Appendix \ref{app_Non-minimal} of the first two claims and in Appendix \ref{IP} of the last proceed by showing explicitly that the statements hold for the blowup of every degenerate Weierstrass model which can underlie a Kulikov model of Type II or Type III.

From now on we will only reserve the term {\it Kulikov Weierstrass model} to a configuration where the special fibers are indeed localised at points in the interior
 of the base components $B^i$ only.

To proceed further, we note that we can characterise these special fibers as follows:
\begin{enumerate}
\item
On an I$_0$ component $Y^i$, all types of minimal Kodaira fibers can occur in the sense of the K3-vanishing orders. The number of singular fibers in codimension one on its base $B^i$ is given by $12 \, {\rm deg}(L_i)$. Here $L_i$ is the line bundle on the base  component  $B^i$ defined in (\ref{Lidef1}). For rational elliptic components intersecting
an I$_{n>0}$ component, only $12-n$ of the singular fibers correspond to physical 7-branes in F-theory.
\item
On an I$_{n_i}$ component $Y^i$ with $n_i >0$, there can only occur two types of codimension-one singular fibers\footnote{The vanishing order for $f$ or $g$ is to be replaced by $\infty$ if $f_i \equiv 0$ or $g_i \equiv 0$, respectively. Note that either $f_i$ or $g_i$ must be non-vanishing as claimed around (\ref{aibiconstr}).}:
\bea
&\text{$A_k$-type fibers:} \qquad   {\rm ord}_{Y_0}(f, g,\Delta)|_{\cal P} &=(0,0,k+1)  \label{Akordersdef}     \\
&\text{$D_k$-type fibers:}  \qquad  {\rm ord}_{Y_0}(f, g,\Delta)|_{\cal P}& =(2,3,2+k)  \label{Dkordersdef}   \,,
\eea
and the number of $D$-type fibers is given by $2 \, {\rm deg}(L_i)$.
The total number of codimension-one singular fibers is given by
\bea
12  {\rm deg}(L_i) - n_i ({\rm deg}(L_i) - 2)  \,.
\eea
The number of physical 7-branes on $B^i$ is obtained by subtracting $n_{j}$ for the intersection of $Y^i$ with each adjacent I$_{n_j}$ component $Y^{j}$.
\end{enumerate}

To derive these claims, it is instructive to parametrise the Weierstrass data in a slightly different way compared to (\ref{fgDeltaK3van}), namely  to consider a fixed base component $B^i$ and to write 
\bea
f = e_i^{l_i} \check f_i + f_i \,, \qquad \quad g= e_i^{m_i} \check g_i + g_i  
\eea
for $l_i  > 0$ and $m_i > 0$ such that $\check f_i$ and $\check g_i$ contain no overall powers in $e_i$ and furthermore $f_i$ and $g_i$ are independent of $e_i$.\footnote{If $f$ itself is independent of  $e_i$, then $\check f_i \equiv 0$, and similarly for $g$ as well as for $\Delta$, which is expanded in (\ref{Deltaexpansioni}).}
By definition, $f_i$ and $g_i$ coincide with the restrictions 
to $B^i$ introduced in (\ref{Lidef1}).
In terms of these quantities the discriminant $\Delta = 4 f^3 + 27 g^2$ becomes
\bea \label{Deltaexpansioni}
\Delta  = e_i^{s_i} \check \Delta_i +    \Delta_i   \,, \qquad \quad \Delta_i = 4 f_i^3 + 27 g_i^2 \in H^0(B^i, L_i^{12})  \,,
\eea
for some $s_i > 0$, where  $\check \Delta_i$ has no overall powers of $e_i$. 
Since, as argued before, the codimension-zero fibers of $Y^i$ can only be of Type I$_{n_i}$, we know that $f_i \not\equiv 0$  or $g_i \not\equiv 0$ (or both).
Let us consider the resulting possibilities in turn:

First, if $f_i \not\equiv 0$, but $g_i \equiv 0$, then $\Delta_i = 4 f_i^3 \not\equiv 0$. This means that $\Delta$ contains no overall factor of $e_i$ so that the generic codimension-zero fiber
of $Y^i$ is of Type I$_0$. 
The case $g_i \not\equiv 0$, but $f_i \equiv 0$, is analogous.
The second  possibility to consider is that both $f_i \not\equiv 0$ and $g_i \not\equiv 0$. If  $f_i$ and $g_i$ are such that the combination $\Delta_i = (4 f_i^3 + 27 g_i^2)  \not\equiv 0$,
we see again that the generic fiber of $Y^i$ is of Kodaira Type I$_0$.

In all these configurations, the K3-vanishing orders at special points on $B^i$ can then in principle take any value below the bound for non-minimality, according the Kodaira classification.
Furthermore the total number of singular fibers in codimension-one of the I$_0$ component is given 
by the degree of $\Delta_i$, which is computed 
as $12 \,  {\rm deg}(L_i) \in \{0,12,24\}$ for the three possible such surfaces.  However, for a rational elliptic component (for which ${\rm deg}(L_i) =1$), not all of the 12 singular fibers correspond to 7-branes in F-theory. This is because when such a surface intersects an I$_{n>0}$ component, $n$ of the $12$ singular fibers on the rational elliptic component are due to the codimension-zero degeneration of the other component  and only $12-n$ singular fibers correspond to physical 7-branes in the theory.

The remaining possibility to consider is the case where $f_i$ and $g_i$ cancel  in the combination $\Delta_i = 4 f_i^3 + 27 g_i^2 \equiv 0$. In this situation, $Y^i$ has codimension-zero fibers of Type I$_{n_i >0}$ for $n_i =s_i$.
A complete cancellation between $f^3_i$ and $g_i^2$ is possible only if 
\be
f_i  = - 3  h_i^2   \,, \qquad  g_i = 2 h_i^3        \qquad  \text{for} \,\, h_i \in H^0(B^i, L_i^2) \,.  
\ee
The fibers over special points of $B^i$ are now very restricted. To see this we compute the discriminant as 
\bea \label{DeltaInpos}
\Delta  = \Delta_i +  12 f_i  e_i^{l_i}  \left( f_i \check f_i  + e_i^{l_i} (\check f_i)^2   \right) + 54 g_i   \check g_i e^{m_i} + (4 e^{3 l_i} \check f_i^3 + 27 e^{2 m_i} \check g_i^2 )  = e_i^{n_i} \check \Delta_i \,,
\eea
where we recall that $\Delta_i \equiv  0$. The K3-vanishing orders of the discriminant on $Y^i$ are then read off from $\check\Delta_i |_{B^i}$, which must be determined from
(\ref{DeltaInpos}).
Its total degree is given by
\bea
{\rm deg}(\check\Delta_i |_{B^i}) &=&{ \rm deg}( \Delta |_{B^i}) - n_i (B^i \cdot B^i) = 12  {\rm deg}(L_i) - n_i ( {\rm deg}(L_i)-2)   \\
&=&  \begin{cases}  2 n_i & {\rm for} \quad   {\rm deg}(L_i) =0      \\   12 + n_i   & {\rm for} \quad  {\rm deg}(L_i) =1   \\ 24   & {\rm for} \quad  {\rm deg}(L_i) =2   \end{cases}
\eea
and counts the number of singular fibers in codimension-one of $Y^i$.  
Here $B^i \cdot B^i$ computes the self-intersection of the compact curve $B^i = \{e_i =0\}$ within the base ${\cal B}$ of the elliptic threefold ${\cal Y}$.
We have used that the line bundle $L_i$ is the restriction of $-K_{\cal B}$ to the curve $B^i$ and applied the Riemann-Roch theorem, i.e. $B^i \cdot (K_{\cal B} + B^i) = 2g -2 = -2$, to deduce that 
\be \label{degLi}
{\rm deg}(L_i) = -K_{\cal B} \cdot B^i = B^i\cdot B^i + 2  \,.
\ee

Taking into account that $f_i  = - 3  h_i^2$ and $g_i = 2 h_i^3 $, one finds explicitly that
\bea
\check \Delta_i |_{B^i} = h_i^k  \, \Delta_i'' \,,
\eea
where  $\Delta_i''$ contains no overall powers of $h_i$, and where $k \geq 2$. 
In fact, the value $k=2$ can occur if there are non-trivial cancellations among the different terms for special forms of $\check f_i$ and $\check g_i$ and for $l_i = m_i$. 
At each of the zeroes of $h_i$, the K3-vanishing orders are of $D$-type, (\ref{Dkordersdef}).
Since $h_i \in H^0(B^i, L_i^2)$, there is always an even number of such fibers. More precisely, the number of $D$-type fibers is given by
$2 \, {\rm deg}(L_i) \in \{0,2,4\}$. 
Note that if two or more of the roots of $L_i$ coincided, the fiber would be of non-minimal type, which cannot occur in a Kulikov model.

Apart from the $D$-type singularities, additional codimension-two singularities arise at the zeroes of $\Delta''_i$ away from the vanishing locus of $h_i$.
These are of $A$-type (\ref{Akordersdef}).
The total number of physical 7-branes on an I$_{n_i>0}$ surface $Y^i$  is given by 
\bea
{\rm deg}(\check\Delta_i |_{B^i}) - n_{i+1} - n_{i-1}
\eea
where the subtraction accounts for the singular fibers at the intersections with the adjacent surface components $Y^{i+1}$ and/or $Y^{i-1}$ of type I$_{n_{i\pm1}}$.

The observation that only $A$- or $D$-type singularities can occur on a component with I$_{n_i >0}$ codimension-zero fibers resonates with the physics intuition 
from F-theory: The Weierstrass function of the complex structure $\tau$ of the elliptic fiber diverges on a component with I$_{n_i >0}$ fibers as $j(\tau) \to i \infty$.
Identifying $\tau$ with the 
 the axio-dilaton of F-theory implies that the latter 
 diverges as $ \tau  \to i \infty$,
corresponding to a local weak coupling regime. This behaviour is known to be incompatible with any fibers of Type II, III, IV or II$^\ast$, III$^\ast$  and IV$^\ast$ in the sense of K3-vanishing orders.
Note furthermore that, unlike on elliptic fibrations with non-degenerate codimension-zero elliptic fibers, we can find singularities of type $D_k$ including the values $k = 0,1,2,3$.
The physics interpretation is again related to the local weak coupling nature of such surface components: Singularities of type $D_k$ are interpreted as O7-planes with $k$ mutually local 7-branes on top, and it is well-known that away from the strict weak coupling limit $ \tau  \to i \infty$ such configurations split up into mutually non-local branes if $k<4$ \cite{Sen:1996vd}, explaining their absence for 
elliptic fibrations other than those with I$_{n>0}$ degenerations over generic points.

\subsection{Components of Kulikov Weierstrass models}   \label{sec_TypeIIIellipticclass}

In the previous section we have characterised the possible surface components of a Kulikov Weierstrass model.
We now explain how to combine these different types of components into a Kulikov Weierstrass model of Type III.

It is useful to structure the discussion according to the possible values for the self-intersection of the central base components $B^i$. 
 The Riemann-Roch theorem (\ref{degLi})  together with   (\ref{Lide2})     implies that
 the base $B^i$ of a component $Y^i$ can be a $(-k_i)$ curve (i.e. $B^i\cdot B^i =-k_i$) for $k_i=0,1,2$, and
 \bea \label{degLianddegf}
{\rm deg}(L_i) = 2- k_i     \quad    \Longrightarrow  \quad   ({\rm deg}(f_i),{\rm deg}(g_i)) = ( 4(2-k_i), 6(2-k_i))     \qquad k_i=0,1,2   \,.
 \eea
We have furthermore seen in the previous section that the surface components $Y^i$ have the following properties:
 \begin{itemize}
\item
$k_i=2$: \\
The generic fibers of $Y^i$ are either of type I$_{n_i}$ with $n_i > 0$, and only $A$-type singularities from the branes in codimension-one appear, or $Y^i$ is a trivial fibration of smooth elliptic curves. 
\item
$k_i = 1$: \\
 If the generic fibers are of Type I$_0$, this is a rational elliptic surface (with general minimal Kodaira fibers in codimension one), or else the generic fibers can be of Type I$_{n_i>0}$ with precisely two $D$-type singularities and in addition only $A$-type singularities occurring in codimension one.
\item
$k_i=0$: \\
For generic  I$_0$ fibers, we recover a K3 surface with general minimal Kodaira fibers in codimension-two, while for I$_{n_i >0}$ fibers in codimension zero, there must appear four $D$-type singularities and in addition only $A$-type singularities  in codimension one.
\end{itemize}
We summarize the above characterization of different component types in Table~\ref{tb:ComponentTypes}. 
\begin{table}
  \centering\renewcommand\arraystretch{1.2} 
 \begin{tabular}{|c||*{2}{c|}}\hline
\backslashbox{$k_i$}{I$_{n_i}$} &   ~~~~~${n_i=0}$~~~~~ &   ~~~~~${n_i>0}$~~~~~ \\\hline\hline
$2$ & $-$ & $A_k$    \\\hline
$1$ & minimal Kodaira & $A_k$, $D_k$(2)\\\hline
$0$ &   minimal Kodaira    & $A_k$, $D_k$(4) \\\hline
\end{tabular}
   \caption{Allowed singularity types of special fibers in the central surface components $Y^i$ of a Kulikov Weierstrass model. Note that the first column distinguishes the three types of surface components $Y^i$ based on the self-intersection $-k_i:=B^i\cdot B^i$ of the base components and the first row refers to the Kodaira type $I_{n_i}$ of the generic fibers. The third column uses the notation of (\ref{Akordersdef}), (\ref{Dkordersdef}).  The numbers in the parentheses count the $D$-type singularities. For $n_i>0$, also singularities of Type $D_k$ for $0 \leq k \leq 3$ can arise.}  \label{tb:ComponentTypes}
  \end{table}

These three types of surfaces $Y^i$ with $k_i=0,1,2$ can be combined as follows in a Kulikov Weierstrass model:

First, since the total degree of $(f, g)$ on the base curve $B_0 = \cup_{i=0}^P B^i$ must add up to $(8,12)$, it is clear from (\ref{degLianddegf}) along with $B_i \cdot B_i = - k_i$ that there can either be precisely two components with $B^i \cdot B^i =-1$ or one component with $B^i \cdot B^i = 0$, and in both cases an additional number of components with $B^i \cdot B^i =-2$.
Second, we will show below that configurations containing a base component with $B^i \cdot B^i =0$ do not correspond to Kulikov Weierstrass models of Type III.

This leaves us with configurations with precisely two $(-1)$ base curves. For these, 
 the base $B_0 = \cup_{i=0}^P B^i$ of $Y_0$ must form a {\it chain} of rational curves,
 \beq\label{Bdecompositionmaintext}
B^0 - B^1 - \cdots - B^{P-1} - B^P \,, 
\eeq
with 
$(-1)$ curves at both ends and $(-2)$ curves in between, where the number of the latter curves, $P-1$, is arbitrary.

The rationale behind this claim is that a component with $B^i \cdot B^i =-1$ can never intersect more than one other components.
To see this, note that if a $(-n)$ curve intersects a $(-1)$ curve in a point, contracting the $(-1)$ curve turns the $(-n)$ curve into a $-(n-1)$ curve.\footnote{Such a contraction of a $(-1)$ curve is always possible
without inducing singularities in the base.}
Hence if a $(-1)$ base curve $B^i$ intersected e.g. two $(-2)$ base curves, $B^{i-1}$ and   $B^{i+1}$, contracting $B^i$ would turn them into two intersecting $(-1)$ curves and therefore increase the total degree of $f$ and $g$ on $B_0$. It thus follows that the $(-1)$ curve components can only intersect one additional curve component of $B_0$.
This argument also shows that in a configuration with $(-1)$ curves, every $(-2)$ curve can only intersect two additional curves because by suitable contractions of the $(-1)$
curves we can eventually transform the $(-2)$ curve into a $(-1)$ curve, which by assumption could only intersect one other curve. 
As a consequence, the only configuration with a $(-1)$ base component leads to a chain of surfaces $Y^i$ where the end surfaces $Y^0$ and $Y^P$ are Weierstrass models over $(-1)$ curves and the intermediate components $Y^i$ are fibered over $(-2)$ curves.

Let us discuss in more detail  the remaining possibility of such a chain of surfaces.
If the surface components $Y^0$ and $Y^P$ are both dP$_9$ surfaces, then the associated Kulikov Weierstrass model is of Type II if all intermediate components have generic I$_0$ fibers\footnote{ In particular it must be possible to bring the Kulikov model into Type II.a form, according to the classification reported in Section \ref{subsec_TypeIIamth}.} (including the case without any intermediate fibers, i.e. $P=1$) and of Type III if $P >1$ and all intermediate surface components have I$_{n>0}$ fibers in codimension zero.\footnote{Recall from Section \ref{sec_Possiblecomp} that we can always assume, without loss of generality,
that there are no special fibers at intersections of the base components, and in fact define a Kulikov Weierstrass model to have this property. Then a Kulikov Weierstrass model with two rational elliptic end components and $P=1$ necessarily has an I$_0$ fiber at the intersection and is hence of Type II.a, rather than of Type III.}
These are the only allowed possibilities in order for the resolution $X_0$ of $Y_0$ to occur as the central fiber of a Kulikov model: Otherwise the resolved model would exhibit both
rational and elliptic curves as the intersection curves of different surface components, in contradiction with the definition of a Kulikov model.

\begin{figure}[t!]
\centering
\includegraphics[width=6.5cm]{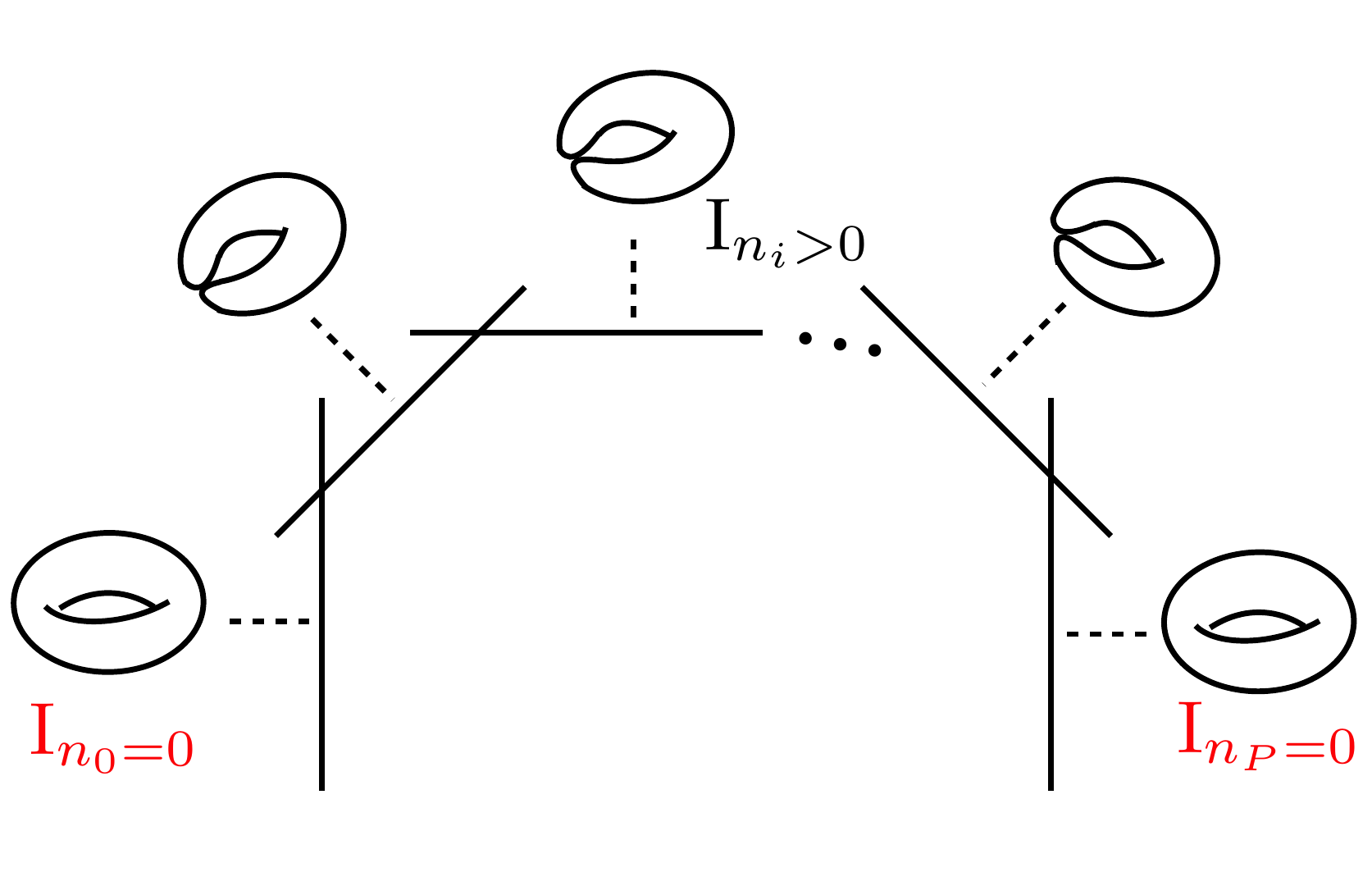}
\quad \quad\quad\quad    \includegraphics[width=6.5cm]{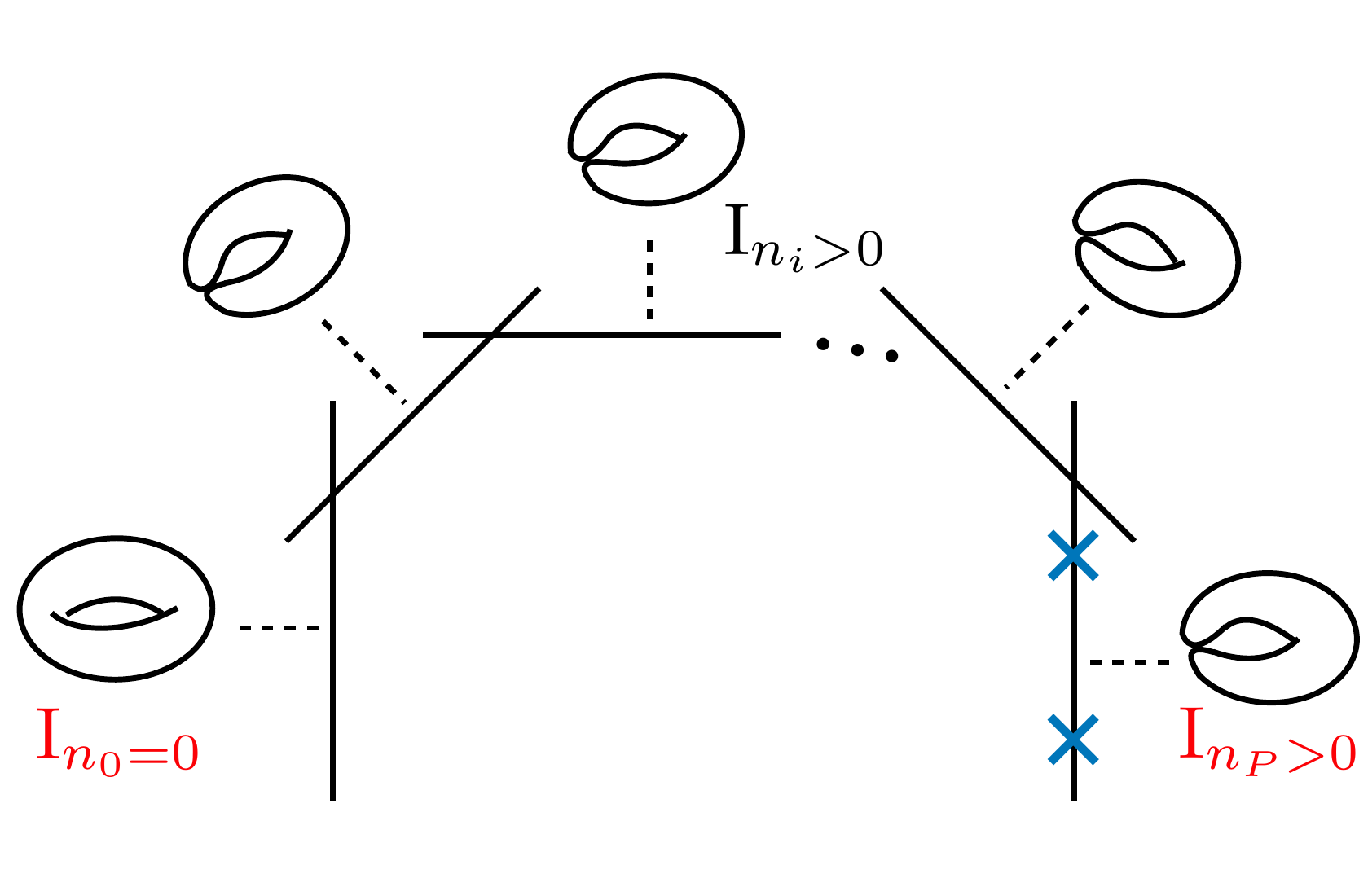}
\caption{
The two classes of Weierstrass models underlying elliptic Kulikov Type III.a degenerations. The blue crosses denote $D$-type singularities in codimension one.}
\label{fig:TypeIIIageneral}
\end{figure}

\begin{figure}[t!]
\centering
\includegraphics[width=9cm]{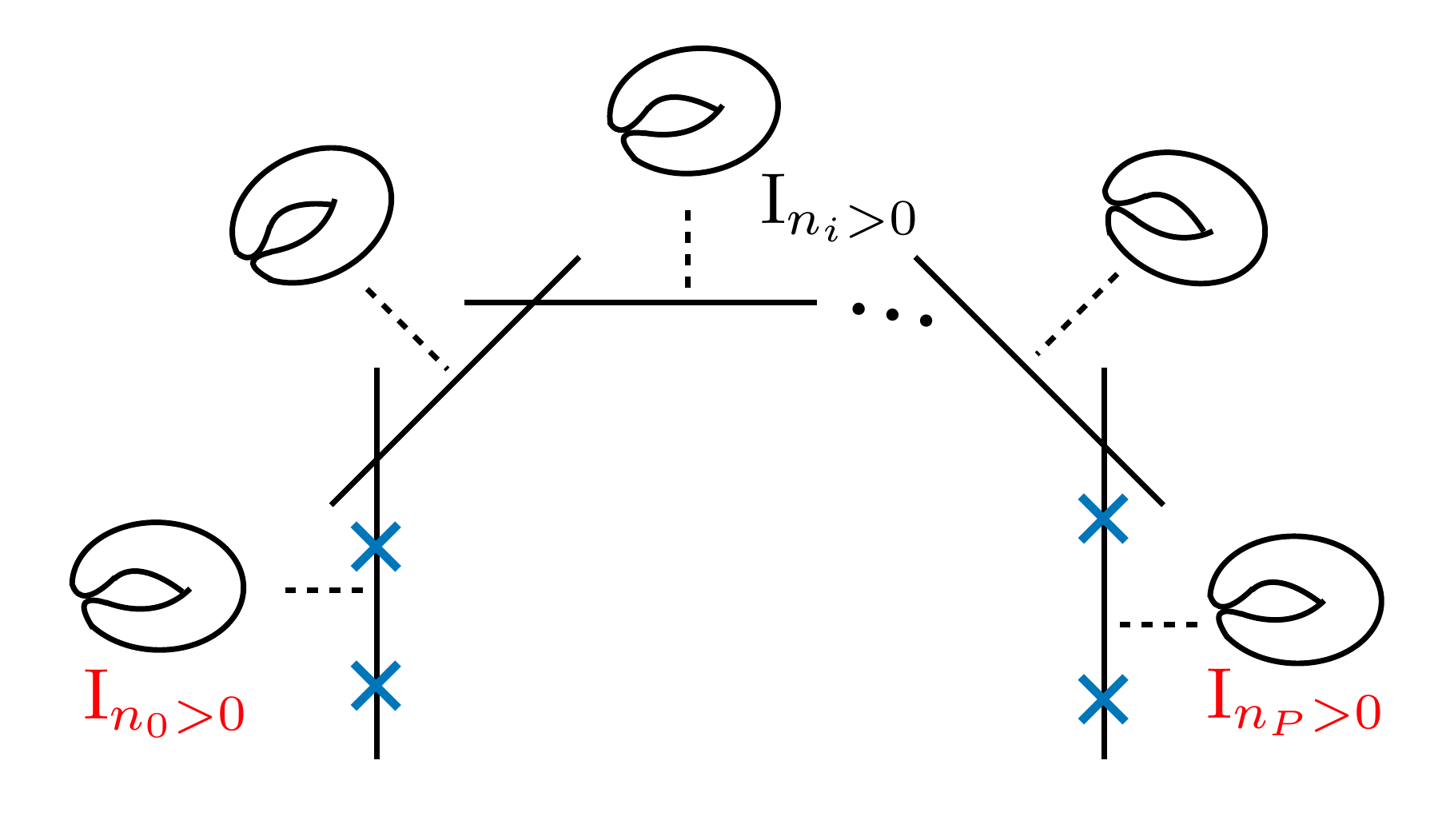}
\caption{
Weierstrass model underlying elliptic Kulikov Type III.b degenerations. The blue crosses denote $D$-type singularities in codimension one.}
\label{fig:TypeIIIbgeneral}
\end{figure}

We therefore arrive at the following characterisation of the family of Weierstrass models $\cal Y$ underlying a family $\cal X$ of elliptic K3 surfaces of Kulikov Type III, which can always be attained possibly upon base change, as well as blowups and blowdowns:

\begin{whitebox}
\begin{enumerate}
\item
{\bf Type III.a degenerations:} \\
The central fiber $Y_0$ of $\cal Y$ degenerates as a chain of surfaces $Y_0 = \cup_{i=0}^P Y^i$ with $P \geq 1$. If $P>1$, one or both of the end components $Y^0$ and $Y^P$ are rational elliptic, i.e. dP$_9$, surfaces, while the other end component not of rational elliptic type (if present) has codimension-zero fibers of Kodaira Type I$_{n>0}$ and precisely 2 singular $D$-type fibers (and possibly others of $A$-type) in codimension one; the middle components all have codimension-zero fibers of type I$_{n_i}$ with $n_i > 0$
 and only codimension-one singularities of $A$-type.
If $P =1$, only one of the end components is rational elliptic. 
 See Figure \ref{fig:TypeIIIageneral} for an illustration.

\item
{\bf Type III.b degenerations:} \\
The central fiber $Y_0$ of $\cal Y$ degenerates as a chain of surfaces $Y_0 = \cup_{i=0}^P Y^i$ with $P \geq 1$. All 
components  $Y^i$ have codimension-zero fibers of type I$_{n_i}$ with $n_i > 0$.
The two end components have precisely 2 singular $D$-type fibers (and possibly others of $A$-type) and the middle components
only have codimension-one fibers of $A$-type.
See Figure \ref{fig:TypeIIIbgeneral}.

\end{enumerate}
\end{whitebox}

As remarked at the beginning of this section, this classification is, to the best of our understanding, consistent with the mathematical results obtained in \cite{Brunyantethesis,alexeev2021compactifications}. We will show in the companion paper \cite{HetFpaper} that the physics interpretation of F-theory on such degenerating K3 surfaces
is very different for models of Type III.a and Type III.b (see Section \ref{sec_Ftheoryinterpret} for a brief summary).

It remains to explain why configurations with a self-intersection zero base component cannot occur as Weierstrass models of Type III Kulikov models.
We will proceed in two steps.  First, let us assume that the Weierstrass model over such a curve is the only 
component of $Y_0$. We will argue that such configurations can never correspond to a Type III Kulikov model upon resolution of the singularities in the elliptic fibers.
In the second step we extend this conclusion to more general geometries including base curves with $B^i \cdot B^i =0$.

Consider therefore the surface component over a self-intersection zero base component and assume that it represents the only component of $Y_0$.
Two possibility can arise: If the generic fiber is of Kodaira Type I$_0$, then $Y_0$ is 
 a K3 surface which, by our previous arguments, must have minimal Kodaira fibers; the associated semi-stable degeneration is then of Kulikov Type I and the degeneration
 lies at finite distance.
If, by contrast, the generic fibers are of Kodaira Type I$_n$ with $n>0$, then $Y_0$ must have 
$4$ $D$-type singularities.
The Weierstrass model for such $Y_0$ can always be brought into the form 
\bea \label{Weierstrass-hlimit}
f = - 3 h^2 + u^a \eta \,, \qquad g = - 2 h^3 + u^b \rho  
\eea
for $\eta \in H^0(B_0,{\cal O}_{B_0}(8))$ and $\rho \in H^0(B_0,{\cal O}_{B_0}(12))$ and for $a$, $b$ positive integers.
The resulting discriminant takes the form
\bea
\Delta = 108 h^4 u^a \eta - 36 h^2 u^{2 a} \eta^2 + 4 u^{3 a} \eta^3 - 
 108 h^3 u^b \rho + 27 u^{2 b} \rho^2  \,.
\eea
After suitable blowups the singularities in the fiber of $Y_0$ are resolved.
The generic fiber of the resolved surface $X_0$ is of Kodaira type I$_{n>0}$, but for generic choices of $h \in H^0(B_0,{\cal O}_{B_0}(4))$, the fibers undergo a non-trivial monodromy as one encircles one of the simple zeroes of $h=0$ on $B_0$.
Indeed, according to the classification of elliptic fibrations (see in particular \cite{Grassi:2011hq} for an account of this point), the I$_n$ fibers at $u=0$ are of so-called {\it non-split type} unless the ratio
\bea
\left.\frac{g}{f}\right\vert_{u=0} \sim h  
\eea
factorises.
For a generic choice of $h \in H^0(B_0,{\cal O}_{B_0}(4))$ this means that a $\mathbb Z_2$ monodromy interchanges some of the surface components $X^i$ of $X_0$ whose fibers represent the exceptional
curves   resolving the local codimension-zero singularity in $Y_0$. Taking into account the action of this  $\mathbb Z_2$ monodromy, the fibers of a non-split I$_n$-fibration intersect globally like the nodes of the affine Dynkin diagram of Sp$(\floor*{\frac{n}{2}})$. 

In the simplest case of $n=2$, we recover the Type II.b models described already in Section \ref{subsec_TypeIIamth}:
The two components $X^1$ and $X^2$ of the resolution $X_0$ intersect in a double cover of $B_0$ ramified in four points (corresponding to the four $D$-type singularities on $Y_0$, i.e. the four zeroes of $h$), which is an elliptic curve.\footnote{For surfaces with only 2 $D$-type singularities the distinction between non-split and split I$_{n_i}$ is not important for us because the intersection of the components $X^i$ after the resolution intersect at worst in a double cover of $B_i$ with 2 ramification points, which is still a rational curve.}

For I$_n$ surfaces with $n>2$, the double locus still contains such smooth elliptic curves, this time from the intersection of the 
nodes at the two ends of the affine diagram of Sp$(\floor*{\frac{n}{2}})$ with their respective neighbouring nodes.
At the same time, the interior nodes intersect their neighbours in a single cover of $B_0$ and hence a rational curve.
This shows that such configurations are not yet in Kulikov form since the double curves must be either all elliptic (Type II) or all rational (Type III).
From a physics perspective, however, we expect a similar behaviour as in the simpler case of $n=2$ and therefore conjecture that all such non-split single component I$_n$ degenerations
must lead to Kulikov Type II models, after suitable base changes and birational transformations. The explicit analysis of these operations is an interesting problem which we leave for future work.

We are left with the question if also split I$_{n>0}$ components can occur. This requires that the four zeroes of $h=0$ must collide pairwise, i.e. $h = \psi^2$ for $\psi \in H^0(B_0,{\cal O}_{B_0}(2))$.
However, at the zeroes of $\psi$ the K3-vanishing orders of $f$ and $g$ become $(4,6)$, indicating a non-Kodaira type behaviour. Possibly after suitable base change this induces non-minimal fibers also for the 3-fold ${\cal Y}$. Blowing up the base to remove these eventually leads to multi-component degenerations with only minimal fibers, as explained above.

The upshot is that single component degenerations fibered over a $(0)$-curve do not lead to Kulikov Type III degenerations. 
This conclusion does not change if the base includes additional $(-2)$ components. The only possible Type III degenerations would necessitate a split I$_{n>0}$ fibration over the $(0)$ curve and hence a collision of two or more  $D$-type fibers in codimension-one. This requires a blowup leading to a degeneration with two $(-1)$ base curves instead of the self-intersection zero base curve.

\subsection{Type III models as blowups of non-minimal Weierstrass models} \label{sec_BlowupWeier}

Degenerations leading to Type III and Type II.a Kulikov models
are constructed as blowups of Weierstrass models with suitably non-minimal Kodaira fibers.
Indeed, it is well known \cite{Aspinwall:1997ye} that after a  sequence of blowups 
a Weierstrass model with vanishing orders beyond $(4,6,12)$  can be brought into the form of a union of components each without non-minimal Kodaira singularities. 
The question, however, is how to characterise the non-minimal singularities responsible for degenerations of Kulikov Type II or Type III.

In Appendix \ref{app_Non-minimal} we prove the following 

\begin{theorem}   \label{ref-Theorem1}
Consider a one-parameter family $\widehat {\cal Y}$ of Weierstrass models whose generic members $\hat Y_u$ are K3 surfaces elliptically fibered over a family $\hat B_u$ of base curves, with only minimal Kodaira fibers. Suppose the central fiber $\hat Y_0$ of the family exhibits only minimal Kodaira singularities except for the singularities over at least one point $\cal P$ on the base $\hat B_0$ where\footnote{Note that if $m>0$ or $n>0$, then automatically $k=0$.}
\be \label{nonKodairasing-k}
{\rm ord}_{\widehat{\cal Y}}(f,g,\Delta)|_{\cal P} =  (4+m,6,12 + k) \quad {\rm or}  \quad (4,6 +n,12 + k)    \,, \qquad m \geq 0, \quad n\geq 0, \quad k\geq 0
\ee
and suppose that this criterion continues to be satisfied after arbitrary base change in the parameter $u$ of the family. Then
 \begin{itemize}
\item
for $ k\geq 1$ the central fiber of the family
can be blown up, possibly up to base change, such that the resulting degenerating Weierstrass model $Y_0$ forms the central fiber of a family ${\cal Y}$ associated with a Kulikov model of Type III.a or Type III.b as characterised above, while
\item
for $k=0$, the blowups give rise to a Kulikov model of Type II.
\end{itemize}

\end{theorem}

Indeed, for the class of non-minimal fibers as appearing in the Theorem \ref{ref-Theorem1}, we systematically perform the blowups in Appendix \ref{app_Non-minimal} and observe that the resulting degenerate surface is precisely of Kulikov form Type III ($k>1$) and Type II ($k=0$).
In the case of a Type II model ($k=0$), the general theorems reviewed in Section \ref{subsec_TypeIIamth} imply the existence of a birational transformation
which brings the model into Type II.a form.

For Kulikov models of Type III, it is interesting to see how the three possibilities - Type III.a models with one or two rational elliptic end components and Type III.b models - 
are distinguished at the level of the original Weierstrass model $\hat Y_0$.
This will be explained in detail in Appendix \ref{App-WeakCoupling} and illustrated in the concrete examples of Sections \ref{sec_ExIII.a1}, \ref{sec_ExIII.a2} and \ref{sec_ExIII.b}.
Kulikov models of Type III.a with precisely one rational end component and Kulikov models of Type III.b both require that the original Weierstrass model
$\hat Y_0$ not only exhibits at least one non-minimal singularity, but that in addition the Weierstrass sections $f|_{u=0}$ and $g|_{u=0}$ form perfect squares or cubes, more precisely 
\bea   \label{fghathform}
f |_{u=0}= -3 \hat h^2   \,  \qquad g |_{u=0}= 2 \hat h^3 
\eea
for a polynomial 
\bea
\hat h = s^2 (H_0 t^2 + H_1 s t  + H_2 s^2) \,,   \qquad H_i \in \mathbb C \,.
\eea
Here $[s : t]$ denote homogenous coordinates on the base of the Weierstrass model.  
Without additional tuning of the $u$-dependent terms, such degenerations lead to a model of Type III.a with only one rational elliptic component.
A Type III.b degeneration requires a further tuning of the $u$-dependent terms in addition to (\ref{fghathform}).

In order to make a precise statement about such a tuning, let us first define $f_-$ and $g_-$ as the part of the Weierstrass sections consisting only of the terms of degrees $d_f \leq 4$ and, respectively, $d_g \leq 6$ in $s$, whose vanishing orders in $u$ are precisely ${(4-d_f)P}$ and ${(6-d_g)P}$. Here, $P$ denotes the minimum of those vanishing orders divided by $4-d_f$ and $6-d_g$ for the terms with degrees $d_f <4$ and $d_g<6$, respectively; in fact $P$ turns out to be the number of required blowups (see Appendix~\ref{App-WeakCoupling} for more details). Then, the extra requirement is that
\be   \label{fminusgminus}
f_- = - 3 t^4 (L_0 u^{2P} t^2 + L_1 u^P s t + L_2 s^2)^2 \,,   \quad 
g_- = 2 t^6 (L_0 u^{2P} t^2 + L_1 u^P s t + L_2 s^2)^3    \,,
\ee
for $L_i \in \mathbb C$. 

The converse of Theorem \ref{ref-Theorem1} is also true, in the following sense proven in Appendix \ref{Class5-TypeI}:
\begin{theorem} \label{ref-Theorem2}
Let $\mathcal X$ be a Type II or III Kulikov model of elliptic K3 surfaces and $\mathcal Y$ be the associated Weierstrass model, whose central fiber $Y_0$ decomposes into $P+1$ components, 
$Y_0 = \cup_{i=0}^P Y^i$, each elliptically fibered itself over the component $B^i$ of the chain of base curves~\eqref{Bdecompositionmaintext}. Let us define $\widehat {\mathcal Y}$ as the blowdown of $\mathcal Y$ along its base, obtained by contracting the $P$ base curves $B^P, \ldots, B^1$ in turn, and $\hat Y_0$ as the central fiber of $\widehat {\mathcal Y}$. 
Then
\begin{itemize}
\item
for $P\geq 1$, the surface $\hat Y_0$ has a non-minimal fiber of the form (\ref{nonKodairasing-k});
\item
for $P=0$, the defining sections of the Weierstrass model of the family $\widehat {\mathcal Y}$ are of the form
\beq\label{fgh}
f_u= -3h^2 + u \eta \,, \qquad g_u= -2h^3 + u \rho \,, 
\eeq
where $h$, $\eta$ and $\rho$ are sections of degree 4, 8 and 12, respectively, $h$ has four distinct zeroes, and $u$ is the parameter of the family.
\end{itemize}
Furthermore, $\hat Y_0$ can never have a non-minimal fiber with
${\rm ord}_{\widehat{\cal Y}}(f,g,\Delta)|_{\cal P} =  (4+m,6+n,12 + k)$ for $m>0$ and $n>0$ (which implies that also $k>0$), or a non-minimal fiber which can be brought into this form only by base change.
\end{theorem}

The last statement in Theorem \ref{ref-Theorem2} reflects the following observation:
If, possibly after base change, the vanishing order can be brought into the form 
\be \label{supernonmin}
{\rm ord}_{\widehat{\cal Y}}(f,g,\Delta)|_{\cal P} =  (4+m,6 +n,12 + k) \, \qquad  m >0 \quad {\bf and}  \quad n>0
\ee
(dubbed Class 5 models in Appendix \ref{app_Non-minimal}), the non-minimal fibers can be removed by blowups, but the resulting central fiber
$Y_0$ necessarily exhibits surface components with codimension-zero fibers which are not of Kodaira type I$_n$.
This is shown in Appendix \ref{construction}.
As we have explained, such families are not yet the Weierstrass families of semi-stable degenerations.
However, the semi-stable reduction theorem and Kulikov's theorem guarantee the existence of base changes and birational transformations which bring this family
 into Kulikov form.
 
In Section \ref{subsec_Class5ex},  we will demonstrate  for concrete examples how non-minimal degenerations of the form (\ref{supernonmin}) can either lie at finite distance, in which case the blowups can be transformed to a Type I Kulikov model, or at infinite distance.
In other words, depending on the details of the model, Weierstrass models with degenerations of the form  (\ref{supernonmin}) can be base changed and blown up into Type II/III Weierstrass models or into Type I Weierstrass degenerations.

After performing the minimal base changes and blowups required to remove all non-minimal singularities, some of the special singular fibers in codimension-one may coincide with some intersection points 
 \be
 B^{i,i+1} = B^i \cap  B^{i+1} \,.
 \ee 
When this happens, the Kodaira type of the special fibers is ambiguous.
However, as anticipated already in Section \ref{sec_Possiblecomp}, one can always find a suitable base change for the original Weierstrass model $\widehat{\cal Y}$ such that after performing the required blowups removing all non-minimal singularities no such special singular fibers occur at any intersection $B^{i,i+1}$. More precisely, in Appendix \ref{IP} we prove Propositions \ref{prop8} and \ref{prop9}, which we summarise here as 

\begin{theorem} \label{nointersectiontheorem}
Consider the blowup ${\cal Y}$ of a degenerate Weierstrass model $\widehat{\cal Y}$ as in Theorem \ref{ref-Theorem1} and suppose that for some $p \in {0,\ldots, P-1}$
special fibers collide at the intersection locus $B^{p,p+1}$, i.e. suppose that the object $\Delta'$ defined in (\ref{fgDeltaK3van}) vanishes at some intersection point, $\Delta'|_{B^{p,p+1}} =0$.
Then there exists a positive integer number $k$ such that after base change $u \to u^k$ of the parameter of the original Weierstrass model $\widehat{\cal Y}$ and after performing $\bar P = k P$ blowups, the new central fiber of the family is a chain of $\bar P+1$ surfaces, $\bar Y_0 = \cup_{\bar i=0}^{\bar P} \bar Y^{\bar i}$, each fibered over base $\bar  B^{\bar i}$, such that the new family $\overline{\cal Y}$ is the Weierstrass family for a Type II or Type III Kulikov model of the same class as  ${\cal Y}$ with the following properties\footnote{Property {\it ii)} is a general property of base change and hence holds for any positive integer $k$, while {\it i)} holds precisely for $k = n k_0$ for all $n \in \mathbb N_0$ and with
$k_0 := \prod_{p=0}^{P-1} k_0^{(p)}$,
where $k_0^{(p)}$ are the smallest positive integers with the defining property~\eqref{k0pdef}, given explicitly as~\eqref{k0p-practical}.}:
\begin{enumerate}[label={\roman*)}]
\item
 In the central element $\bar Y_0$ there are no special fibers at any intersection loci $\bar B^{\bar i, \bar i+1}$.
\item
The base change and blowups do furthermore not change the types of special singular fibers over the totality of all points in the interior of the base components.
Specifically, in $Y_0$ a singular fiber is located over the point in $B^i\setminus B^{i-1,i} \cup B^{i,i+1}$ with a generic coordinate value for $[e_{i-1}:e_{i+1}]$ if and only if on $\bar Y_0$ a fiber of the same Kodaira type is located over the point in  $\bar B^{ki}\setminus \bar B^{ki-1,ki} \cup \bar B^{ki,ki+1}$ with the same coordinate value for $[\bar e_{ki-1}:\bar e_{ki+1}]$.

\end{enumerate}

\end{theorem}

This means that we can always assume, as we in fact do for a Kulikov Weierstrass model, that all special fibers reside over points in the interior of any of the base components and the type of singularity associated with such fibers
is furthermore an invariant under base change. We will encounter this phenomenon in the explicit examples of Sections \ref{sec_ExIII.a1} and \ref{sec_ExIII.a2}.

 \subsection{Outlook: Affine algebras and F-theory interpretation of the infinite distance limits}  \label{sec_Ftheoryinterpret}

In the companion paper \cite{HetFpaper}, we interpret the geometric results of this work from the point of view of F-theory compactified on an elliptic K3 surface
undergoing an infinite distance degeneration of its complex structure. Our motivation is to test the Emergent String Conjecture \cite{Lee:2019wij}, according to which
all consistent quantum gravity theories must either decompactify to a higher dimensional theory at the infinite distance boundaries of moduli space or asymptote to a weakly coupled fundamental string theory.
Let us give a brief outlook on the main results here:

First, as noted already, limits of Type II.a and II.b have been well familiar in the F-theory/string theory literature since the work of \cite{Morrison:1996pp,Aspinwall:1997ye}.
Limits of Type II.a are known as stable degeneration limits in which F-theory is dual to a compactification of the heterotic string on a torus $T^2_{\rm het}$ of asymptotically large
volume modulus $T_{\rm het} \to i \infty$. Such degenerations are hence decompactification limits from 8d to 10d. Limits of Type II.b, on the other hand, correspond to weak coupling limits in which the theory asymptotes to a perturbative compactification of Type IIB string theory on a torus orientifold with vanishing string coupling $\tau \to i \infty$. This is an emergent string limit in the language of \cite{Lee:2019wij}.

The interpretation of the Type III limits, by contrast, is more involved and crucially rests on the geometric understanding of the degenerating geometry as developed in the current paper.

For Type III.a limits, the key observation \cite{HetFpaper} is that the intersection between the rational elliptic end component, say $Y^0$, and its neighbour $Y^1$ hosts a tower of massless states
which can be identified with bound states of particles corresponding to the imaginary root $\delta$ of the affine Lie algebra $\hat E_{9-n_1}$. Here $Y^1$ has generic fibers of Kodaira Type I$_{n_1>0}$
and we are assuming, by Theorem \ref{nointersectiontheorem}, that all special fibers in codimension one are localised away from the intersection points of the base components.
In particular, out of the 12 singular elliptic fibers of the rational elliptic surface $Y^0$, only $12-n_1$ such fibers are localised over interior points of $B^0$, while the remaining $n_1$ special fibers are accounted for by the intersection with the generic I$_{n_1}$ fibers of $Y^1$.
The $12-n_1$ interior fibers are in one-to-one correspondence with the position of 7-branes in F-theory whose total $SL(2,\mathbb Z)$ monodromy is identical to the monodromy of a brane stack with affine Lie algebra
$\hat E_{9-n_1}$. The formation of such affine algebras via mutually non-local 7-branes was explained in \cite{DeWolfe:1998eu}. In the infinite distance limit prior to the blowup, these $12-n_1$ branes coincide and their collision is responsible for the non-minimal Kodaira singularity in the originial Weierstrass family (called $\widehat{\cal Y}$ in Theorem \ref{nointersectiontheorem}). The blowups artificially separate
the branes in the auxiliary geometry $Y_0$ of the Weierstrass Kulikov model. Similar remarks apply to the other rational elliptic end component, if present.

From the point of view of the dual M-theory, the towers of particles are due to M2-branes wrapping the transcendental elliptic curve $\gamma$ mentioned around (\ref{calibratedvolumegammai}). This transcendental cycle is constructed by combining the vanishing $(1,0)$ cycle in the elliptic fiber with a 1-cycle on the base that pinches at the intersection point $B^{0,1} = B^0 \cap B^1$. In F-theory language the resulting tower of massless particles is obtained from $(1,0)$ string junctions encircling the intersection $B^{0,1}$ arbitrarily often. 
Being associated with the imaginary root $\delta$, they play the role of a Kaluza-Klein tower in a dual heterotic formulation so that the
infinite distance limits encodes a partial decompactification from 8d $\to$ 9d \cite{HetFpaper}.

If both end components $Y^0$ and $Y^P$ are rational elliptic surfaces, the asymptotic symmetry algebra can be read off as
\bea
G_{\infty} = H \oplus (\hat E_{9-n_1} \oplus  \hat E_{9-n_{P-1}})/\sim  \,,
\eea
where the quotient indicates that the imaginary root for both affine components are identified. If only a single end component is rational elliptic, the second affine factor is to be omitted.
The Lie algebra factor $H$ is the gauge algebra associated with the stacks of coincident 7-branes corresponding to the special singular fibers over isolated points of the non-rational elliptic components.
Here we are making crucial use of Theorem \ref{nointersectiontheorem} in two respects: First, if branes coincide at intersection points $B^{p,p+1}$, the Kodaira fibers of the Weierstrass model are ambiguous and hence it is impossible to read off their contribution to $H$. This makes it necessary to perform a suitable base change and to focus on such Weierstrass models in which the special fibers are located only in the interior of the components.
Second, the notion of a non-abelian symmetry algebra $H$ from enhancements in the interior of the base components is well-defined only if the latter are invariant under base change; fortunately this is indeed guaranteed by Theorem \ref{nointersectiontheorem}.

Combined with the idea that the towers from the imaginary root of the affine algebra are Kaluza-Klein states, we conclude that the non-abelian gauge algebra in 9d is of the form
\bea
G_{\rm 9d} = H \oplus (E_{9-n_1} \oplus  E_{9-n_{P-1}})  \,.
\eea
Note that the same reasoning can also be applied to Type II.a Kulikov models: Here the asymptotic symmetry algebra is the double loop algebra 
$(\hat E_9 \oplus \hat E_9)/\sim$, whose two imaginary roots $\delta_1$ and $\delta_2$ explain the dual decompactification from 8d to 10d.

Degenerations of Type III.b, on the other hand, are weak coupling limits since the elliptic fiber degenerates generically to an I$_{n_i>0}$ fiber on each 
surface component. However, unlike in their generic counter-parts of Kulikov form Type II.b, the Type IIB orientifold is to be taken in the limit of infinite complex structure for the Type IIB torus, $U_{\rm IIB} \to i \infty$.
Therefore the degeneration describes a full decompactification limit from 8d to 10d \cite{HetFpaper}.

\section{Examples}   \label{sec_Examples}

In this section we illustrate, via various examples, the explicit construction of Kulikov Type III degenerations from families of K3 Weierstrass models
with non-minimal fibers in codimension one.
Our starting point is a Weierstrass model with base $\IP_{[s:t]}^1$ of the form
\beq\label{original}
y^2 = x^3 + f_u (s, t) x z^4 + g_u(s,t) z^6 \,,
\eeq
which describes a family $\widehat{\mathcal Y}$ of K3 surfaces $\hat Y_u$ degenerating for $u=0$. 
For convenience we will drop the subscript in $f_u$ and $g_u$.
We will engineer non-minimal singularities of the form (\ref{nonKodairasing-k}) and explicitly perform the blowups necessary to remove all non-minimal degenerations.
The general procedure for the blowups is detailed in Appendix \ref{nomen}.

\subsection{Type III.a model with two rational elliptic surfaces}  \label{sec_ExIII.a1}

We begin with the family $\widehat{\cal Y}$ of Weierstrass models parametrised by
\be
\begin{split}  \label{III.aex1}
f &= -3  (  \hat h^2  + s^8   + 2 s^3 t^5 u^2)   \,, \cr
\qquad    g&=2  (\hat h^3 + 3 s^5 t^7 u^2)   
\end{split}
\ee
with
\be  \label{hathdefintition}
\hat h= s^2 (s^2 + t^2)   \,.
\ee
For generic values of $u \neq 0$, the family exhibits a  Kodaira Type III$^\ast$ singularity at  $s=0$ (corresponding to the Lie algebra $E_7$),
which for $u = 0$ becomes non-minimal, with vanishing orders
\bea \label{nonmins1}
{\rm ord}_{\widehat {\cal Y}}(f,g,\Delta) |_{s=u=0}= (4,6,14)  \,.
\eea
To remove the non-minimality, we perform a  blowup of the form 
\bea
u  &\longrightarrow& e_{0}\, e_{1} \,,\\
s &\longrightarrow& s \,e_1 \,, 
\eea
together with a rescaling 
\be   \label{rescalingwithe1}
(f,g,\Delta) \to (e_1^{-4} f, e_1^{-6} g,  e_1^{-12} \Delta) \,,
\ee
followed by a second blowup
\bea
e_{1} &\longrightarrow& e_{1}\, e_{2} \,,\\
s &\longrightarrow& s \,e_2 \,
\eea
and the analogous rescaling by powers of $e_2$.
As a result, the non-minimal singularity at $s=0$ is removed and the central fiber of the family $\widehat{\cal Y}$ is replaced by
 a chain of three intersecting surfaces $Y^0 - Y^1 - Y^2$, each fibered over a curve $B^i = \{e_i =0 \}$.
The  local coordinates on the base which do not have any common zeroes include the combinations
 \bea \label{SRI-III.b}
 (e_0, e_2) \,, (t,e_1) \,, (t,e_2) \,, (s,e_0)  \,, (s,e_1)  \,, (s,t) \,.
 \eea
By slight abuse of notation, we denote the Weierstrass sections after the blowup by the same symbols and compute these as
\be
\begin{split}
f & = -3 s^3 (2 e_1^4 e_2^8 s^5 + 2 e_1^2 e_2^4 s^3 t^2 + s t^4 + 2 e_0^2 e_1 t^5)    \,, \cr
g &=  2 s^5 (e_1^6 e_2^{12} s^7 + 3 e_1^4 e_2^8 s^5 t^2 + 3 e_1^2 e_2^4 s^3 t^4 + 
   s t^6 + 3 e_0^2 e_1 t^7)  \,,   \cr
\Delta  &=   e_1^2 \Delta'  \,,
\end{split}
\ee
where $ \Delta'$, which we do not display explicitly, is defined as in~\eqref{fgDeltaK3van}, so that it does not contain any overall factors of $e_i$.
The generic fibers over the base components $B^i$ are of Kodaira Type I$_{n_i}$ for $n_0 =0$, $n_1 = 2$, $n_2 =0$.
This  realises the Weierstrass model associated with a Type III.a Kulikov model with rational elliptic surfaces on both ends of the chain.

To analyse the singularities in codimension one, we compute the restrictions $f_i = f|_{e_i=0}$, $g_i  =g|_{e_i=0}$ and $ \Delta_i '=  \Delta'|_{e_i =0}$ as
   \begin{align}
 & f_0 =  -3 (2 e_1^4 + 2 e_1^2 t^2 + t^4)   \,, \quad    &&g_0 = 2 (e_1^2 + t^2)^3   \,,  \quad   &&\Delta_0' = -108 e_1^2 d_8(t,e_1)     \,, \\
 &  f_1 =  -3    \,,  \quad   &&g_1 = 2   \,,  \quad   &&\Delta_1' = -324 e_0^4       \,, \\
& f_2 = -3 s^3 (2 e_1 + s)    \,, \quad     &&g_2 = 2 s^5 (3 e_1 + s)    \,, \quad   &&\Delta_2' = - 108 s^9 (8 e_1 + 3 s)     \,,
\end{align}
 where we made use of  (\ref{SRI-III.b}) and the freedom to rescale to one those coordinates which cannot vanish simultaneously with $e_i$ on the three components.
 Furthermore, $d_8(t,e_1)$ denotes a polynomial of degree $8$ in $t$ and $e_1$ without overall factors in them.
From these expressions, one reads off the K3-vanishing orders (\ref{K3vanishingorderdef}) at special points on the various components.
 There are 8 I$_1$ singularities at generic points on $B_0$.
On $B^2$, one finds an $E_7$ singularity at $s=0$ and a Kodaira Type I$_1$ fiber at $8e_1 +3s =0$. 
However, what hampers an interpretation of the model is that at the point $e_0 =e_1=0$, two singular fibers from $B^0$ and four singular fibers from $B^1$ coalesce. 
As stressed in Section \ref{sec_BlowupWeier}, such singularities at intersections of adjacent components defy an unambiguous attribution of Kodaira vanishing orders.

To remedy this, we follow the strategy summarised in Theorem \ref{nointersectiontheorem} and apply a base change to the original Weierstrass model (\ref{III.aex1}). As it turns out, the minimal required choice 
which removes the singular fibers at component intersections is
\be
u \to u^3 \,.
\ee
Indeed, after six blowups 
one arrives at a chain of 7 surface components with generic Kodaira fibers
\bea
{\rm I}_0 - {\rm I}_4 - {\rm I}_8 - {\rm I}_6 - {\rm I}_4 - {\rm I}_2 - {\rm I}_0   \,,
\eea
where the lefthand component with base $\bar B^0$ contains the point $t=0$ and the righthand component with base $\bar B^6$ contains the point $s=0$.\footnote{Following the notation  of the Appendix, we put bars above the symbols denoting objects in the base-changed configuration and its blowup. } 
By explicit analysis one confirms that the structure of singular fibers on the end components (away from the adjacent fibers) is unchanged compared to the original model, as guaranteed by  Theorem \ref{nointersectiontheorem}, and that 
the singular fibers at $e_0=e_1=0$ in the original model get resolved into three I$_2$ fibers on the base component $B^2$.

According to the discussion in Section \ref{sec_Ftheoryinterpret}, 
the end components $\bar Y^0$ and $\bar Y^6$ contribute an affine Lie algebra factor $\hat E_5$ and, respectively, $\hat E_7$, while the branes on the intermediate component $\bar Y^2$ contribute a Lie algebra factor 
$A_1^{\oplus 3}$. Altogether  
the non-abelian part of the symmetry algebra of the degenerate model is therefore
\bea
G_{\infty} =   A_1^{\oplus 3} \oplus (\hat E_7 \oplus \hat E_5)/\sim \,.
\eea
The degeneration, as probed by F-theory, describes a partial decompactification limit from 8d to 9d with a 9d non-abelian gauge algebra\footnote{The Lie algebra $E_5$ is by definition equal to $D_5$.} \cite{HetFpaper}
\bea
G_{\rm 9d} = A_1^{\oplus 3}  \oplus   E_7 \oplus D_5   \,.
\eea

\subsection{Type III.a model with one rational elliptic surface}   \label{sec_ExIII.a2}

If we drop the term $s^8$ in (\ref{III.aex1}), i.e. for
\be
\begin{split}  \label{III.aex2}
f &= -3  (  \hat h^2 + 2 s^3 t^5 u^2)   \,, \cr
\qquad    g&=2  (\hat h^3 + 3 s^5 t^7 u^2)   \,,
\end{split}
\ee
the Weierstrass data takes the special form (\ref{fghathform})  in the limit $u=0$ (for $H_1 =0$ and $H_0 = H_2=1$),
while the  
singularity at $s=0$ remains non-minimal as in  (\ref{nonmins1}). Since the sectors of $f$ and $g$ of low degrees in $s$ do not lead to the form (\ref{fminusgminus}), we therefore expect a Type III.a Kulikov model with only one rational elliptic surface component.

Indeed, after two blowups one finds the new Weierstrass model given by
\be
\begin{split}
f & = -3 s^3 (e_1^4 e_2^8 s^5 + 2 e_1^2 e_2^4 s^3 t^2 + s t^4 + 2  e_0^2 e_1 t^5 )  \,, \cr
g &= 2 s^5 (e_1^6 e_2^{12} s^7 + 3 e_1^4 e_2^8 s^5 t^2 + 3 e_1^2 e_2^4 s^3 t^4 + 
   s t^6 + 3 e_0^2 e_1 t^7 )   \,,   \cr
\Delta  &=  e_0^2 e_1^2    \Delta' \,.  
\end{split}
\ee
The generic fibers over the base components $B^i$ are now of Kodaira Type I$_{n_i}$ for $n_0 =2$, $n_1 = 2$, $n_2 =0$, as expected.

From the restrictions $f_i = f|_{e_i=0}$, $g_i  =g|_{e_i=0}$ and $\Delta_i' = \Delta'|_{e_i =0}$,
  \begin{align}
 & f_0 =  -3 (e_1^2 + t^2)^2   \,, \quad    &&g_0 = 2 (e_1^2 + t^2)^3   \,,  \quad   &&\Delta_0' = - 648 e_1 t^5 (e_1^2 + t^2)^3     \,, \\
 &  f_1 =  -3    \,,  \quad   &&g_1 = 2   \,,  \quad   &&\Delta_1' = -324 e_0^2       \,, \\
& f_2 = -3 s^3 (2 e_1 + s)    \,, \quad     &&g_2 = 2 s^5 (3 e_1 + s)    \,, \quad   &&\Delta_2' = - 108 s^9 (8 e_1 + 3 s)     \,,
\end{align}
we next compute the K3-vanishing orders on the various components.
On $B^0$ we deduce the presence of an $A_4$ singularity at $t=0$ and two $D_1$-singularities at $e_1 \pm i t =0$.
On $B^2$, one finds an $E_7$ singularity at $s=0$ and a  Kodaira Type I$_1$ fiber at $8e_1 +3 s =0$. 
At $e_0 =e_1=0$ one singular fiber from $B^0$ and two singular fibers from $B^1$ coalesce.
To determine the unambiguous interpretation of the Kodaira fibers we perform the same minimal base change $u \to u^3$ as in the previous model,
followed by six blowups. The pattern of generic fibers for the resulting chain of surface components becomes
\bea
{\rm I}_6 - {\rm I}_7 - {\rm I}_8 - {\rm I}_6 - {\rm I}_4 - {\rm I}_2 - {\rm I}_0   \,.
\eea
The change of the codimension-zero Kodaira type  on the left end component, 
\beq
{\rm I}_2~~\text{on}~~Y^0 \quad \longrightarrow \quad  {\rm I}_6~~\text{on}~~\bar Y^0 \,,
\eeq
of course reflects the base change $u \to u^3$.
Apart from this, the singular fibers on the end components (away from the adjacent components) are unaltered and the three coalescing singular fibers at $e_0 =e_1=0$ in the original model
are resolved into three I$_1$ fibers over the three points $e_1^3 + 2 e_3^3 =0$ on the component $\bar B^2$.
The non-abelian part of the asymptotic symmetry group is therefore
\bea
G_{\infty} =   A_4   \oplus \hat E_7 \,,
\eea
whose maximal Lie sub-algebra
\bea
G_{\rm 9d} =   A_4  \oplus E_7
\eea
is the non-abelian part of the gauge algebra in the 9d limit \cite{HetFpaper}.

\subsection{Type III.b example}   \label{sec_ExIII.b}

Upon further tuning the low $s$-degree sectors of $f$ and $g$, we can modify the infinite distance limit such that it becomes of Kulikov Type III.b form.
Concretely, consider now\footnote{Note that we view~\eqref{III.bex1} as a further tuning of~\eqref{III.aex2} even though the former involves extra terms added to the latter. This is because the low $s$-degree sectors $f_-$ and $g_-$ are related by~\eqref{fminusgminus} only after these extra terms are introduced.}
\be
\begin{split}  \label{III.bex1}
f &= -3  (  \hat h^2 + 2 s^3 t^5 u^2 + s^2  t^6 u^4)   \,, \cr
\qquad    g&=2  (\hat h^3 + 3 s^5 t^7 u^2 +  3 s^4 t^8 u^4 + s^3 t^9 u^6)   \,.
\end{split}
\ee
This realises the tuning (\ref{fminusgminus}) for $L_0=0$, $L_1 =L_2=1$ and $P=2$.
For generic values of $u \neq 0$, the singularity at $t=0$ is of Kodaira Type I$_5$ (corresponding to Lie algebra $A_4$) and at $s=0$ of Kodaira Type I$^\ast_4$ (corresponding to $D_{8}$).
For $u=0$,  non-minimal singularity at $s=0$ enhances to
\bea
{\rm ord}_{\widehat {\cal Y}}(f,g,\Delta) |_{s=u=0}= (4,6,15)  \,.
\eea
After the $P=2$ blowups, we obtain
\be
\begin{split}
f &= -3 s^2 (e_1^4 e_2^8 s^6 + 2 e_1^2 e_2^4 s^4 t^2 + s^2 t^4 + 
   2 e_0^2  e_1 s t^5 + e_0^4 e_1^2 t^6 ) \,, \cr
g &=2 s^3 (e_1^6 e_2^{12} s^9 + 3 e_1^4 e_2^8 s^7 t^2 + 3 e_1^2 e_2^4 s^5 t^4 + 
   s^3 t^6 + 3 e_0^2 e_1 s^2 t^7  + 3 e_0^4 e_1^2 s t^8  + e_0^6 e_1^3 t^9 ) \,,   \cr
   \Delta  &=  e_0^2 \, e_1^3 \, e_2^4     \,  \Delta'   \,.  
\end{split}
\ee
The generic fibers over the base components $B^i$ are of Kodaira Type I$_{n_i}$ for $n_0 =2$, $n_1 = 3$, $n_2 =4$.
From the restrictions 
  \begin{align}
 & f_0 =  -3 (e_1^2 + t^2)^2   \,, \quad    &&g_0 = 2 (e_1^2 + t^2)^3   \,,  \quad   &&\Delta_0' = -648 t^5 (e_1^2 + t^2)^3    \,, \\
 &  f_1 =  -3    \,,  \quad   &&g_1 = 2   \,,  \quad   &&\Delta_1' = -648       \,, \\
& f_2 = -3 s^2 (e_1 + s)^2    \,, \quad     &&g_2 = 2 s^3 (e_1 + s)^3    \,, \quad   &&\Delta_2' = -648  s^{10} (e_1 + s)^3   \,,
\end{align}
one finds, on $B^0$, an $A_4$ singularity at $t=0$ and two $D_1$-singularities at $e_1 \pm i t =0$,
and on $B^2$,  a $D_8$ singularity at $s=0$ and a $D_1$ singularity at $e_1 +s =0$. 
This is in agreement with the general expected form of a Type III.b Kulikov Weierstrass model.

\subsection{K3-Non-mimality and base change}   \label{sec_K3nonmin}

It is interesting to contrast this example with an initial configuration where a similar blowup procedure results in a chain of surfaces where all 3-fold vanishing orders are minimal,
but where the K3-vanishing orders remain non-minimal. As explained generally in Proposition \ref{prop6}, in this case the original model can be modified, by base change, into a degeneration
whose blowup is free of such pathologies and which falls into the classification scheme of Section \ref{sec_TypeIIIellipticclass}.

Concretely, consider the degeneration
\be
\begin{split}   \label{degen-ex-basechange}
f &= -3 s^3 (s^5 + 2 s^3 t^2 + s t^4 + t^5 u^3)   \cr
g &= 2 s^4 (s^8 + 3 s^6 t^2 + 3 s^4 t^4 + s^2 t^6 + t^8 u^5)   \,,
\end{split}
\ee
which for $u =0$ is of the form $f = -3 \hat h^2$ and $g = 2 \hat h^3$ for $\hat h = s^2 (s^2 + t^2)$ and hence again of the form (\ref{fghathform}).

To remove the non-minimal singularity at $s=u=0$, we perform two blowups and arrive at a chain of surfaces
$Y_0 = \cup_{i=0}^2 Y^i$ whose Weierstrass sections are given by
\be
\begin{split}  
f&=-3 s^3 (e_1^4 e_2^8 s^5 + 2 e_1^2 e_2^4 s^3 t^2 + s t^4 + e_0^3 e_1^2 e_2 t^5 )  \,, \cr
g &=2 s^4 (e_1^6 e_2^{12} s^8 + 3 e_1^4 e_2^8 s^6 t^2 + 3 e_1^2 e_2^4 s^4 t^4 + 
   s^2 t^6 + e_0^5 e_1^3 e_2 t^8 )  \,, \cr
\Delta &= e_0^3   e_1^2 e_2    \Delta'  \,. 
\end{split}
\ee
Since there remain no points of non-minimal 3-fold vanishing orders, no further blowups are possible.
However, we notice that the K3-vanishing orders at the point $s=0$ on $B^2$ take the non-Kodaira form
\bea
{\rm ord}_{Y_0}(f,g,\Delta) |_{s=e_2=0}= (4,6,10)   \,.
\eea
This follows from the form of the restrictions
\bea
f|_{e_2=0} = -3 s^4   \,, \qquad g|_{e_2=0} = 2 s^6   \,, \qquad     \Delta' |_{e_2 =0} = -108s^{10} (-2 e_1 + 3 s)    \,.
\eea

We are tempted to interpret these K3-vanishing orders as indicative of a $\hat E_7$ singularity at $s=0$.
At the same time, the codimension-one singularities on the end component $Y^2$ clearly do not fit into the classification scheme of Section \ref{sec_TypeIIIellipticclass}.
To resolve this puzzle, we go back to the original degeneration (\ref{degen-ex-basechange}) and apply a base change
$u \to u^2$. This realises the general procedure described in the proof of Proposition \ref{prop6} presented in Appendix \ref{proofs}. 
After the base change, a succession of 5 blowups is required to remove the 3-fold non-minimalities.
Concretely, this yields the Weierstrass data
\be
\begin{split}  
f &=-3 s^3 (e_1^4 e_2^8 e_3^{12} e_4^{16} e_5^{20} s^5 + 
   2 e_1^2 e_2^4 e_3^6 e_4^8 e_5^{10} s^3 t^2 + s t^4 + 
   e_0^6 e_1^5 e_2^4 e_3^3 e_4^2 e_5 t^5)   \,, \cr
g &= 
2 s^4 (e_1^6 e_2^{12} e_3^{18} e_4^{24} e_5^{30} s^8 + 
   3 e_1^4 e_2^8 e_3^{12} e_4^{16} e_5^{20} s^6 t^2 + 
   3 e_1^2 e_2^4 e_3^6 e_4^8 e_5^{10} s^4 t^4 + s^2 t^6 + 
   e_0^{10} e_1^8 e_2^6 e_3^4 e_4^2 t^8) \,, \cr
 \Delta &=  e_0^6   e_1^5 e_2^4 e_3^3 e_4^2    \Delta'   \,.  
\end{split}
\ee
The resulting degenerate surface $Y_0 = \cup_{i=0}^5 Y^i$ is of Type III.a, with $Y^i$ having generic fibers of type
I$_6$ -- I$_5$ --  I$_4$ --  I$_3$ --  I$_2$ -- I$_0$.
On the rational elliptic end component $Y^5$, at $s=e_5=0$, the K3-vanishing orders   ${\rm ord}_{Y_0}(f,g,\Delta) |_{s=e_5=0}= (4,4,8)$ indicate a standard $E_6$ enhancement, along with an additional pair of I$_1$ enhancements away from the intersection with $Y^4$. Since $Y^5$ intersects $Y^4$ in an I$_2$ fiber, according to the logic of Section \ref{sec_Ftheoryinterpret} it contributes a factor of $\hat E_7$ to the total symmetry algebra, whose non-abelian part is
\bea
G_{\infty} = \hat E_7 \oplus A_4   \,.
\eea
This realises a decompactification limit to 9d with non-abelian gauge algebra $E_7 \oplus A_4$ \cite{HetFpaper}.

\subsection{Strictly non-mimimal degenerations of Kulikov Type I or II}  \label{subsec_Class5ex}

We now present examples of degenerations of the form (\ref{supernonmin}) and illustrate the birational transformations required to turn them into Kulikov form.
We will present examples where the respective Kulikov models are of Type I (finite distance) or of Type II (infinite distance).

As our first example, consider the family of K3 surfaces with Weierstrass data
\be
\begin{split}   \label{degen-Class5}
f &=  s^4 t^3 ( s + u \,  t ) \,,  \cr
g &=   u \, s^5 t^6 ( s  + u \,  t  ) \,.
\end{split}
\ee
For $u \neq 0$, the K3 surface exhibits a Kodaira Type II$^\ast$ ($E_8$) singularity at $s =0$ and a Kodaira Type III$^\ast$ ($E_7$) singularity
 at $t=0$. For $u =0$, the $E_8$ singularity becomes non-minimal, with 3-fold vanishing orders
\bea
{\rm ord}_{\widehat {\cal Y}}(f,g,\Delta) |_{s=u=0}= (5,7,14)  \,.
\eea
The non-minimality is removed after two blowups and we obtain a degenerate surface defined by the Weierstrass data
\be
\begin{split}    \label{Class5blowupnK1}
f &=e_1 e_2 s^4 t^3 (e_2 s + e_0 t)   \,, \cr
g &= e_0 e_1 s^5 t^6  (e_2 s + e_0 t) \,,
 \cr
 \Delta &=   e_1^2 s^{10} t^9 (e_2 s + e_0 t)^2 (4 e_1 e_2^4 s^3 + 4 e_0 e_1 e_2^3 s^2 t  + 
   27 e_0^2 t^3)      \,.
\end{split}
\ee
The two end components $Y^0$ and $Y^2$ are both rational elliptic surfaces, but unlike in a Type II or III Kulikov model, the middle component has generic fibers of Kodaira Type II,
as can be read off from the vanishing orders ${\rm ord}_{\widehat {\cal Y}}(f,g,\Delta) |_{e_1=0}= (1,1,2)$. 
The K3-vanishing orders on the end component $B^2$ of the base reveal an $E_8$ singularity at $s=0$, where ${\rm ord}_{\rm K3}(f,g,\Delta) |_{s=e_2=0}= (\infty,5,10)$, and an $E_7$ singularity at $t=0$ on $B^0$ with ${\rm ord}_{\rm K3}(f,g,\Delta) |_{t=e_0=0}= (3,\infty,9)$.
On $B^1$, the discriminant reveals altogether 4 singular fibers in codimension one, two of which sit at the intersection point $B^0 \cap B^1$ (together with an additional special fiber brought in from $B^0$), while the other two combine into a Kodaira Type II fiber in the interior of $B^1$.

  We first blow down the two end components $Y^0$ and $Y^2$ by setting $e_0 \equiv 1$ and $e_2 \equiv 1$.
 The resulting degenerate Weierstrass model still exhibits a Type II singularity over generic points of the single base component $B^1$.
 To remove this singularity, we perform a base change $e_1 \to e_1^6$, followed by a rescaling $(f,g,\Delta) \to (e_1^{-4} f, e_1^{-6} g,e_1^{-12} \Delta)$, which does not affect the Calabi-Yau
 condition.\footnote{Note that after blowing down $B^0$ and $B^2$, the coordinate $e_1$ is simply the parameter of the family of Weierstrass models and hence defines a trivial class on the non-compact Calabi-Yau threefold given by this family.}
 As a result, we obtain an elliptic fibration which we still denote, by abuse of notation, as $Y^1$ over base $B^1$ and which is described by the central fiber of the family of Weierstrass models
  \be
\begin{split}  
f &=e_1^2  s^4 t^3 (s + t)   \,, \cr
g &=  s^5 t^6  (s + t) \,,
 \cr
 \Delta &= s^{10} t^9 (s + t)^2 (4 e_1^6 s^3 + 4 e_1^6 s^2 t + 
   27 t^3 )      \,.
\end{split}
\ee
The vanishing orders ${\rm ord}_{Y^1}(f,g,\Delta) |_{s=0}= (\infty,5,10)$ indicate an $E_8$ singularity, but now it is
at $t=e_1=0$ that the singularity is non-minimal since ${\rm ord}_{{\widehat{\cal Y}}'}(f,g,\Delta) |_{t=e_1=0}= (5,6,12)$ in the blown-down configuration $\widehat{\cal Y}'$.
The appearance of this non-minimal singularity could be anticipated from the fact that in the 3-component surface $Y_0$, a total of 3 singular fibers are localised at the intersection
$B^0 \cap B^1$; blowing down $B^0$, which away from this point already exhibits an $E_7$ singularity at $t=0$, hence leads to a rank-3 enhancement of $E_7$, which necessarily implies a non-minimal fiber.

\begin{figure}[t!]
\centering
\includegraphics[width=15cm]{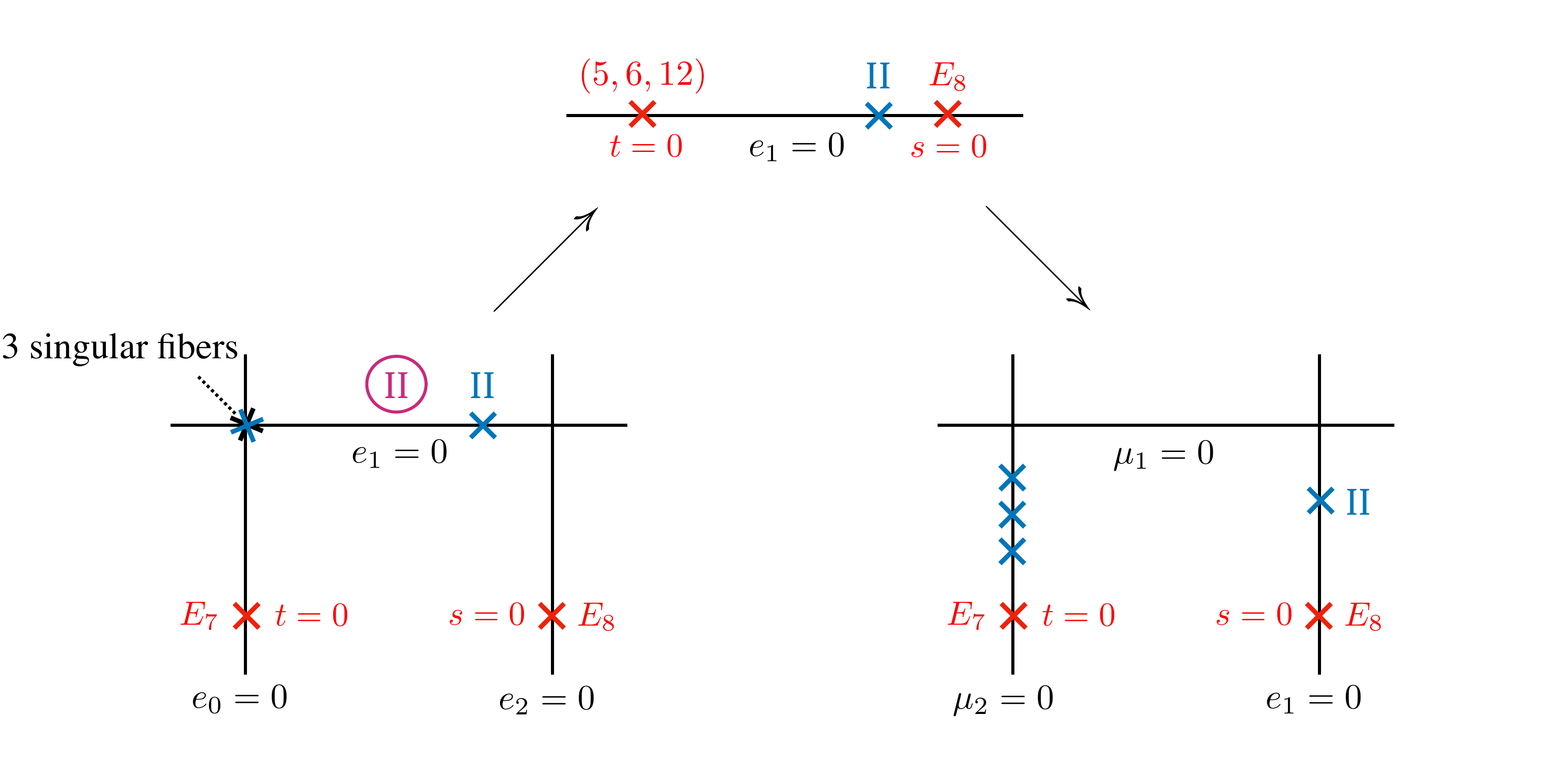}
\caption{
Blowup of degeneration (\ref{degen-Class5}) into the non-Kulikov Weierstrass model (\ref{Class5blowupnK1})(left), followed by blowdowns and base change (top) and final blowup into a Type II Kulikov Weierstrass model (right). \label{Fig:E8bey46}}
\end{figure}

According to Theorem \ref{ref-Theorem1}, the non-minimal degeneration must lead to a Weierstrass model associated with a Kulikov model of Type II.
Indeed, after blowing up at $t = e_1=0$ by performing $(t,e_1) \to (t \mu_1, e_1 \mu_1)$ and rescaling, and then also performing $(t, \mu_1) \to (t \mu_2, \mu_1 \mu_2)$ and rescaling, one arrives at the Weierstrass model
  \be
\begin{split}  
f &=  e_1^2 \mu_1 s^4 t^3 (s + \mu_1 \mu_2^2 t)    \,, \cr
g &=  s^5 t^6 (s + \mu_1\mu_2^2t)
 \cr
 \Delta &=  s^{10} t^9 (s + \mu_1 \mu_2^2t)^2 (27 t^3 + 4  e_1^6 \mu_1^3 s^3+ 4 e_1^6  \mu_1^4 \mu_2^2  s^2 t )     \,.
\end{split}
\ee
We recognise two rational elliptic surfaces at the end, each intersecting a trivial fibration of smooth elliptic curves, which is birational to a Type II.a Kulikov model.
The special singular fibers in the various configurations of this model are depicted in Fig.~\ref{Fig:E8bey46}.

\begin{figure}[t!]
\centering
\includegraphics[width=15cm]{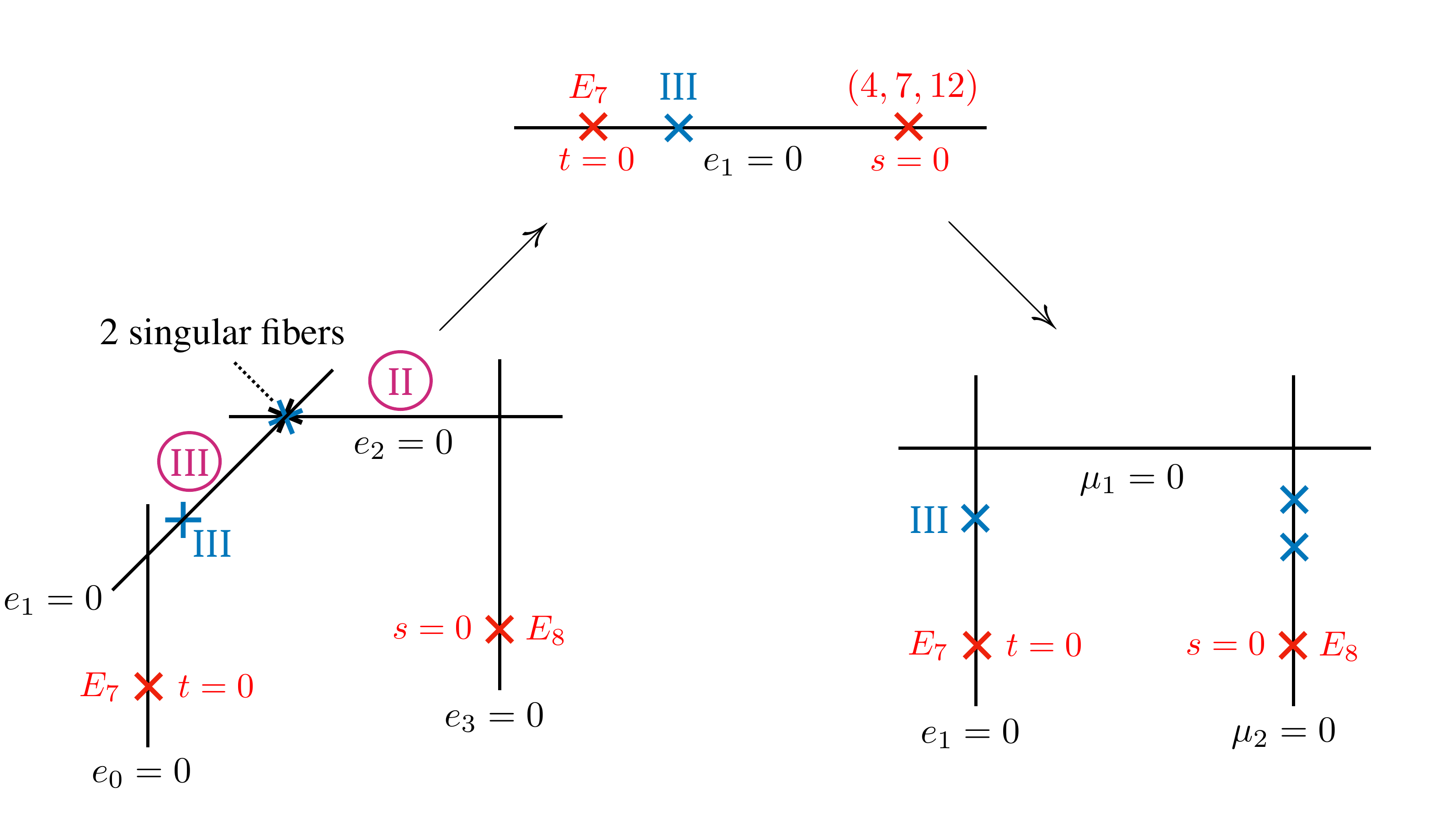} 
\caption{
Blowup of degeneration (\ref{Class5degen2}) into the non-Kulikov Weierstrass model (\ref{Class5blowupnK12})(left), followed by blowdowns and base change (top) and final blowup into a Type II Kulikov Weierstrass model (right).  \label{Fig:E7bey46}}
\end{figure}

A similar, but more complicated example is to start with
  \be
\begin{split}  \label{Class5degen2} 
f &=  s^4 t^3 (s +  u\, t)\,, \cr
g &= u\,  s^5 t^5  ( s^2 +  u \, s t + u^2 \, t^2 ) 
\end{split}
\ee
with an $E_7$ singularity at $t=0$ and an $E_8$ singularity at $s=0$ for $u \neq 0$. The latter enhances to a non-minimal singularity with
\bea
{\rm ord}_{\widehat {\cal Y}}(f,g,\Delta) |_{s=u=0}= (5,8,15)  \,.
\eea
This time, three blowups are required to remove the non-minimality, leading to
 \be
\begin{split}  \label{Class5blowupnK12}
f &=  e_1 e_2 e_3 s^4 t^3 (e_2 e_3^2 s + e_0 t )    \,, \cr
g &= e_0  e_1^2 e_2 s^5 t^5 (e_2^2 e_3^4 s^2 + e_0 e_2 e_3^2 s t + 
   e_0^2 t^2)   \,,
 \cr
 \Delta &=   e_1^3 e_2^2  \Delta'     \,.   
\end{split}
\ee
We observe Kodaira Type III and Type II fibers over generic points of $B^1$ and $B^2$, respectively. Apart from the original $E_7$ and $E_8$ singularities at $t=0$ and $s=0$ on the end components, respectively,
we find 2 singular fibers at the intersection $B^1 \cap B^2$ along with 3 more singular fibers on $B^1$ away from this point.
To bring the model in Kulikov form, we blow down all components except $B^1$ by setting $e_0 \equiv 1$, $e_2 \equiv 1$ and $e_3 \equiv 1$ and perform a base change $e_1   \to e_1^4$.
After rescaling  $(f,g,\Delta) \to (e_1^{-4} f, e_1^{-6} g,e_1^{-12} g)$, we obtain
 \be
\begin{split}  
f &=   s^4 t^3 (s + t)    \,, \cr
g &= e_1^2 s^5 t^5 (s^2 + s t + t^2)    \,.
\end{split}
\ee
The non-minimal singularity ${\rm ord}_{\widehat {\cal Y}'}(f,g,\Delta) |_{s=e_1=0}= (4,7,12)$ indicates that a blowup to a Type II Kulikov model must be possible. Indeed, two blowups lead to a chain of surfaces birational to a Type II.a Kulikov model. See Fig.~\ref{Fig:E7bey46} for the pictorial description of the special fibers in the various configurations.

As our final example, we illustrate how to bring a degeneration with strictly non-minimal singularities into Kulikov Type I form.
To this end, we start with
 \be  \label{812ex1}
\begin{split}  
f &=   u^8  t^8 - 3 s^8 \,, \cr
g &=u^{12}  t^{12} - 2 s^{12} \,,  \cr
 \Delta &=  t^8 u^8 (108 s^{16} - 108 s^{12} t^4 u^4 - 36 s^8 t^8 u^8 + 31 t^{16} u^{16})       \,,
\end{split}
\ee
which for $u\neq 0$ exhibits a Kodaira I$_8$ singularity at $t=0$ and Kodaira I$_1$ fibers at 16 additional points. In the limit $u =0$, the degeneration suffers from a non-minimal singularity with vanishing orders
\bea
{\rm ord}_{\widehat{\cal Y}}(f,g,\Delta) |_{s=u=0}= (8,12,24)  \,.
\eea
We perform one blowup and rescale the Weierstrass data $(f,g,\Delta) \to (e_1^{-4} f, e_1^{-6} g, e_1^{-12} \Delta)$, leading to
 \be
\begin{split}  
f &= e_1^4 (e_0^8 t^8 -3 s^8)      \,, \cr
g &= e_1^6 (e_0^{12} t^{12} - 2 s^{12})  \,, \cr
 \Delta &= e_0^8 e_1^{12} t^8 (108 s^{16} - 108 e_0^4 s^{12} t^4  - 36 e_0^8 s^8 t^8  + 
   31 e_0^{16} t^{16} )      \,.
\end{split}
\ee
At this stage non-minimal fibers arise over generic points of $B^1$. Next, we perform a blowdown of $B^0$ by setting $e_0 \equiv 1$ and rescale $(f,g,\Delta) \to (e_1^{-4} f, e_1^{-6} g, e_1^{-12} \Delta)$. This then leads us to a trivial family of Weierstrass models
in which all dependence on the parameter $e_1$ has dropped out:
\be
\begin{split}  
f &= t^8  -3 s^8   \,, \cr
g &= t^{12} - 2 s^{12} \,, \cr
 \Delta &= t^8 (108 s^{16} - 108 s^{12} t^4 - 36 s^8 t^8 + 31 t^{16})    \,.
\end{split}
\ee
We recover the same singularity structure as in the original family~\eqref{812ex1} away from $u = 0$.
The family of K3 surfaces is therefore of Kulikov Type I and hence the degeneration $u\to 0$ in (\ref{812ex1}) lies at {\it finite} distance despite the appearance of a non-minimal singularity.

\section{Conclusions and Outlook} \label{sec_Conclusions}

In this article we have obtained a refined classification of the infinite distance limits in the complex structure moduli space of elliptic K3 surfaces.
As established in the seminal works \cite{MumfordToroidal,Kulikov1,Kulikov2,PerssonPink,FriedmanMorrison}, all complex structure degenerations of K3 surfaces can be brought into the form of a Kulikov model, by birational degenerations and base changes if necessary. Degenerations of 
Kulikov Type I lie at finite distance in moduli space and those of Type II and Type III at infinite distance.
For elliptically fibered K3 surfaces, the Type II Kulikov models admit two canonical forms \cite{Clingher:2003ui}, which we reviewed in 
Section \ref{sec_KulikovTypeII}: In limits of Type II.a, the K3 surface degenerates into a pair of rational elliptic surfaces intersecting over an elliptic curve, while Kulikov models of Type II.b correspond to a union of two rational fibrations
intersecting over a common bisection.

The main result of this article is that a similarly canonical form exists also for the Weierstrass models associated with Type III Kulikov models, i.e. for the geometry
obtained by blowing down all exceptional curves in the elliptic fiber.
The refined classification holds modulo birational equivalences and base change and is, to the best of our understanding, in agreement with the independent recent analysis in \cite{alexeev2021compactifications,Brunyantethesis}.

In Section \ref{sec_TypeIIIellipticclass} we have explained that all degenerate Weierstrass models underlying Kulikov models of Type III correspond to chains of surface components, $Y_0 = \cup_{i=0}^P Y^i$. Each surface $Y^i$ is a Weierstrass model whose generic fibers can only be of
Kodaira Type I$_{n_i}$. For the intermediate components $Y^i$,  $i= 1, \ldots, P-1$, in the chain, the generic fibers are of Type I$_{n_i >0}$. The classification then distinguishes between the generic fiber types
for the end components $Y^0$ and $Y^P$: If at least one of the end components has generic fibers of Type I$_0$, then the model is said to be of Type III.a, depicted in Figure \ref{fig:TypeIIIageneral}, while if both end components 
have generic fibers of Type I$_{n_i>0}$, the model is of Type III.b as illustrated in Figure \ref{fig:TypeIIIbgeneral}.

We have shown that the canonical Weierstrass forms of the Type II and Type III elliptic Kulikov models can be engineered as blowups  of degenerate Weierstrass models in which
the fiber either degenerates everywhere, or in which a non-minimal singularity occurs over one or more isolated special points of the base.
In the first case, the resulting degeneration is of Type II.b, while the correspondence between the non-minimal singularities and the remaining canonical forms is the content of Theorems \ref{ref-Theorem1} and \ref{ref-Theorem2}. Furthermore, it can always be arranged that the special fibers are localised exclusively over interior points of the components of the base, as guaranteed by Theorem \ref{nointersectiontheorem}. This is important in order to establish a unique interpretation of the Kodaira type of the special fibers.

The proofs for these claims are collected in the technical appendices, which systematically explain the blowup procedure and the relation between the degrees of non-minimality and the refined Kulikov type of the Weierstrass models after blowup in full generality.
We have illustrated the different types of Kulikov Weierstrass models and the blowups leading to them via concrete examples in Section \ref{sec_Examples}. In particular we have exemplified, in Section \ref{subsec_Class5ex}, how strictly non-minimal Weierstrass singularities
can in some cases lie at finite distance (the birational model being of Kulikov Type I), and in others at infinite distance.

Part of the motivation behind our present analysis is to understand the physics of F-theory compactified to eight dimensions on an elliptic K3 surface, in infinite distance limits of the
complex structure. 
The Swampland Distance Conjecture \cite{Ooguri:2006in} predicts that at infinite distance in moduli space, an infinite tower of particles should become asymptotically massless.
According to the Emergent String Conjecture \cite{Lee:2019wij}, the towers must be either Kaluza-Klein states, signalling a decompactification on the boundary of moduli space to a higher
dimensional theory, or excitations of a unique asymptotically weakly coupled emergent critical string.
An explicit confirmation of this conjecture in the complex structure moduli space requires a detailed account of the geometry of infinite distance degenerations.
In the companion paper \cite{HetFpaper}, briefly summarised here in Section \ref{sec_Ftheoryinterpret}, we build on
the geometric understanding obtained in this work to establish a satisfactory physics picture for the infinite distance limits of F-theory compactifications to eight dimensions:
Complex structure degenerations of Type II.a and II.b correspond to dual decompactifications from 8d to 10d \cite{Morrison:1996na,Morrison:1996pp} and, respectively, to weak coupling limits of the Type IIB string theory in 8d \cite{Aspinwall:1997ye,Sen:1996vd,Clingher:2012rg}.
Kulikov models of Type III.a, on the other hand, are found to encode partial decompactifications from 8d to 9d, as signalled by an affine enhancement of the symmetry algebra.
Finally, models of Type III.b represent weak coupling limits superimposed with a limit of infinite complex structure of the Type IIB orientifold torus; such limits are decompactification limits from 8d to 10d.
That the canonical forms of the Type II and Type III Kulikov models are in such clear correspondence with the possible types of infinite distance limits of F-theory compactifications to eight dimensions is a very non-trivial confirmation of the Emergent String Conjecture.

\subsection*{\bf Acknowledgements}
We thank Wolfgang Lerche for collaboration on the companion paper  \cite{HetFpaper} and numerous important discussions. We furthermore gratefully acknowledge insightful exchanges and correspondence with Adrian Clingher and with Antonella Grassi, whom we also thank for comments on the draft.
The work of S.-J.L. is supported by IBS under the project code, IBS-R018-D1.
The work of T.W. is supported in part by Deutsche Forschungsgemeinschaft under Germany's Excellence Strategy EXC 2121 Quantum Universe 390833306.

\appendix
\section{Non-minimal Degenerations and their Blow-ups}  \label{app_Non-minimal}

In this technical appendix we explain how K3 degenerations leading to non-minimal Kodaira fibers can be blown up into Kulikov Weierstrass models.
We begin in Appendix \ref{nomen} with a careful definition of our notation and collect the main properties of the degenerate Weierstrass models obtained from non-minimal singularities via blowups.
We must distinguish five classes (\ref{class1-5}) of non-minimal enhancements. The blowup chains required to remove the non-minimal singularities for each class are discussed in detail in Appendix \ref{construction}.
In Appendix \ref{proofs} we present the proofs of our claims summarised in Appendix \ref{nomen}.

\subsection{Nomenclature and summary}\label{nomen}

Our starting point is a Weierstrass model over $\IP_{[s:t]}^1$,
\beq\label{original}
y^2 = x^3 + f_u (s, t) x z^4 + g_u(s,t) z^6 \,,
\eeq
which describes a family $\widehat{\mathcal Y}$ of K3 surfaces $\hat Y_u$ degenerating for $u=0$. 
Note that we even allow for degenerations under which the generic elliptic fiber of $Y_0$ becomes singular. 
In this case, an appropriate base change, $u\to u^k$, followed by a rescaling of the Weierstrass sections 
\be
(f_u, g_u) \to (\lambda^4 f_u,  \lambda^6 g_u)
\ee
for suitable $\lambda$, can turn the generic elliptic fiber of $\hat Y_0$ into a Type I$_{n_0}$ fiber with $n_0 \geq 0$. Without loss of generality, we can therefore assume
that the generic fiber of $Y_0$ is of this type.

Unless noted otherwise, the subscript $u$ in $f_u$ and $g_u$ will often be omitted and we write instead
\bea
f (s,t;u)&:=& f_u(s,t) \,, \\
g (s,t;u)&:=& g_u(s,t) \,,
\eea
with discriminant 
\beq
\Delta(s,t;u) := \Delta_u(s,t)=4f^3 + 27 g^2\,.
\eeq
We start by assuming that the elliptic fiber at $s=0$ of the generic surface $\hat Y_{u\neq0}$ is of a minimal Kodaira type with K3-vanishing orders
\beq\label{generic-van}
{\rm ord}_{\hat Y_{u}} (f,g,\Delta)|_{s=0}:=(v_{s=0}(f_u), v_{s=0}(g_u), v_{s=0}(\Delta_u)) = (a,b,c) \,,
\eeq
for a fixed generic $u \neq 0$.
Explicitly, we expand
\bea\label{foriginal}
f &=& s^a \,\sum_{i=0}^{8-a} \,\mathcal{F}_{i} (u) \,s^i\, t^{8-a-i}  \,,\\ 
g &=& s^b \,\sum_{j=0}^{12-b} \,\mathcal{G}_{j} (u) \,s^j\, t^{12-b-j}  \,,\\ 
\Delta &=& s^c \,\sum_{k=0}^{24-c} \,\mathcal{D}_{k} (u) \,s^k\, t^{24-c-k} \,,
\eea
with expansion coefficients $\mathcal{F}_i$, $\mathcal{G}_j$ and $\mathcal{D}_k$. 
Let us suppose, however, that for $u=0$ the fiber type at $s=0$ becomes non-minimal, with 3-fold-vanishing orders\footnote{We implicitly assume that the fibers are minimal away from $s=0$ on the curve $u=0$; otherwise, we simply repeat the set of operations that we describe in this appendix to the other non-minimal enhancements.}
\beq\label{non-minimal-general}
{\rm ord}_{\widehat{\mathcal Y }} (f,g,\Delta)|_{(u,s)=0} := (4+\alpha, 6+\beta, 12+\gamma)\,, \quad \alpha, \beta, \gamma \geq 0\,.
\eeq
Here, the vanishing orders are read off from the perspective of the 3-fold $\widehat {\mathcal Y}$, by setting both $u$ and $s$ to $\mu$ and extracting the leading powers of $\mu$ as in (\ref{orders6d-2}).

In this appendix, we will analyze how configurations containing such a non-minimal fiber in the limit can be systematically and explicitly transformed into an appropriate Kulikov Weierstrass model without non-minimality. To this end, we will have to go through a certain chain of birational transformations (upon first performing an appropriate base change if needed). 

\subsubsection*{Blowup Chain}\label{ChainP}
In order to remove the non-minimal singularity~\eqref{non-minimal-general} we perform a chain of $P$ blowups. Here, the $p^{\rm th}$ blowup with $1\leq p \leq P$ is the operation that transforms the sections $f$, $g$ and $\Delta$ as
\bea
f (s,t; e_0, \ldots, e_{p-1})&\longrightarrow& f (s,t; e_0, \ldots, e_{p})\,, \\
g(s,t; e_0, \ldots, e_{p-1}) &\longrightarrow& g(s,t; e_0, \ldots, e_{p}) \,, \\
\Delta(s,t; e_0, \ldots, e_{p-1}) &\longrightarrow& \Delta (s,t; e_0, \ldots, e_p) \,,
\eea
by replacing the coordinates as
\bea \label{rep}
e_{p-1} &\longrightarrow& e_{p-1}\, e_{p} \,,\\
s &\longrightarrow& s \,e_p \,, 
\eea
and then dividing the triplet of sections by $e_{p}^4$, $e_p^6$ and $e_{p}^{12}$, respectively. For convenience, we rename the original parameter $u$ as $e_0$ after the first blowup, i.e., the coordinate replacement~\eqref{rep} for the very first blowup is given as 
\bea
u &\longrightarrow& e_0 \,e_1 \,.
\eea
After performing the chain of $P$ blowups as described above, the Weierstrass sections are expanded as
\bea\label{fp}
f &=& s^a \,\sum_{i=0}^{8-a} \,\mathcal{F}_{i} (e_0, \cdots, e_P) \,s^i\, t^{8-a-i}  \,,\\ \label{gp}
g &=& s^b \,\sum_{j=0}^{12-b} \,\mathcal{G}_{j} (e_0, \cdots, e_P) \,s^j\, t^{12-b-j}  \,,\\ 
\Delta &=& s^c \,\sum_{k=0}^{24-c} \,\mathcal{D}_{k} (e_0, \cdots, e_P) \,s^k\, t^{24-c-k} \,,
\eea
where, by abuse of notation, we keep using the same symbols for the sections $(f, g, \Delta)$, as well as for their expansion coefficients $(\mathcal F_i, \mathcal G_j, \mathcal D_k)$, before and after the blowup operations.  
In our later analysis a particularly important role will be played by the vanishing orders of $(\mathcal F_i, \mathcal G_j, \mathcal D_k)$, which we denote as 
\bea\label{ordmu}
\mu_{p,i}&:=& v_{e_p=0} (\mathcal F_i) \,,\\ \label{ordnu}
\nu_{p,j} &:=&v_{e_p=0} (\mathcal G_j) \,,\\ \label{ordrho}
\rho_{p,k} &:=& v_{e_p=0} (\mathcal D_k) \,.   
\eea
This indicates that the expansion coefficients after the blowups take the following forms,
\bea\label{expfp}
\mathcal{F}_i (e_0, \cdots, e_P) &=& A_{i}  \prod_{p=0}^P e_p^{\mu_{p,i}} + \cdots \,,\\ \label{expgp}
\mathcal{G}_j (e_0, \cdots, e_P)&=& B_{j}  \prod_{p=0}^P e_p^{\nu_{p,j}} + \cdots \,,\\ \label{expDp}
\mathcal{D}_k (e_0, \cdots, e_P)&=& C_{k}  \prod_{p=0}^P e_p^{\rho_{p,k}} + \cdots \,,
\eea
where $A_{i}$, $B_{j}$ and $C_{k}$ are complex constants and the ellipses only contain monomials of a strictly higher degree in each of the $P+1$ coordinates $e_p$ than the leading monomials. 

One can see from the definition of the blowup operations that the vanishing orders~\eqref{ordmu},~\eqref{ordnu} and~\eqref{ordrho} obey the relations
\bea \label{recmu}
\mu_{p,i} &=& \mu_{0,i} - p(4-a-i) \,, \\ \label{recnu} 
\nu_{p,j} &=& \nu_{0,j} -p(6-b-j)\,, \\ \label{recrho}
\rho_{p,k} &=& \rho_{0,k} - p(12-c-i)\,. 
\eea
In view of these recursion relations, it is natural to split the index set $\cI$ labeling $\mathcal F_i$, 
\beq
\cI = \{\, i \;|\; A_i \neq 0 \,\} ~\subseteq~  [0, 8-a]\,,
\eeq
into the following three subsets   
\bea\label{group-i}  \nn
\cI_{-} &=& \{\, i \in \cI \;|\; 0\leq i < 4-a \,\}\,,\\ 
\cI_0 &=& \{\,i \in \cI \;|\; i=4-a \,\}\,,\\ \nn 
\cI_+ &=& \{\, i\in \cI\;|\; 4-a<i \leq 8-a  \,\} \,, 
\eea
and similarly, the index set $\cJ$ labeling $\mathcal G_j$,
\beq
\cJ = \{\, j\;|\; B_j \neq 0  \,\} ~\subseteq~[0,12-b]\,, 
\eeq
into 
\bea\label{group-j} \nn 
\cJ_{-} &=& \{\,  j \in \cJ \;|\; 0\leq j <6-b\,\}\,,\\
\cJ_0 &=& \{\,j \in \cJ \;|\;  j= 6-b \,\}\,,\\ \nn
\cJ_+ &=& \{\, j \in \cJ \;|\; 6-b<j \leq 12-b  \,\} \,.
\eea
Here, the index subsets $\cI_-$ and $\cJ_-$ may be empty if $a \geq 4$ and $b \geq 6$, respectively. However, since the singular fiber at $s=0$ of a generic surface $\hat Y_{u\neq0}$ is of a minimal Kodaira type, we cannot have both $a \geq 4$ and $b \geq 6$ at the same time, and hence, 
\beq
\cI_- \neq \emptyset \qquad \text{or} \qquad \cJ_- \neq \emptyset \,.
\eeq
For $i \in \cI_-$ (respectively, $i \in \cI_0$ and $i \in \cI_+$) 
the relation~\eqref{recmu} implies that 
the leading exponents $\{\mu_{p,i}\}_{p=0}^P$ of the leading monomials of each $\mathcal F_i$ in~\eqref{expfp} are monotonically decreasing in $p$  (respectively, constant and monotonically increasing), i.e., 
\bea
\mu_{0,i} > \mu_{1,i} > \cdots > \mu_{P, i} \quad &\text{for}& i \in \cI_- \,,\\
\mu_{0,i} = \mu_{1,i} = \cdots = \mu_{P, i}  \quad &\text{for}& i \in \cI_0 \,,\\
\mu_{0,i} < \mu_{1,i} < \cdots < \mu_{P, i}  \quad &\text{for}& i \in \cI_+ \,.
\eea
Similarly, the exponents $\nu_{p,j}$ and $\rho_{p,k}$ appearing in $\mathcal G_j$ and $\mathcal D_k$ exhibit an analogous monotonic behaviour, with transitions  at $j=6-b$ and $k=12-c$, respectively.

Let us now note that in order for the $(p+1)^{\rm th}$ blowup to be required the exponents of $e_p$ for $p \leq P-1$ must obey
\bea\label{ineqmu} 
\mu_{p,i} \geq 4-a-i \quad  &\text{for}&i \in \cI_-  \,, \\ \label{ineqnu} 
\nu_{p,j} \geq 6-b-j \quad &\text{for}&j \in \cJ_-  \,. 
\eea
Furthermore, since the blowup procedure is to end at the $P^{\rm th}$ step, there must exist either $i_0\in \cI_-$ with 
\beq\label{final-i}
0 \leq \mu_{P,i_0} < 4-a-i_0 \,, 
\eeq
or $j_0 \in \cJ_-$ with 
\beq\label{final-j}
0\leq \nu_{P, j_0} < 6-b-j_0 \,.
\eeq 

For brevity, the $P+1$ rational base components after the $P$ blowups will be denoted as 
\beq
B^p :=\{e_p =0 \}\,,\quad \text{for}~ 0\leq p \leq P\,, 
\eeq
and the $P$ intersection points of the adjacent component pairs as
\beq
B^{p,p+1} : = B^p \cap B^{p+1} \,,\quad\text{for}~0\leq p <P \,.
\eeq
Note that without loss of generality we will assume 
\beq
P >1\,,
\eeq 
since degenerations can be defined up to base changes.\footnote{Should the blowups end already after the first step, we can modify the initial Weierstrass data e.g. by the order $2$ base change $u \to u^2$  so that the non-minimal fiber of the base-changed configuration requires $P>1$ blowups.} 

\subsubsection*{Classification Scheme}\label{scheme}
In order to distinguish the geometries after the blowups from the ones before, we will use symbols without hats for the former. Specifically, we denote by $\mathcal Y$ the 3-fold model obtained by applying the $P$ blowups to $\widehat{\mathcal Y}$, and by $Y_0$ the total transform of $\hat Y_0$. 

While by ~\eqref{non-minimal-general} the 3-fold-vanishing orders of $(f, g)$ at $s=0$ are initially non-minimal, 
\beq\label{nonmin}
{\rm ord}_{\widehat{\mathcal Y}} (f)|_{(u,s)=0} = 4 +\alpha \geq 4  \quad \text{and}\quad {\rm ord}_{\widehat{\mathcal Y}} (g)|_{(u,s)=0} =6+\beta \geq 6  \,, 
\eeq
the $P$ blowups render them minimal,
\beq
{\rm ord}_{\mathcal Y} (f)|_{(e_P,s)=0}  < 4 \quad \text{or}\quad {\rm ord}_{\mathcal Y} (g)|_{(e_P,s)=0}  <6 \,, 
\eeq
as reflected by~\eqref{final-i} and~\eqref{final-j}.
For an exhaustive and systematic analysis of non-minimal fibers subject to~\eqref{non-minimal-general}, we first group them into the following five classes, 
\beq\label{class1-5}
\begin{array}{ll}
\textrm{Class 1} & \qquad \alpha = 0\,,~\beta =0\,,~\gamma =0 \,,  \\
\textrm{Class 2} & \qquad   \alpha > 0\,,~\beta =0\,,~\gamma =0 \,,\\
\textrm{Class 3} & \qquad    \alpha = 0\,,~\beta > 0\,,~\gamma =0 \,, \\
\textrm{Class 4} & \qquad    \alpha = 0\,,~\beta =0\,,~\gamma > 0 \,,\\
\textrm{Class 5} & \qquad \alpha > 0\,,~\beta >0\,,~\gamma >0 \,.   
\end{array}
\eeq
Note that certain non-minimal fibers with either $\alpha$ or $\beta$ vanishing may in fact turn into a Class 5 fiber by base changes of a large enough order $k$,
\beq
u \to u^k \,. 
\eeq 
In order for  the classification scheme to be invariant under the initial base change, we therefore classify non-minimal fibers of this kind as Class 5. 
With this definition, the Weierstrass section $f$ in any model with non-minimality of Class 1, 3, or 4 ($\alpha=0$) at $s=0$ contains a non-trivial term of bi-degree $(4,4)$ in $[s:t]$, i.e., 
\beq\label{Amiddle}
A_{4-a} \neq 0 \,,  
\eeq
and similarly, the section $g$ in models with a Class 1, 2, or 4 fiber ($\beta=0$) have
\beq\label{Bmiddle}
B_{6-b} \neq 0 \,.
\eeq

\subsubsection*{Summary}\label{summary}
The aim of this appendix is to provide an explicit method for constructing Kulikov Weierstrass models of Type II/III associated with non-minimal fibers and to describe the resulting brane configurations in the internal space of F-theory. To this end, our main focus will be on non-minimal fibers of Class 1--4, which will indeed be shown to result in a Type II or III Kulikov model $\mathcal X$ via the chain of $P$ blowups defined in section~\ref{ChainP}. In fact, the 3-fold model~$\mathcal Y$ obtained by such base blowups turns out to be the Weierstrass model of~$\mathcal X$, as defined in~\eqref{XtoYdef}. In particular, the degenerate K3 surface $Y_0$ in the family $\mathcal Y$,
\beq\label{decompY}
Y_0 = \cup_{p=0}^P Y^p \,,
\eeq
results from contracting the exceptional fibral curves of the degeneration $X_0$ described by $\mathcal X$.\footnote{\label{non-generic-limit}The degenerate surface $Y_0$ may in principle contain a non-minimal fiber from the K3 surface perspective, in which case $Y_0$ is not (the fibral blow-down of) a Kulikov model from the mathematical point of view nor a suitable F-theory background from the physical point of view. While such a situation only arises from a highly non-generic limit, if it occurs one has to further modify the model by base change and/or additional blowups to eventually reach a Kulikov Weierstrass model; we discuss the details of this procedure in Proposition~\ref{prop6}. For simplicity of presentation, however, most of the time we will regard the model $\mathcal Y$ and the fibral blowdown of $\mathcal X$ identical, as is indeed the case for a generic limit.}

By construction, the base curves $B^p$ of the surface components $Y^p$ in the decomposition~\eqref{decompY} form a chain,
\beq\label{Bchain}
B^0 - B^1 - \cdots - B^{P-1} - B^P \,. 
\eeq
Furthermore, as will be shown in Appendix~\ref{construction}, the codimension-zero fibers over $B^p$ will be of Type I$_{n_p}$ and hence, the surfaces $Y^p$ will be referred to as  
\beq\label{In-comp}
Y^p\;:\; {\rm I}_{n_p}\text{-components} \,. 
\eeq
We will then proceed to also show that the vanishing orders $n_p$ are constrained as
\bi
\item Class 1--3 \beq\label{np-C123} n_p = 0 \quad \text{for}~~ 0 \leq p \leq P \,; \eeq
\item Class 4 \beq\label{np-C4} n_p > 0 \quad \text{for}~~ 0 < p < P \,. \eeq
\ei
The constraints~\eqref{np-C123} and~\eqref{np-C4} indicate that the associated Kulikov models are of Type II and III for non-minimal fibers of Class 1--3 and Class 4, respectively.  

For fibers of Class 5, however, while the blowups still lead to a decomposition~\eqref{decompY} over the base~\eqref{Bchain}, the generic fibers over the components $B^p$ in the middle (i.e., for $0<p<P$) can never be of Type I$_n$. Of course, this does not imply that all Class 5 fibers necessarily result in a Type I Kulikov model. Rather, as exemplified in section~\ref{subsec_Class5ex}, some Class 5 fibers can be transformed into Kulikov Weierstrass models of  Type I, while others cannot. In the latter case, however, it must be possible to bring the model into a degeneration of Class 1--4. This will be explained in Appendix~\ref{Class5-TypeI}. In any event our study of non-minimal fibers at infinite distance does not lack any generality if we focus only on Class 1--4 limits, justifying our restriction in this appendix.

In the following we will present our main claims concerning the brane configuration of the degeneration~\eqref{decompY} associated with the non-minimal fibers, which can occur at infinite distance, i.e., those of Class 1--4. In the interest of readability, all the proofs are relegated to Appendix~\ref{proofs}, to which we refer for details.

First, in the geometry of the decomposition~\eqref{decompY}, we observe that fibers of Type I$_n$ are guaranteed to arise not only at codimension-zero as in~\eqref{In-comp} but also at the
intersection loci
\beq
B^{p,p+1} = B^p \cap B^{p+1} \,,\quad p=0, \cdots , P-1 \,,
\eeq 
of the adjacent components in the chain.   

\vspace{.2\baselineskip}
{\proposition\label{prop1} The adjacent surface components $Y^{n_p}$ and $Y^{n_{p+1}}$ intersect in an $I_{n_{p,p+1}}$ fiber over $B^{p,p+1}$. Furthermore, for Class 1--3 fibers, $n_{p,p+1}=0$ for all $p=0, \cdots, P-1$.}

{\proof.} See Proof~\ref{proof1} in Appendix~\ref{proofs}.\\

\noindent 
As will shortly become clearer, the fibration $Y^p$ over $B^p$ will exhibit rather distinct features depending on whether $0<p<P$ or $p=0, P$. We will call the former fibrations {\it middle} components and the latter {\it  end} components. When necessary, we will use the terminology {\it left (right, resp.) end} for $p=0$ (resp., $p=P$), having in mind the chain structure~\eqref{Bchain}. 

\vspace{.2\baselineskip}
{\proposition\label{prop2} For blowups of Class 4 degenerations, the generic codimension-zero $I_{n_p}$ fibers are split (in the sense of the generalised Kodaira classification \cite{Grassi:2000we,Grassi:2011hq}) for the middle components, while both split and non-split forms are allowed at either end.} 

{\proof.} See Proof~\ref{proof2} in Appendix~\ref{proofs}.\\

By now we have discussed the geometry of the elliptic fibers in $Y_0$ characterized by the 3-fold vanishing orders~\eqref{orders6d-2}. However, in the context of F-theory, the actual $7$-brane configuration responsible for the 8d effective physics is encoded in the K3-vanishing orders~\eqref{K3vanishingorderdef} of the special fibers. Upon analyzing the latter, we arrive at

\vspace{.2\baselineskip}
{\proposition\label{prop3} The special fibers located on the middle components can only be of Kodaira Type $I_n$ in the sense of the K3-vanishing orders~\eqref{K3vanishingorderdef}.}  \\

\noindent Note that we refer to such special fibers as giving rise to $A_{n-1}$-singularities in the main text, see  (\ref{Akordersdef}).

{\proof.} See Proof~\ref{proof3} in Appendix~\ref{proofs}.\\

\vspace{.2\baselineskip}
{\proposition\label{prop4} The total number $x_p$ of special fibers (i.e. of 7-branes in F-theory) on the component $B^p$ is constrained as follows:
\bea\label{claim1}
&&x_0 + (n_1-n_0) = 12 \,,\\ \label{claim2}
&&x_P + (n_{P-1}-n_P) = 12\,,\\ \label{claim3}
&&x_p + (n_{p-1} - n_{p}) + (n_{p+1} - n_p) = 0 \,, \quad \forall p = 1, \cdots, P-1 \,.
\eea
}

{\proof.} See Proof~\ref{proof4} in Appendix~\ref{proofs}.\\

\noindent 
In fact, the same set of equations as in Proposition~\ref{prop4} holds even for Class 5 degenerations, for which the definition of $n_p$ is generalized as the maximal overall powers of $e_p$ in $\Delta$. As an immediate corollary of Proposition~\ref{prop4} we also note

\vspace{.2\baselineskip}
{\corollary\label{coro1} For Class 1--3, the fibrations $Y^p \to B^p$ in the middle are trivial $I_0$ fibrations and the two end components, $Y^0$ and $Y^P$, are rational elliptic surfaces.} \\

\noindent The reason is that, with $n_p=0$ for all $p=0, \cdots, P$, the number of branes $x_p$ on each middle component is obtained via~\eqref{claim3} as
\beq
x_p = 0 \quad \text{for}~p=1, \cdots, P-1\,, 
\eeq
which indicates that the generic smooth fibration over $B^p$ is in fact a direct product globally. \\

\noindent 
As a consequence of Proposition~\ref{prop3}, there can never arise special fibers on any of the middle components that violate the condition
\beq\label{phys-nm-cond}
 {\rm ord}_{Y_0}(f)|_\cP <4 \quad \text{or}\quad {\rm ord}_{Y_0}(g)|_\cP <6 \,
 \eeq 
because all special fibers on the middle components are of Kodaira Type I$_n$ in the sense of the K3-vanishing orders.
On the end components, by contrast, such special fibers can a priori not be excluded.
We will refer to those special fibers that do violate~\eqref{phys-nm-cond} as being {\it physically non-minimal} (motivated by the fact that the special fibers correspond to the 7-branes in F-theory), while the special fibers that violate the analogous condition for the 3-fold vanishing orders will still simply be referred to as non-minimal. Regarding this notion of physical non-minimality, we formulate

\vspace{.2\baselineskip}
{\proposition\label{prop6} After performing the blowups required to remove all non-minimal fibers, there may in principle remain physically non-minimal special fibers on the right-end component, either at $s=0$ or at $s\neq0$. These physically non-minimal fibers can then be removed as follows:
\bi
\item [1.]
Physically non-minimal special fibers at $s=0$ in $B^P$ can be avoided by
 applying an appropriate base change to the original single-component model and performing a chain of blowups.
\item [2.]
Physically non-minimal special fibers away from $s=0$ in $B^P$ can eventually be removed by an additional set of blowups. 
\ei
}

{\proof.} See Proof~\ref{proof6} in Appendix~\ref{proofs}.

\subsection{Construction of the Kulikov Weierstrass models as blowups}\label{construction}
As already summarized in Appendix~\ref{summary}, for non-minimal degenerations of Class 1--4, the resulting K3 surface
\beq
Y_0 = \cup^P_{p=0} Y^p \,,
\eeq 
constructed via the $P$ blowups of section~\ref{ChainP}, is the fibral blowdown of the central fiber $X_0$ of a Kulikov model.\footnote{We emphasize once again that there may arise a subtlety in a highly non-generic limit as explained in footnote~\ref{non-generic-limit}.}  In this section we will investigate
the blown-up geometry associated with each of the five non-minimality classes in turn. For Class 1--4, we will thereby confirm that the surfaces $Y^p$ are I$_{n_p}$-components with $n_p$ subject to the constraints~\eqref{np-C123} and~\eqref{np-C4}.

\subsubsection*{Class 1: $(4,6,12)$}\label{class1}

We first recall the expansion~\eqref{fp} of the Weierstrass section $f$ and the expansion coefficients $\mathcal F_i$ therein, where the leading monomials in $e_p$ define the exponents $\mu_{p,i}$ as~\eqref{expfp}. It then follows, essentially from the relations~\eqref{recmu}, that these exponents obey
\bi
\item for $p=0$: 
\beq\label{mu0}
\mu_{0, i} 
\left\{ 
\begin{array}{ l l }
> 4-a-i > 0  & \quad \textrm{if }~ i \in \cI_- \\
= 0 & \quad \textrm{if }~ i \in \cI_0 \\
\geq  0 & \quad \textrm{if }~ i \in \cI_+ \,,
\end{array}
\right.
\eeq

\item for $0< p <P$: 
\beq\label{mu0P}
\mu_{p, i} 
\left\{ 
\begin{array}{ l l }
> 0 & \quad \textrm{if }~ i \in \cI_- \\
= 0 & \quad \textrm{if }~ i \in \cI_0  \\
> 0 & \quad \textrm{if }~ i \in \cI_+ \,,
\end{array}
\right.
\eeq

\item for $p=P$: 
\beq\label{muP}
\mu_{P, i} 
\left\{ 
\begin{array}{ l l }
\geq 0 & \quad \textrm{if }~ i \in \cI_- \\
= 0 & \quad \textrm{if }~ i \in \cI_0  \\
>  0 & \quad \textrm{if }~ i \in \cI_+ \,.
\end{array}
\right.
\eeq
\ei
Similarly, the exponents $\nu_{p,j}$ defined through the leading monomials of $\mathcal G_j$ in~\eqref{expgp} satisfy the following set of analogous conditions:
\bi
\item for $p=0$: 
\beq\label{nu0}
\nu_{0, j} 
\left\{ 
\begin{array}{ l l }
> 6-b-j > 0  & \quad \textrm{if }~ j \in \cJ_- \\
= 0 & \quad \textrm{if }~ j \in \cJ_0  \\
\geq  0 & \quad \textrm{if }~ j \in \cJ_+ \,,
\end{array}
\right.
\eeq

\item for $0< p <P$: 
\beq\label{nu0P}
\nu_{p, j} 
\left\{ 
\begin{array}{ l l }
> 0 & \quad \textrm{if }~ j \in \cJ_- \\
= 0 & \quad \textrm{if }~ j \in \cJ_0  \\
> 0 & \quad \textrm{if }~ j \in \cJ_+ \,,
\end{array}
\right.
\eeq

\item for $p=P$: 
\beq\label{nuP}
\nu_{P, j} 
\left\{ 
\begin{array}{ l l }
\geq 0 & \quad \textrm{if }~ j \in \cJ_- \\
= 0 & \quad \textrm{if }~ j \in \cJ_0  \\
>  0 & \quad \textrm{if }~ j \in \cJ_+ \,.
\end{array}
\right.
\eeq
\ei

Note that the constraints~\eqref{mu0} for $\mu_{0,i}$ imply that 
\beq
a+i + \mu_{0,i} 
\left\{ 
\begin{array}{ l l }
>4 & \quad \textrm{if }~ i \in \cI_- \cup \cI_+ \\
= 4 & \quad \textrm{if }~ i \in \cI_0  \,, 
\end{array}
\right.
\eeq
which, together with the expansion~\eqref{fp} and the leading behaviour~\eqref{expfp}, tells us that the 3-fold-vanishing order of $f$ at $s=0$, 
\beq
{\rm ord}_{\widehat{\mathcal Y}} (f)|_{(u,s)=0} = 4+\alpha = 4 \,,
\eeq
is achieved by the bi-degree $(4,4)$ sector, 
\beq\label{f04-a}
\mathcal F_{4-a} s^4 t^4 = A_{4-a} s^4 t^4 + \cdots \,,
\eeq
while all other terms of $f$ have a higher 3-fold-vanishing order. 
Analogously, we learn from the constraints~\eqref{nu0} for $\nu_{0,j}$ that the 3-fold-vanishing order of $g$,
\beq
{\rm ord}_{\widehat{\mathcal Y}} (g)|_{(u,s)=0}= 6+\beta = 6 \,,
\eeq
is achieved only by the bi-degree $(6,6)$ sector,
\beq\label{g06-b}
\mathcal G_{6-b} s^6 t^6 = B_{6-b} s^6 t^6  + \cdots \,. 
\eeq
Here, the coefficients $A_{4-a}$ and $B_{6-b}$ in~\eqref{f04-a} and~\eqref{g06-b} are non-zero as addressed in~\eqref{Amiddle} and~\eqref{Bmiddle}, respectively. Therefore, the expansions of $f$ and $g$ in each $e_p$ have a non-trivial constant sector, i.e., 
\bea
f_p:=f|_{e_p=0} \not\equiv 0 \,,\quad g_p:=g|_{e_p=0} \not\equiv 0 \,,\qquad \text{for}~~ p=0,\cdots,P\,.
\eea

Focusing now on the discriminant $\Delta$, note that since its vanishing order is
\beq
{\rm ord}_{\widehat{\mathcal Y}} (\Delta)|_{(u,s)=0}= 12+\gamma = 12 \,,
\eeq 
the order $12$ term arising from the cube of~\eqref{f04-a} and the one from the square of~\eqref{g06-b} cannot cancel in $\Delta$, i.e., 
\beq\label{nocancel}
4A_{4-a}^3 + 27 B_{6-b}^2 \neq 0 \,.
\eeq 
We then observe that the expansion of $\Delta$ in each $e_p$ is also equipped with a non-vanishing constant sector, i.e., 
\beq\label{Deltapnon0}
\Delta_p := \Delta|_{e_p=0} \not\equiv 0 \,,\quad \text{for}~~ p =0, \cdots, P \,.
\eeq
This follows from 
\beq\label{follows}
\Delta_p = (4f^3 + 27 g^2) |_{e_p=0} = 
\left\{ 
\begin{array}{ l l }
(4A_{4-a}^3 + 27 B_{6-b}^2) s^{12} t^{12} + \cO(s^{13})  & \quad \textrm{for }~ p =0 \\
(4A_{4-a}^3 + 27 B_{6-b}^2) s^{12} t^{12} & \quad \textrm{for }~ 0<p<P   \\
(4A_{4-a}^3 + 27 B_{6-b}^2) s^{12} t^{12} + \cO(t^{13})& \quad \textrm{for }~ p =P \,,
\end{array}
\right.
\eeq
which, when combined with~\eqref{nocancel}, indicates that the coefficient of $s^{12}t^{12}$ is non-vanishing in $\Delta_p$, guaranteeing the non-triviality~\eqref{Deltapnon0}. We thus learn that $\Delta$ cannot acquire an overall common factor of $e_p$ and hence that $Y^p$ is an I$_0$-component, i.e., 
 \beq
 n_p = 0 \quad \text{for}~~ 0 \leq p \leq P \,. 
 \eeq

\subsubsection*{Class 2: $(>4, 6, 12)$}\label{class2}
The key ideas used in section~\ref{class1} for analyzing fibers of Class 1 apply also to the analysis of Class 2 fibers. In particular, the conditions on the exponents $\mu_{p,i}$ and $\nu_{p,j}$, as summarized in~\eqref{mu0} through~\eqref{nuP}, hold almost unaltered, with the only change being that the vanishing of $\mu_{p,i}$ for $i \in \cI_0$ should be modified to 
\beq\label{mupi>0}
\mu_{p,i} > 0 \,, \quad \forall p~~\text{and}~~ i\in \cI_0 \,. 
\eeq
The reason is that the initial vanishing order of $f$ at $s=0$,
\beq
{\rm ord}_{\widehat{\mathcal Y}} (f)|_{(u,s)=0}= 4 +\alpha > 4 \,, 
\eeq
is strictly bigger than $4$ for Class $2$ non-minimality. It thus follows that the positivity~\eqref{mupi>0} holds for $p=0$, which then propagates via~\eqref{recmu} to all $p >0$. 

This in turn modifies~\eqref{follows} to
\beq\label{follows2}
\Delta_{p} = (4f^3 + 27 g^2) |_{e_p=0} = 
\left\{ 
\begin{array}{ l l }
27 B_{6-b}^2 s^{12} t^{12} + \cO(s^{13})  & \quad \textrm{for }~ p =0 \\
27 B_{6-b}^2 s^{12} t^{12} & \quad \textrm{for }~ 0<p<P   \\
27 B_{6-b}^2 s^{12} t^{12} + \cO(t^{13})& \quad \textrm{for }~ p =P \,,
\end{array}
\right.
\eeq
where the coefficient $B_{6-b}$ on the RHS is a non-zero constant as discussed in~\eqref{Bmiddle}.
Therefore, we have
\beq
\Delta_p \not\equiv 0 \,,
\eeq
and hence $\Delta$ cannot acquire an overall factor of $e_p$ for any $p =0, \cdots, P$, indicating that the surfaces $Y^p$ are I$_0$-components.

\subsubsection*{Class 3: $(4, >6, 12)$}
In much the same way as how the expression~\eqref{follows} for the discriminant of a Class 1 configuration had to change for Class 2 to~\eqref{follows2}, we now have
\beq\label{follows3}
\Delta_{p} = (4f^3 + 27 g^2) |_{e_p=0} = 
\left\{ 
\begin{array}{ l l }
4 A_{4-a}^3 s^{12} t^{12} + \cO(s^{13})  & \quad \textrm{for }~ p =0 \\
4 A_{4-a}^3 s^{12} t^{12} & \quad \textrm{for }~ 0<p<P   \\
4 A_{4-a}^3 s^{12} t^{12} + \cO(t^{13})& \quad \textrm{for }~ p =P \,. 
\end{array}
\right.
\eeq
where $A_{4-a}$ is a non-zero constant in~\eqref{Amiddle}. We thus learn that $Y^p$ are I$_0$-components.

\subsubsection*{Class 4: $(4, 6, >12)$}\label{sec-class4}
With the vanishing orders of $f$ and $g$ given as
\bea
{\rm ord}_{\widehat{\mathcal Y}} (f)|_{(u,s)=0} &=& 4+\alpha = 4 \,,\\
{\rm ord}_{\widehat{\mathcal Y}} (g)|_{(u,s)=0} &=& 6+\beta = 6 \,,
\eea
the arguments in section~\ref{class1} for Class 1 persist up until~\eqref{g06-b}.
In particular, the Weierstrass sections $f$ and $g$ are non-trivial on the component curves $B^p$
\bea
f_p=f|_{e_p=0} \not\equiv 0 \,,\quad g_p=g|_{e_p=0} \not\equiv 0 \,,\qquad \text{for}~~ p=0,\cdots,P\,,
\eea
which already implies that the surfaces $Y^p$ are I$_{n_p}$-components. 

However, since $\Delta$ vanishes to a higher order as
\beq
{\rm ord}_{\widehat{\mathcal Y}} (\Delta)|_{(u,s)=0}= 12+\gamma > 12 \,,
\eeq
the condition~\eqref{nocancel} changes to
\beq
4 A_{4-a}^3 + 27 B_{6-b}^2 = 0 \,,
\eeq
so that the order $12$ terms in $\Delta$ may cancel. 
Therefore, the discriminant expression~\eqref{follows} for Class 1 simplifies to
\beq
\Delta_p = (4f^3 + 27 g^2) |_{e_p=0} = 
\left\{ 
\begin{array}{ l l }
\cO(s^{13})  & \quad \textrm{for }~ p =0 \\
0 & \quad \textrm{for }~ 0<p<P   \\
\cO(t^{13})& \quad \textrm{for }~ p =P \,, 
\end{array}
\right.
\eeq
guaranteeing the positivity of the vanishing orders, 
\beq
v_{\mathcal P_{\rm gen}}(\Delta_p) >0 \quad \text{for}~~0<p<P \,,
\eeq
where $\mathcal P_{\rm gen}$ is a generic point in $B^p$. 
We thus conclude that 
 \beq
 n_p > 0 \quad \text{for}~~ 0 < p < P \,,  
 \eeq
as claimed in~\eqref{np-C4}.

\subsubsection*{Class 5: $(>4, >6, >12)$}
Non-minimal fibers of Class 5 are distinguished from those of the previous four classes in that the chain of $P$ blowups does not lead to a (blown-down) Kulikov model. Specifically, as it turns out, we have
\beq\label{claim5}
f_p \equiv 0\,,\quad  g_p \equiv 0 \,,\quad \Delta_p \equiv 0 \,, \qquad \text{for}~~p=1, \cdots, P-1\,,
\eeq
and therefore none of the surfaces $Y^p$ in the middle are I$_n$-components.  

In order to show~\eqref{claim5}, we first recall from section~\ref{class2} that with the vanishing order of $f$ given as
\beq
{\rm ord}_{\widehat{\mathcal Y}} (f)|_{(u,s)=0} = 4+\alpha >4 \,, 
\eeq
the exponents $\mu_{p,i}$ should obey~\eqref{mu0},~\eqref{mu0P}, and~\eqref{muP}, except that the vanishing of $\mu_{p,i}$ for $i \in \cI_0$ is modified to 
\beq
\mu_{p,i} > 0 \,, \quad \forall p~~\text{and}~~ i\in \cI_0 \,. 
\eeq
Similarly, since the vanishing order of $g$ is assumed to satisfy 
\beq
{\rm ord}_{\widehat{\mathcal Y}} (g)|_{(u,s)=0} = 6+\beta >6 \,, 
\eeq
the constraints~\eqref{nu0},~\eqref{nu0P}, and~\eqref{nuP} still apply to $\nu_{p,j}$, with the modification 
\beq
\nu_{p,j} > 0 \,, \quad \forall p~~\text{and}~~ j\in \cJ_0 \,. 
\eeq
We thus deduce the following positivity properties, 
\beq
\mu_{p, i} > 0 \,,\quad \forall i\,,\quad \text{and}\quad   \nu_{p,j} > 0 \,, \quad \forall j\,,\qquad \text{for}~~p=1, \cdots, P-1 \,, 
\eeq
implying that $f$, $g$, and $\Delta$ identically vanish in each of the middle components $B^p$ as claimed. 

\subsection{Proofs}\label{proofs}
In the following we will present the proofs of the propositions stated in section~\ref{summary}. 

{\PROOF~\label{proof1}{\rm\bf{[Proposition~\ref{prop1}].}}}
Let us first note that for non-minimal degenerations of Class 1--4,
either of the following two equations holds:
\beq\label{ordf=4}
{\rm ord}_{\widehat{\mathcal Y}} (f)|_{(u,s)=0} = 4+\alpha = 4 \,\quad \text{or}\quad 
{\rm ord}_{\widehat{\mathcal Y}} (g)|_{(u,s)=0} = 6+\beta = 6 \,.
\eeq
We suppose here that the former condition is obeyed (under the latter condition the argument proceeds in an analogous manner). Since we may assume without loss of generality that the number $P$ of the required blowups is 
\beq
P \geq 2 \,,
\eeq 
the first blow up by itself should not have removed the non-minimality at $s=0$. As stated in section~\ref{class1}, for every $i \in \cI_-$ we thus have
\beq
\mu_{1,i} \geq 4 - (a+i) > 0 \quad  \Longrightarrow \quad \mu_{0,i} + a + i = \mu_{1,i} +  4 > 4 \,, 
\eeq
and hence the vanishing order~\eqref{ordf=4} of $f$ is achieved by the sector with $i \in \cI_0$, i.e., 
\beq\label{F4-a}
\mathcal F_{4-a} s^4 t^4 =  A_{4-a} s^4 t^4 + \cdots \,, 
\eeq
with $A_{4-a}$  a non-zero constant. Since the term~\eqref{F4-a} does not depend on each $e_p$, the Weierstrass section $f$ has 3-fold-vanishing order $0$ at $B^{p,p+1}$ for $p=0, \cdots, P-1$. Therefore, the elliptic fiber there is of Type I$_{n_{p,p+1}}$, where $n_{p,p+1}$ is subject to 
\beq
{n_{p,p+1} \geq n_p + n_{p+1} }\,,
\eeq
since the fiber at $B^{p,p+1}$ is obtained by the collision of an I$_{n_p}$ and an I$_{n_{p+1}}$ fiber. 

Let us now suppose that 
\beq
n_p = 0 = n_{p+1}\,,
\eeq
i.e., that we start with a non-minimal fiber of Class 1--3, not of Class 4. Then, the middle degree sectors,
\beq
\mathcal F_{4-a} s^4 t^4 \quad \text{and}\quad \mathcal G_{6-b} s^6 t^6 
\eeq
of $f$ and $g$, respectively, lead to the bi-degree $(12,12)$ sector in $\Delta$ of the form,
\beq
\mathcal D_{12-c} (e_0, \cdots, e_P) s^{12} t^{12} \,, 
\eeq
where the coefficient $\mathcal D_{12-c}$ contains the following non-trivial sector independent of each $e_p$, 
\beq
\mathcal D_{12-c}(0, \cdots, 0) =
\left\{ 
  \begin{array}{ l l }
    4A_{4-a}^3 + 27 B_{6-b}^2 & \quad \textrm{for Class 1} \\
    27 B_{6-b}^2  & \quad \textrm{for Class 2} \\ 
    4A_{4-a}^3 & \quad \textrm{for Class 3} \,.  
  \end{array}
\right.
\eeq
We therefore conclude that  
\beq
n_{p,p+1} = 0 \,,
\eeq
i.e. that the fiber at $B^{p,p+1}$ is of Type $I_0$ for $p=0, \cdots, P-1$.
\hfill\(\Box\)

{\PROOF~\label{proof2}{\rm\bf{[Proposition~\ref{prop2}].}}}
We start by noting that the leading exponents $\mu_{p, i}$ of $e_p$ in the expansion coefficients $\mathcal F_i$ are subject to the positivity condition
\beq\label{c1}
\mu_{p, i} > 0 \quad \text{for}~~ 0< p<P\,,~ i\not\in \cI_0 \,. 
\eeq
This is manifest from the fact that $\{\mu_{p, i}\}_{p=0}^P$ is an increasing (resp., decreasing) sequence for $i \in \cI_-$ (resp., $i \in \cI_+$). On the other hand, for $i\in \cI_0$ we have  
\beq\label{c2}
\mu_{p, 4-a} = 0 \quad \forall p \,,
\eeq
since the non-minimality is of Class 4. 
Similarly, the exponents $\nu_{p,j}$ obey 
\beq\label{c3}
\nu_{p, j} > 0 \quad \text{for}~~ 0< p<P \,, ~j\not\in \cJ_0 \,,
\eeq
and 
\beq\label{c4}
\nu_{p,6-b} = 0 \quad \forall p \,.
\eeq
Combining the constraints~\eqref{c1}--\eqref{c2}, it follows for $0<p<P$ that
\bea
f_p &=& A_{4-a} s^4 t^4 \,, \\
g_p &=& B_{6-b} s^6 t^6 \,, 
\eea
where $A_{4-a}$ and $B_{6-b}$ are non-vanishing constants, and hence, that 
\beq\label{goverf}
\left(\frac{g}{f}\right)\bigg|_{e_p =0} = \frac{B_{6-b}}{A_{4-a}} s^2 t^2 \,.
\eeq
Since this ratio is a perfect square, each middle component $B^p$ supports a codimension-zero I$_{n_p}$ fibers of split form. However, the expression~\eqref{goverf} for the ratio does not hold in general along the end components, for which both split and non-split forms are allowed in principle. \hfill\(\Box\)

{\PROOF~\label{proof3}{\rm\bf{[Proposition~\ref{prop3}].}}}
Let us recall that we have either~\eqref{c1} and~\eqref{c2} for $\mu_{p, i}$ (if $\alpha=0$) or~\eqref{c3} and~\eqref{c4} for $\nu_{p,j}$ (if $\beta=0$). As the argument proceeds in a symmetric fashion, we will assume that the former holds with $\alpha=0$, i.e., 
\beq
{\rm ord}_{\widehat{\mathcal Y}} (f)|_{(u,s)=0} = 4+\alpha = 4 \,.
\eeq
 Then, for $0<p<P$, we have 
\beq
f_p = A_{4-a} s^4 t^4  \,,
\eeq
which vanishes nowhere in $B^p$. Therefore, at every point $\mathcal P$ in the middle component $B^p$ the K3-vanishing order of $f$ vanishes, i.e., 
\beq
{\rm ord}_{Y_0} (f)|_{\mathcal P} = 0 \,,\quad \forall \mathcal P \in B^p \,, ~0<p<P\,,
\eeq
and hence, only I$_n$ enhancements can occur for the $8$-dimensional gauge theory.  \hfill\(\Box\)

{\PROOF~\label{proof4}{\rm\bf{[Proposition~\ref{prop4}].}}}
The discriminant $\Delta$ factorizes as
\beq\label{factDelta}
\Delta = \prod_{q=0}^P e_q^{n_q}  \Delta' \,,
\eeq
with $\Delta'$ admitting no overall factors of $e_q$ for any $q$. Here, the exponents $n_q$ appear since $Y^q$ are I$_{n_q}$-components for Class 1--4 models, but we can define $n_q$ even for Class 5 degenerations. 

In studying the branes localized in a fixed base component $B^p$, it suffices to analyze the terms in $\Delta$ that come precisely with the minimal power of $e_p$, i.e., with $e_p^{n_p}$. Each such term originates either from a cubic product of the terms in $f$ or a quadratic product of the terms in $g$. Let us focus first on terms of the former type, given by   
\beq\label{AAA}
4 A_{i_i} A_{i_2} A_{i_3} \Big(\prod_{q=0}^P e_q^{\mu_{q,i_1} + \mu_{q, i_2} + \mu_{q,i_3}} \Big) s^{3a+i_1 + i_2 + i_3} t^{24-3a-i_1-i_2-i_3} \,,
\eeq
for a triple of indices $(i_1, i_2, i_3)$ subject to 
\beq\label{mumumu}
\mu_{p, i_1} + \mu_{p,i_2} + \mu_{p,i_3} = n_p \,.
\eeq
First, for the left-end component with $p=0$, the points on the rational curve $B^0$ are parametrized by the homogeneous coordinates $[t:e_1]$, and the terms of the form~\eqref{AAA} in $\Delta$ are of
 homogeneous degree 
\bea
(24-3a-i_1-i_2-i_3) + (\mu_{1, i_1} + \mu_{1,i_2}+\mu_{1,i_3}) &=& (\mu_{0,i_1} + \mu_{0,i_2} + \mu_{0,i_3}) + 12 \\  \label{n0+12}
&=& n_0 + 12 \,,
\eea
where we have used the recursion relation~\eqref{recmu} in the first step and the constraint~\eqref{mumumu} in the second.\footnote{Note that we are implicitly assuming that the triple products that involve subleading terms in the expansion~\eqref{expfp} remain subleading in $\Delta$. This may not be the case if cancellations occur amongst products of the form~\eqref{AAA} from $f^3$ that only involve the leading terms and analogous products from $g^2$. However, all that we have used in deriving the homogeneous degree~\eqref{n0+12} is the recursion relation~\eqref{recmu} and the constraint~\eqref{mumumu}, which continue to hold for the subleading terms. The argument therefore goes through even if cancellations occur in $\Delta$. With this understanding, and to be explicit, we will henceforth only consider cross products of leading terms whenever we need to consider such products.} One can also see that, for a similar reason, the same homogeneous degree~\eqref{n0+12} arises for the terms that originate from $27 g^2$. Since the discriminant $\Delta$ has an overall factor of $e_1^{n_1}$ as in~\eqref{factDelta}, we thus conclude that the zeroes of $ \Delta'$ in $B^0$ are counted as
\beq
x_0 = (n_0+12) - n_1  \,,
\eeq
as claimed by~\eqref{claim1}. 
Similarly, we can count the zeroes of $\Delta'$ on the right-end component $B^P$ as 
\beq
x_P = (n_P + 12)-n_{P-1} \,,
\eeq
as in~\eqref{claim2}. 
Finally, if $0<p<P-1$, the homogenous degree of $\Delta'$ in the coordinates $e_{p\pm1}$ of $B^p$ is read off, e.g., from the cubic products~\eqref{AAA}, as
\bea
\sum_{\delta \in \{-1, 1\}}\left( \sum_{k=1}^3 \mu_{p+\delta , i_k}   - n_{p+\delta}\right)
&=& 2\sum_{k=1}^3 \mu_{p,i_k} - (n_{p-1}+n_{p+1}) \\ 
&=& 2n_p - n_{p-1}-n_{p+1} \,,
\eea
where, once again, the relation~\eqref{recmu} and the constraint~\eqref{mumumu} have been used in turn. We thus learn that $\Delta'$ has 
\beq
x_p = 2n_p - n_{p-1} - n_{p+1}\,,
\eeq
zeroes in the middle components $B^p$, thereby confirming~\eqref{claim3}. \hfill\(\Box\)

{\PROOF~\label{proof6}{\rm\bf{[Proposition~\ref{prop6}].}}}
We prove the proposition in the following two steps.

\paragraph{{\it Step 1\,}:} Let us start by performing the base change
\beq\label{basec}
u \to u^{(4-a)!(6-b)!} 
\eeq
where $a$ and $b$ are the K3-vanishing orders~\eqref{generic-van} of the Weierstrass sections $f$ and $g$ for a generic $u \neq 0$.\footnote{To be precise, we are assuming $a\leq 4$ and $b \leq 6$ in~\eqref{basec}. For $a>4$ the base change should be replaced by $u \to u^{(6-b)!}$, and similarly for $b>6$, by $u \to u^{(4-a)!}$.} 
Since the leading exponents $\mu_{p,i}$ and $\nu_{p,j}$ obey the relations~\eqref{recmu} and~\eqref{recnu}, after the base change~\eqref{basec}, there must exist either $i_0 \in \cI_-$ with $\mu_{P,i_0}=0$ or $j_0 \in \cJ_-$ with $\nu_{P,j_0}=0$. In the rest of this proof we will assume the former (a similar argument can be applied in the latter case). 

The Weierstrass section $f$ restricted to the right-end component thus behaves as
\beq\label{fPI0}
f_P := f|_{e_P=0} =  \sum_{i \in \cI_-^{(0)}}\mathcal F_i s^{a+i} t^{8-a-i} + \mathcal F_{4-a}|_{e_P=0} s^4 t^4 \,,
\eeq
where $\cI_-^{(0)} \subset \cI_-$ is the index subset with the defining property $\mu_{P, i} = 0$. Importantly, this subset $\cI_-^{(0)}$ is non-empty since $i_0 \in \cI_-^{(0)}$ by assumption. Therefore, the K3-vanishing order of $f_P$ at $s=0$ is strictly less than $4$, and hence, the fiber there is no longer physically non-minimal. 

\paragraph{{\it Step 2\,}:} While the absence of potential physical non-minimality is guaranteed at $s=0$ thanks to the base change~\eqref{basec}, it may still be present away from $s=0$ on the right-end $B^P$. We thus suppose that the fiber is physically non-minimal at a point $\mathcal P_* \in B^P$, 
\beq\label{P*}
\mathcal P_* : \{s= s_* e_{P-1} \} \,, \quad \text{for}~~s_* \neq 0\,,
\eeq 
with $[s:e_{P-1}]$ being the homogeneous coordinates of the rational component $B^P$. 

Let us now recall that $f_P$ and $g_P$ are sections of degree $4$ and $6$ line bundles $L_P^{\otimes 4}$ and $L_P^{\otimes 6}$, respectively, where $L_P$ is of degree $1$ on $B^P$. We then note that $f_P$ is non-trivial since it acquires at least one non-trivial term from the summand with $i_0 \in \cI_-^{(0)}$ in~\eqref{fPI0}. As the maximal power of $s$ appearing in~\eqref{fPI0} is $4$, the following necessary condition for physical non-minimality, 
\beq
{\rm ord}_{Y_0} (f)|_{\mathcal P_*} \geq 4 \,,
\eeq 
must be saturated and $f_P$ must take the form, 
\beq\label{fPP*}
f_P = A (s- \sigma_* e_{\downarrow} t)^4 t^4 \,. 
\eeq
Here, $A$ and $\sigma_*$ are non-zero constant and the latter is determined in terms of $s_*$ via
\beq\label{s*}
\sigma_* e_\downarrow t  = s_* e_{P-1} \,,  
\eeq
where $e_\downarrow$ denotes the product of $e_p$,
\beq
e_{\downarrow} = \prod_{p=0}^P e_p^{P-p} \,, 
\eeq
with the exponents decreasing in $p$. 
We also observe from~\eqref{fPP*} that the physical non-minimality assumption at $\mathcal P_*$ implies  
\beq
a=0
\eeq
for the initial K3-vanishing order of $f$
and 
\beq\label{alpha0}
\alpha=0
\eeq 
for the class distinction, i.e. Class 2 is ruled out. In fact the global behavior of $f$ can be specified as
\beq\label{globalf}
f = A( s-  \sigma_* e_\downarrow t)^4 t^4 + \sum_{i=5}^8 A_i (e_\rightarrow)^{\mu_{0,i}}(e_{\uparrow})^{i-4} s^{i} t^{8-i} \,, 
\eeq
where  $e_\uparrow$ and $e_\rightarrow$ are defined as
\bea
e_\uparrow &=& \prod_{p=0}^P e_p^p \,,\\
e_\rightarrow &=& \prod_{p=0}^P e_p \,. 
\eea
Note that, for simplicity, any contributions to $f$ from the higher-order terms in~\eqref{expfp} have been neglected in~\eqref{globalf} and the constant $A$ can been further specified as $A_4$. 

If the initial non-minimality is of Class 1 or 4 with $\beta = 0$ on top of~\eqref{alpha0}, $g_P$ is also non-trivial in presence of the non-trivial contribution,
\beq
\mathcal G_{6-b} s^6 t^6 \,,
\eeq
and hence, a similar conclusion can be drawn for $g_P$, 
\beq\label{gPP*}
g_P = B (s-\sigma_* e_\downarrow t )^6 t^6 \,,
\eeq
where $B$ is a non-zero constant. This implies that
\beq
b=0\,,
\eeq
and also leads to the global behavior of $g$, 
\beq
g = B( s-  \sigma_* e_\downarrow t)^6 t^6 + \sum_{j=7}^{12} B_j (e_\rightarrow)^{\nu_{0,j}}(e_{\uparrow})^{j-6} s^{j} t^{12-j} \,, 
\eeq
again modulo higher-order terms in~\eqref{expgp}. 
In case the initial non-minimality was of Class 3 with $\beta >0$ instead, the global form~\eqref{gPP*} would still be valid, except that $B$ is not a constant but rather 
\beq
B=B_{6} (e_\rightarrow)^{\nu_0} \,,
\eeq
where $\nu_0$ is the common exponent of $e_p$ in $\mathcal G_6$. However, this modification does not affect the argument that will follow, and hence, we will assume for simplicity of the notation that $\beta =0$.  

All in all, we have so far seen that the Weierstrass sections, modulo higher-order terms, should take the form  
\bea
f &=& A_4( s-  \sigma_0 e_\downarrow t)^4 t^4 + \sum_{i=5}^8 A_i (e_\rightarrow)^{\mu_{0,i}}(e_{\uparrow})^{i-4} s^{i} t^{8-i} \,,\\ 
g &=& B_6(s-\sigma_0 e_\downarrow t)^6 t^6 + \sum_{j=7}^{12} B_j (e_\rightarrow)^{\nu_{0,j}} (e_{\uparrow})^{j-6} s^j t^{12-j} \,.
\eea
We will now prove that the physical non-minimality at $\mathcal P_*$, 
\beq
[s:e_{P-1}] = [s_*:1]\,,
\eeq 
can eventually be removed by an additional chain of blowups. To be explicit, we will start by introducing a new pair of homogeneous coordinates for $B^P$:
\beq
[s:e_{P-1}] \to [s:v] \,,
\eeq
where the coordinate $v$ is defined as
\beq
v := s-s_* e_{P-1}\,.
\eeq
Then, the Weierstrass section $f$ can be rewritten as
\bea
f &=& A_4 v^4 t^4 + \sum_{i=5}^8 A_i \prod_{p=0}^P e_p^{\mu_{0,i} + (i-4)p} s^i t^{8-i} \,\\ \label{fP+0}
&=& A_4 v^4 t^4 + \sum_{i=5}^8 A_i \prod_{p=0}^{P-2} e_p^{\mu_{0,i} + (i-4)p} s_*^{-K_i} (s-v)^{K_i} e_{P}^{\mu_{0,i} + (i-4) {P}} s^i t^{8-i} \,
\eea
with $K_i := \mu_{0,i} + (i-4)(P-1)$, and similarly, $g$ as
\bea\label{gP+0}
g &=& B_6 v^6 t^6 + \sum_{j=7}^{12} B_j \prod_{p=0}^{P-2} e_p^{\nu_{0,j} + (j-6)p}  s_*^{-L_j} (s-v)^{L_j} e_{P}^{\nu_{0,j} + (j-6) {P}} s^j t^{12-j} \,,
\eea
with $L_j:=\nu_{0, j} + (j-6) (P-1)$.\footnote{Strictly speaking, the expressions~\eqref{fP+0} and~\eqref{gP+0} are not the most appropriate ones for a global algebraic-geometric description since there appear negative powers of the coordinates,
\beq\label{t-thru-(P-2)}
t, e_0, \cdots, e_{P-2}\,,
\eeq
 via $s_*^{-K_i}$ and $s_*^{-L_j}$, respectively, where
\beq
s_* = \sigma_* t e_0^P e_1^{P-1} \cdots e_{P-2}^1 \,.
\eeq
For the purpose of analyzing the behavior of $f$ and $g$ restricted to $B^{P+q}$ for $q=0, \cdots, Q$, however, one may set all the coordinates~\eqref{t-thru-(P-2)} to unity.}

Clearly, we have to blow up at $v=e_P=0$ since $P$ has been made large enough via the initial base change~\eqref{basec}. To make a precise statement, let us first define
\beq
Q:={\rm min} \{Q_f, Q_g\}\,,
\eeq
where $Q_f$ and $Q_g$ are given as
\bea\label{Qf}
Q_f &=& {\rm min}_{i=5}^8 \{\lfloor{\frac{\mu_{0,i} + (i-4)P}{4}}\rfloor\}\,,\\ \label{Qg}
Q_g &=&  {\rm min}_{j=7}^{12} \{\lfloor{\frac{\nu_{0,j} +(j-6)P}{6}}\rfloor\}\,.
\eea
Our claim is that the non-minimality is removed by performing additional $Q$ blowups, where the $q^{\rm th}$ additional blowup for $1 \leq q \leq Q$ is the operation that transforms $f$ and $g$ as
\bea
f(s,t; e_0, \ldots, e_{P+q-1}) &\longrightarrow&  f(s,t; e_0, \ldots, e_{P+q}) \,,\\
g(s,t; e_0, \ldots, e_{P+q-1}) &\longrightarrow&  g(s,t; e_0, \ldots, e_{P+q}) \,,
\eea
by applying the coordinate replacement,
\bea
e_{P+q-1} &\longrightarrow& e_{P+q-1}\,e_{P+q} \,,\\
v  &\longrightarrow& v \,e_{P+q} \,,
\eea
and dividing the resulting $f$ and $g$ by $e_{P+q}^4$ and $e_{P+q}^6$, respectively. At the end of these additional $Q$ blowups, we thus obtain
\bea \label{fP+Q}
f &=& A_4 v^4 + \sum_{i=5}^8 A_i \prod_{p=0}^{P-2} e_p^{\mu_{0,i} + (i-4)p} s_*^{-K_i} (s-v  \prod_{q=1}^Q e_{P+q} )^{K_i}  \prod_{q=0}^Q e_{P+q}^{\mu_{0,i} + (i-4) {P-4q}} s^i t^{8-i} \,, \\ \label{gP+Q}
g &=& B_6 v^6 + \sum_{j=7}^{12} B_j  \prod_{p=0}^{P-2} e_p^{\nu_{0,j} + (j-6)p}  s_*^{-L_j} (s-v \prod_{q=1}^Q e_{P+q} )^{L_j}  \prod_{q=0}^Q e_{P+q}^{\nu_{0,j} + (j-6) {P-6q}}  s^j t^{12-j}\,,   \nonumber
\eea
and the chain~\eqref{Bchain} of base curves further expands to
\beq\label{BchainQ}
B^0 - B^1 - \cdots - B^{P+Q-1} - B^{P+Q}   \,.
\eeq
The self-intersection numbers are  $-2$ for the middle components and $-1$ for the end components $B^0$ and $B^{P+Q}$. Obviously, $f$ and $g$ are non-trivial on each of the additional components $B^{P+q}$. Therefore, the special fibers  on the components $B^{P+q}$ with $q=0, \cdots, Q-1$  can only be of Kodaira Type I$_n$. 

Let us finally turn to the right-end 
component, $B^{P+Q}$, with homogeneous coordinates $[v:e_{P+Q-1}]$. Note that, upon restricting to this end component, the following coordinates,
\beq
s, t, e_0, e_1, \cdots, e_{P+Q-2} \,, 
\eeq
may all be set to unity, and hence,   
the Weierstrass sections~\eqref{fP+Q} and~\eqref{gP+Q} simplify as
\bea\label{fP+Q'}
f_{P+Q} &=& A_4 v^4 + \sum_{i=5}^8 A_i  \sigma_*^{-K_i} e_{P+Q-1}^{\mu_{0,i} + (i-4) {P-4(Q-1)}} \big( e_{P+Q}^{\mu_{0,i} + (i-4) {P-4Q}} \big)|_{e_{P+Q}=0}  \,, \\ \label{gP+Q'}
g_{P+Q} &=& B_6 v^6 + \sum_{j=7}^{12} B_j   \sigma_*^{-L_j}   e_{P+Q-1}^{\nu_{0,j} + (j-6) {P-6(Q-1)}} \big(  e_{P+Q}^{\nu_{0,j} + (j-6) {P-6Q}}\big)|_{e_{P+Q}=0}  \,. 
\eea
Here, the relation~\eqref{s*} has been used to replace $s_*$ by $\sigma_*$ and the summands are non-trivial only for $i \in \cI_+$ with 
\beq
\mu_{0,i} + (i-4)P-4Q = 0 \,,
\eeq
and for $j \in \cJ_+$ with
\beq
\nu_{0,j} + (j-6)P-6Q = 0 \,.
\eeq
Let us define the subsets $\cI_+^{(0)} \subset \cI_+$  and $\cJ_+^{(0)} \subset \cJ_+$ by collecting those $i$ and $j$, respectively. Now, with the base change~\eqref{basec} assumed in the very beginning, it is clear that $Q_f$ and $Q_g$ in~\eqref{Qf} and~\eqref{Qg} could as well have been defined without the floor symbols. Therefore, we learn that either $\cI_+^{(0)}$ or $\cJ_+^{(0)}$ should be non-empty, and hence, that the expressions~\eqref{fP+Q'} and~\eqref{gP+Q'} can be schematically written as
\bea
f_{P+Q} &=& A_4( v^4 +  K e_{P+Q-1}^4 ) \,, \\  
g_{P+Q} &=& B_6( v^6 + L e_{P+Q-1}^6 ) \,, 
\eea
where $K$ and $L$ are constants that cannot both be $0$. This in turn guarantees that no physical non-minimality may arise in $B^{P+Q}$. 
\hfill\(\Box\)

\section{From Kulikov Models to Non-minimal Degenerations}\label{Class5-TypeI}

In Appendix \ref{app_Non-minimal} we have shown that to each non-minimal Weierstrass degeneration $\widehat{\cal Y}$ of Class 1, 2, 3, or 4 one can associate a Kulikov Weierstrass model $\cal Y$, which is the fibral blowdown of a Kulikov model ${\cal X}$ of Type II or III.
We now prove the converse, in the sense that every elliptic Type II or Type III Kulikov model ${\cal X}$, and therefore every Kulikov model associated with a degeneration at infinite distance, is the blowup of a Weierstrass degeneration of Class 1, 2, 3 or 4, or alternatively of a Weierstrass model with degenerate fibers in codimension zero, but without non-minimal fibers in codimension one.

The idea is a straightforward reversal of the arguments of Appendix \ref{app_Non-minimal}: We start by fixing a Kulikov model $\mathcal X$ for a degeneration of Kulikov Type II or III. Let us denote by $\mathcal Y$ the Weierstrass model associated with $\mathcal X$ and by $Y_0$ the central element of the family $\mathcal Y$, which decomposes as 
\beq \label{Ydecomposition}
Y_0 = \cup_{p=0}^P Y^p \,.
\eeq
\noindent
As explained in Appendix \ref{app_Non-minimal}, the base of $Y_0$ is guaranteed to decompose accordingly and in fact takes the form of a chain, 
\beq\label{Bdecomposition}
B^0 - B^1 - \cdots - B^{P-1} - B^P \,.
\eeq
The components $B^p$ are rational curves and, under the assumption that $P>0$,  their self-intersection numbers on the base ${\cal B}$ of $\cY$ are $-2$ for $0<p<P$ and $-1$ for $p=0, P$. Therefore, we can perform a chain of blowdowns starting from the right-end to contract the $P$ base components, $B^P, \ldots, B^1$, which leads to a single-component configuration. The aim of this Appendix is then to prove that this end configuration contains a non-minimal fiber of Class 1, 2, 3 or 4. More precisely, we put forward the following claim: 

\vspace{.2\baselineskip}
{\proposition\label{prop7} Let $\mathcal X$ be a Type II or III Kulikov model of elliptic K3 surfaces and $\mathcal Y$ be the associated Weierstrass model, whose central fiber $Y_0$ decomposes into $P+1$ components as in~\eqref{Ydecomposition}, elliptically fibered itself over the chain of base curves~\eqref{Bdecomposition}. Let us define $\widehat {\mathcal Y}$ as the blowdown of $\mathcal Y$ along the base $\cB$ of the elliptic threefold $\cV$, obtained by contracting the $P$ base curves $B^P, \ldots, B^1$ in turn, and $\hat Y_0$ as the central fiber of $\widehat {\mathcal Y}$. Then, the surface $\hat Y_0$ can never have a non-minimal fiber of Class 5. Specifically, there exists a non-minimal fiber of Class 1, 2, 3 or 4 in $\hat Y_0$ unless $P=0$  in~\eqref{Ydecomposition}. For $P=0$ the Weierstrass model~\eqref{original} of $\widehat {\mathcal Y}$ is given by 
\beq\label{fgh}
f_u= -3h^2 + u \eta \,, \qquad g_u= -2h^3 + u \rho \,, 
\eeq
where $h$, $\eta$ and $\rho$ are sections of degree 4, 8 and 12, respectively, and $h$ has four distinct zeroes.  
} 

\vspace{.2\baselineskip}{\proof.}  
We employ the same notation as in Appendix~\ref{nomen}, although we start here from the $(P+1)$-component configuration $Y_0$ rather than its blown-down version $\hat Y_0$. For instance, the deformation parameter of the model $\widehat{\mathcal Y}$ will be denoted by $u$ and the homogeneous coordinates for the rational base of $\hat Y_0$ will be denoted by $[s:t]$. Furthermore, we use the symbols $e_p$ with $p=0,\ldots,P$ for the coordinates whose vanishing  loci on the base $\cB$ of $\mathcal Y$ describe the curve components $B^p$.

Let us first consider the special cases where the decomposition~\eqref{Ydecomposition} is trivial in the sense that $P=0$, which indicates that $n_0>0$; otherwise, the Kulikov model $\mathcal X$ would be of Type I, contradicting the starting assumption. The unique I$_{n_0}$-component $Y^0$ of the central fiber $Y_0$ is thus described by a Weierstrass model of the form~\eqref{fgh}. Here, it follows that the degree-four section $h$ has four distinct zeroes since otherwise $Y^0$ would develop a non-minimal fiber from the perspective of the K3 surface, which would, in turn, contradict the fact that $Y^0$ is obtained as a fibral blowdown of the central K3 fiber of the Kulikov model $\mathcal X$.

Having characterized the blowndown configuration of a model with $P=0$, let us now focus on the generic cases for which $P\geq 1$. 
Since the $P$ blowdowns transform $\mathcal Y$ to $\widehat {\mathcal Y}$, there must exist a well-defined sequence of reverse operations, i.e., $P$ blowups in the base, which turn $\widehat{\mathcal Y}$ back to $\mathcal Y$. However, these blowups are not necessarily performed at the ``same'' point in the following sense: We first perform $P_1$ blowups, labelled by $p=1, \ldots, P_1$, at the points
\beq\label{defv1}
v^{(1)} :=s-s_*^{(1)} e_{p-2} = 0 \,, 
\eeq
where $[s:v^{(1)}]$ are homogeneous coordinates of the right-end base curve $B^{p-1}$ after the $(p-1)^{\rm th}$ blowup.\footnote{We formally define $e_{-1}:=t$ for the coordinate definition~\eqref{defv1} to be valid also for $p=1$.} We next perform another set of blowups labelled by $p=P_1+1, \cdots, P_2$ at 
\beq\label{defv2}
v^{(2)} :=s-s_*^{(2)} e_{p-2} = 0 \,,   
\eeq
where $[s:v^{(2)}]$ are coordinates of $B^{p-1}$ with $s_*^{(2)} \neq s_*^{(1)}$, and so on. In this way, we perform a total of $P = \sum_{k=1}^m P_k$ blowups, subdivided into $P_k$ blowups at $v^{(k)} = 0$ for $k=1, \cdots, m$, where the coordinates $v^{(k)}$ of $B^{p-1}$ after the $(p-1)^{\rm th}$ blowup are defined as 
\beq
v^{(k)} :=s-s_*^{(k)} e_{p-2} \,.  
\eeq
Here the total of $P$ blowups are performed at $m$ ``different'' points in that 
\beq
s_*^{(k)} \neq s_*^{(k+1)} \,, \quad \text{for}\quad k=1, \cdots, m-1\,.  
\eeq
Notably, this general class of blowup chains reproduces the subclass described in section~\eqref{ChainP} if $m=1$, in which case we may assume without loss of generality that $s_*^{(1)} = 0$. 

Let us first suppose that the blowups are indeed to be repeatedly performed only at the point with $m=1$. If the initial non-minimality at $s=0$ had been of Class 5, according to~\eqref{claim5}, none of the middle components would be an I$_n$-component, which contradicts the fact that $\cY$ is the fibral blowdown of a Kulikov model.\footnote{In principle, we may imagine a certain degenerate surface $\hat Y_0$ containing a Class 5 fiber that leads to~\eqref{Ydecomposition} with $P=1$, so that $Y_0$ may have no middle components. However, in such a model, the right-end component $Y^{P=1}$ can never be an I$_n$-component for the following reason:  Since the initial 3-fold-vanishing orders of $f$ and $g$ at $s=u=0$ are strictly bigger than 4 and 6, respectively, the coefficients $\mathcal F_i$ for $i \in \cI_- \cup \cI_0$ have the leading powers $\mu_{0,i}$ strictly bigger than $4-a-i$, and similarly, $\nu_{0,j}$ are strictly bigger than $6-b-j$ for $j \in \cJ_- \cup \cJ_0$. This means that $\mu_{1,i}$ and $\nu_{1,j}$ are strictly positive for {\it all} index values of $i$ and $j$, i.e., that $f$ and $g$ are both divisible by $e_1$. Therefore, the codimension-zero fibers over $B^1$ cannot be of Type I$_n$.} 

An analogous argument works even when $m>1$, in which case we blow the configuration down in $m$ steps. In other words, we first perform $P_m$ blowdowns, by which the base chain~\eqref{Bdecomposition} reduces to 
\beq\label{Bdecomposition-1}
B^0 - B^1 - \cdots -  B^{P-P_m} \,. 
\eeq
Here, the right-end component $B^{P-P_m}$ supports a non-minimal fiber at $v^{(m)}=0$, where $v^{(m-1)} \neq 0$. However, this does not keep us from blowing the geometry further down, which is precisely the second situation of Proposition~\ref{prop6}. We thus proceed further to perform $P_{m-1}$ blowdowns even though non-minimal fibers arise now at a different point, i.e., at $v^{(m-1)}=0$. The base chain then further reduces to 
\beq\label{Bdecomposition-2}
B^0 - B^1 - \cdots - B^{P-P_m-P_{m-1}} \,. 
\eeq
By repeating the operations of this kind, we end up with a base of the form
\beq\label{Bdecomposition-2}
B^0 - B^1 - \cdots -  B^{P_1} \,,  
\eeq
where each component curve $B^p$ supports fibers of Type I$_{n_p}$ at codimension-zero. Therefore, for the same reason as in the $m=1$ case, we conclude that after performing the remaining $P_1$ blowdowns, the non-minimal fiber at $v^{(1)} = u=0$ cannot be of Class 5. 
\hfill\(\Box\) \\

As an immediate consequence of Proposition~\ref{prop7}, we learn that it suffices to consider Class 1--4 fibers or Weierstrass models of the form (\ref{fgh}) in classifying the effective physics of non-minimal fibers at infinite distance in the complex structure moduli space.

However, Proposition~\ref{prop7} does {\it not} imply that every Class 5 model must necessarily be equivalent to a Type I Kulikov model.
Rather, starting with a Type II or Type III Kulikov model ${\cal X}$ and its Weierstrass model ${\cal Y}$, we can first blow down the latter along the base to a degenerate 
Weierstrass model $\widehat{\cal Y}$, with a non-minimal degeneration of Class 1 -- 4; but depending on the details, it might be possible to apply to this  $\widehat{\cal Y}$ a rescaling of the Weierstrass sections together with a base change in such a way as to artificially enhance the fiber over generic points of the base to a non-I$_n$ fiber. Then performing suitable blowups to remove the non-minimality
may give rise to a degenerate Weierstrass model ${\cal Y}'$ (whose central fiber $Y_0'$ contains components with generic fibers not of Kodaira Type I$_k$) which in turn admit a different blowdown to a model $\widehat{\cal Y}'$ with a Class 5 non-minimal degeneration.

In Section \ref{subsec_Class5ex} we present two examples of Class 5 degenerations which give rise to Type II Kulikov Weierstrass models, by the reverse of this process. 
However, in all such cases it remains true that the Kulikov Type II or Type III Weierstrass model is associated with a more economical degeneration $\widehat{\cal Y}$ with non-minimal fibers of Class 1, 2, 3, or 4, as claimed by 
Proposition~\ref{prop7}.

\section{Special Tunings for Type III Kulikov Models}  \label{App-WeakCoupling}

According to the general classification of Section \ref{sec_TypeIIIellipticclass}, Kulikov Weierstrass model of Type III come in three versions: Type III.a models with two
rational elliptic end components, those with only one rational elliptic end component and finally Type III.b models, where none of the end components
is a rational elliptic surface. 
These three models are therefore distinguished by the value of $n_p$ in the generic I$_{n_p}$ fiber of the components $Y^p$ for $p=0$ and $p=P$. 
 The criterion for a Type III.b model is that for both end components  
\beq\label{wcc}
\text{Type III.b:} \qquad n_0 >0 \quad \text{and} \quad n_P > 0 \,,
\eeq
while for Type III.a models with only one rational component only $n_P >0$ and $n_0 =0$.

In the following we rewrite the criterion~\eqref{wcc} in terms of the explicit tuning of the Weierstrass sections $(f, g)$.  
To this end, we first define the index subsets
\bea \label{I'}
\cI_+^{(0)} &:=&  \{\, i \in \cI_+ \;|\;  \mu_{0,i}=0 \,\}\,,\\ \label{J'}
\cJ_+^{(0)} &:=& \{\, j\in \cJ_+\;|\; \nu_{0,j}=0\,\}\,, 
\eea
in terms of which the restrictions of the Weierstrass sections $f$ and $g$ to the end component $B^{0}$ are given respectively as
\bea
f_0 &=&  \Big(  \mathcal F_{4-a} s^{4} t^4 + \sum_{i \in \cI_+^{(0)}  } \mathcal F_{i} s^{a+i} t^{8-a-i} \Big)\,,\\
g_0 &=&  \Big(  \mathcal G_{6-b} s^{6} t^6 + \sum_{j \in \cJ_+^{(0)}  } \mathcal G_{j} s^{b+j} t^{12-b-j} \Big)\,.
\eea
By abuse of notation, we here denote by $\mathcal F_i$ and $\mathcal G_j$ only the leading monomials in the expansions~\eqref{expfp} and~\eqref{expgp}. 
Similarly, with the index subsets
\bea  
\cI_-^{(0)} &:=&  \{\, i \in \cI_- \;|\;  \mu_{P,i}=0 \,\}\,,\\ 
\cJ_-^{(0)} &:=& \{\, j\in \cJ_-\;|\; \nu_{P,j}=0\,\}\,, 
\eea
we  find on the end component $B^P$ the  restrictions
\bea
f_P &=& \Big( \sum_{i \in \cI_-^{(0)}  } \mathcal F_{i} s^{a+i} t^{8-a-i} + \mathcal F_{4-a} s^{4} t^4  \Big)\,,\\
g_P &=&  \Big( \sum_{j \in \cJ_-^{(0)}  } \mathcal G_{j} s^{b+j} t^{12-b-j} + \mathcal G_{6-b} s^{6} t^6  \Big)\,,
\eea
where, once again, $\mathcal F_i$ and $\mathcal G_j$ represent only the leading monomials therein. 

Given these expressions for the Weierstrass sections, the criterion~\eqref{wcc} can be tested via the equivalent criterion, 
\beq \label{D0, DP}
\Delta_{0} \equiv 0 \quad \text{and}\quad \Delta_P \equiv 0 \,,
\eeq
which, in turn, requires 
\bea
f_0 &=& -3 h^2 \,, \\ 
g_{0} &=& -2 h^3 \,, 
\eea
and 
\bea
f_P  &=& -3 l^2 \,, \\ 
g_P &=& - 2 l^3 \,. 
\eea
Here, $h = h(s, t; e_1, \cdots, e_P)$ and $l = l( s, t; e_0, \cdots, e_{P-1})$ are polynomials of the form,  
\beq
h = s^2 (h_0  t^2 + h_1   s\, t + h_2 s^2) \quad \text{and}\quad l= t^2 (l_0  t^2 + l_1  s\, t + l_2 s^2) \,, 
\eeq
with the coefficient functions given as 
\beq
h_0 = H_0  \,; \qquad 
h_1= H_1   \prod_{p=1}^{P} e_p^{p}  \,;\qquad
h_2=  H_2   \prod_{p=1}^{P} e_p^{2p}  \,, 
\eeq
and 
\beq
l_0 = L_0 \prod_{p=0}^{P-1} e_p^{2(P-p)} \,;\qquad
l_1= L_1   \prod_{p=0}^{P-1} e_p^{P-p}  \,;\qquad
l_2 =  L_2    \,, 
\eeq
for some complex constants $H_{0,1,2}$ and $L_{0,1,2}$. 

To test if a degenerate $\hat Y_0$ with a Class 4 fiber has this weak-coupling property, let us simply rewrite the above set of requirements in terms of the initial Weierstrass data as the following factorizability criteria: 
\bea\label{f+}
f_+:= \mathcal F_{4-a} s^{4} t^4 + \sum_{i \in \cI_+^{(0)}  } \mathcal F_{i} s^{a+i} t^{8-a-i}
&\stackrel{!}{=}&  -3 s^4(H_0 t^2 + H_1 s\,t + H_2 s^2)^2 \,,\\ \label{g+}
g_+:= \mathcal G_{6-b} s^{6} t^6 + \sum_{j \in \cJ_+^{(0)}  } \mathcal G_{j} s^{b+j} t^{12-b-j} 
&\stackrel{!}{=}&-2 s^6(H_0 t^2+H_1 s\,t+H_2 s^2)^3 \,,
\eea
and
\bea \label{f-}
f_- := \sum_{i \in \cI_-^{(0)}  } \mathcal F_{i} s^{a+i} t^{8-a-i} + \mathcal F_{4-a} s^{4} t^4 \
&\stackrel{!}{=}&  -3 t^4(L_0 u^{2P} + L_1 u^P s\,t + L_2 s^2)^2 \,,\\ \label{g-} 
g_- :=  \sum_{j \in \cJ_-^{(0)}  } \mathcal G_{j} s^{b+j} t^{12-b-j} + \mathcal G_{6-b} s^{6} t^6  
&\stackrel{!}{=}&  -2 t^6(L_0 u^{2P} + L_1 u^P s\,t + L_2 s^2)^3 \,. 
\eea
Here, $f_+$ and $g_+$ have been defined to only collect the $u$-independent terms in $f$ and $g$ of $s$-degree not less than $4$ and $6$, respectively. Similarly, $f_-$ and $g_-$ collect terms of $s$-degree not bigger than $4$ and $6$, whose coefficient functions $\mathcal F_i$ and $\mathcal G_i$ have the leading monomials $u^{(4-a-i)P}$ and $u^{(6-b-j)P}$, respectively. Here $P$ is the number of required blowups that can simply be read off as
\beq
P={\rm min}\left[
\{\lfloor\frac{\mu_{0, i}}{4-a-i}\rfloor \;|\; {i\in \cI_-} \}
\cup 
\{\lfloor\frac{ \nu_{0,j}}{6-b-j}\rfloor \;|\;{j\in\cJ_-}\}
\right]  \,.
\eeq

We emphasize first that the criteria~\eqref{f+} and~\eqref{g+} for $n_0>0$ are fulfilled by fine-tuning the Weierstrass coefficients $\mathcal F_i$ and $\mathcal G_j$, respectively, for $i \in \cI_+$ and $j \in \cJ_+$. To see this, let us note that $f_+$ and $g_+$ in general take the form, 
\bea
f_+ = -3 s^4 F_+ (s,t)\,,\\
g_+ = -2 s^6 G_+(s,t) \,,
\eea
where $F_+$ and $G_+$ are homogeneous polynomials of degree $4$ and $6$ in $[s:t]$. In order for them to factorize as
\bea
F_+ \stackrel{!}{=} (H_0 t^2 + H_1 s\,t + H_2 s^2)^2 \,,\\
G_+ \stackrel{!}{=} (H_0 t^2 + H_1 s\,t+ H_2 s^2)^3 \,, 
\eea
the $5$ leading coefficients,
\beq
\mathcal F_{i}|_{u=0} \quad \text{for}\quad  i \in \cI_0 \cup \cI_+ \,,
\eeq
and the $7$ leading coefficients, 
\beq
\mathcal G_j |_{u=0} \quad\text{for}\quad   j \in \cJ_0 \cup \cJ_+ \,,
\eeq
should then be determined in terms only of a triple $(H_0, H_1, H_2)$ of complex numbers, which is a fine tuning.   

On top of that, in order to also fulfill the criteria~\eqref{f-} and~\eqref{g-} for achieving $n_P>0$, an additional fine tuning is required. The lower degree counterparts of $f_+$ and $g_+$ take the form,
\bea
f_- =- 3 t^4 F_- (s,t; u) \,,\\
g_- = -2 t^6 G_-(s,t;u) \,, 
\eea
where $F_-$ and $G_-$ are also polynomials of degree $4$ and $6$ in $[s:t]$. Obviously, they can only factorize as
\bea
F_- \stackrel{!}{=} (L_0 u^{2P} t^2 + L_1 u^P s\,t + L_2 s^2)^2 \,,\\
G_- \stackrel{!}{=} (L_0 u^{2P} t^2 + L_1 u^P s\,t + L_2 s^2)^3 \,, 
\eea
if the $4-a$ leading coefficients, 
\beq
\frac{\mathcal F_i}{u^{(4-a-i)P}}\Big\rvert_{u=0}\quad \text{for} \quad i\in \cI_-\,,
\eeq
and the $6-b$ leading coefficients, 
\beq
\frac{\mathcal G_j}{u^{(6-b-j)P}}\Big\rvert_{u=0}\quad\text{for}\quad j\in \cJ_-\,,
\eeq
are also tuned so that they may be fixed by a triple $(L_0, L_1, L_2)$.\footnote{In fact, of this triple, only $L_0$ and $L_1$ are the free parameters, provided that the $H_0$, $H_1$ and $H_2$ have already been fixed after the first fine tuning in achieving $n_0 > 0$. This is because $L_2$ must equal $H_0$ so that the middle-degree sectors of $f$ and $g$ may be well-defined.}

\section{Special Fibers at Component Intersections}\label{IP}

In the configuration $\mathcal Y$ obtained by the chain of $P$ blowups of Appendix~\ref{ChainP}, some of the special fibers may be located at some of the intersections 
\beq
B^{p,p+1} := B^p \cap B^{p+1} \,,\quad p=0,\ldots, P-1 
\eeq
of adjacent base components. In fact, as will become clearer in the following, in presence of special fibers at intersections $B^{p,p+1}$, there should exist some special fiber both at $e_{p+1}=0$ in $B^p$ and at $e_p=0$ in $B^{p+1}$. Ambiguities thus arise as to how to interpret the Kodaira type of associated with this collision of special fibers, or equivalently the gauge algebra realised on the associated 7-branes in F-theory. However, as it turns out, this seeming ambiguity is resolved by an appropriate base change in $u$. 

To make a precise statement, let us first consider applying the base change
\beq\label{bc-k}
u \to u^k \,,
\eeq
of order $k$ to the initial single-component configuration $\widehat{\mathcal Y}$, which leads to an equivalent single-component configuration, $\widehat{\bar{\mathcal Y}}$. In order to remove the non-minimal fiber in the latter, a chain of $\bar P$ blowups should follow as usual, giving rise to another equivalent configuration, $\bar{\mathcal Y}$. Its central surface decomposes as
\beq
\bar Y_0  = \cup_{\bar p=0}^{\bar P} \bar Y^{\bar p} \,,   
\eeq
where the length of the chain has increased to $\bar P = k P$. The four configurations involved here are schematically described by the following diagram: 
\be 
\renewcommand\arraystretch{2} 
\begin{array}{lllll}
&{\mathcal Y}& &{\bar{\mathcal Y}}&\\ 
 \pi_P \circ \cdots \circ \pi_1 &\Big\downarrow&&\Big\downarrow&\bar\pi_{\bar P} \circ \cdots \circ \bar \pi_1 \\
&\widehat{\mathcal Y}&\xrightarrow{\makebox[2.5cm]{\text{\tiny base change of order $k$}}}&\widehat{\bar{\mathcal Y}}&
\end{array}
\label{bc4}
\ee
where $\pi_p$ denote the blowups at $e_p=s=0$ used for constructing $\mathcal Y$ and $\bar \pi_{\bar p}$ similarly denote the analogous blowups for $\bar {\mathcal Y}$.

Note that we use the same symbols with a bar above them to denote the objects in the base-changed configuration which are analogous to the ones in the original configuration, leaving the information about the order $k$ implicit. 
The major goal of this appendix is then to prove that no special fibers are present at any of the intersections
\beq
\bar B^{\bar p, \bar p+1}:= \bar B^{\bar p} \cap  \bar B^{\bar p+1}\,,\quad \bar p = 0, \ldots, \bar P  -1 \,, 
\eeq
of the new base components, if and only if the order $k$ of the base change~\eqref{bc-k} is divisible by a certain integer $k_0$, which we will define later as~\eqref{k0def} in terms of the geometry of $\mathcal Y$. 

To this end, we begin by recalling that the special fibers are read off from the factored-out discriminant, where the maximal overall powers of the exceptional coordinates in the discriminant are ignored. For the two configurations $\mathcal Y$ and $\bar{\mathcal Y}$, respectively, we thus have their discriminants factorized as
\bea\label{DY}
\Delta &=& \prod_{p=0}^P e_p^{n_p} \Delta'  \,, \\ \label{DYbar}
\bar \Delta &=& \prod_{\bar p=0}^{\bar P} \bar e_{\bar p}^{\bar n_{\bar p}} \bar \Delta' \,, 
\eea
where $\bar e_{\bar p}$ denote the $\bar P +1$ exceptional coordinates for $\bar{\mathcal Y}$. 
Our next task is then to prove two lemmas concerning the structure of these factorizations, in absence and in presence of special fibers at the intersections $B^{p, p+1}$, respectively, of the configuration $\mathcal Y$.  

\vspace{.2\baselineskip}
{\Lemma\label{lemma1} Suppose that there are no special fibers  at $B^{p,p+1}$ for a given $p \in [0, P-1]$. Then the following holds: 
\begin{enumerate}
\item \label{L1-1}The exponents $\bar n_{\bar p}$ in~\eqref{DYbar} for $kp \leq \bar p \leq k(p+1)$ are determined by $n_p$ and $n_{p+1}$ as
\beq \label{nkp+r_abs}
\bar n_{kp + r} = (k-r)n_p + r n_{p+1}  \,,\quad \text{for} \quad r=0, \ldots ,k\,. 
\eeq
\item\label{L1-2} There are no special fibers at $\bar B^{\bar p,\bar p+1}$ for all $\bar p=kp, \ldots, k(p+1)-1$, regardless of the value of $k$. 
\end{enumerate}
}

\vspace{.2\baselineskip}{\proof.}  
\begin{enumerate}
\item 
Let us first note that the absence of special fibers at $B^{p,p+1}$ is equivalent to the presence in $\Delta$ of at least one monomial of the form,
\beq
\cdots e_p^{n_p} e_{p+1}^{n_{p+1}} \cdots \,,
\eeq
whose powers in $e_p$ and $e_{p+1}$ saturate the overall ones in~\eqref{DY}. Note that such a monomial originates either from $f^3$ or $g^2$. Let us assume the former (analogous reasoning works also in the latter case); from the expansions~\eqref{fp} and~\eqref{expfp}, we know that the monomial should take the form 
\beq\label{triple-cross}
A_{i_1} A_{i_2} A_{i_3} \left(\prod_{q=0}^P e_q^{\mu_{q,i_1} + \mu_{q,i_2} + \mu_{q,i_3}} + \cdots\right) s^{3a+i_1+i_2+i_3} t^{24-3a-i_1-i_2-i_3} \,,
\eeq
indicating that
\bea\label{np-const}
\mu_{p, i_1} + \mu_{p,i_2} +\mu_{p,i_3} &=& n_p \,,\\ \label{np+1-const}
\mu_{p+1, i_1} + \mu_{p+1,i_2} +\mu_{p+1,i_3} &=& n_{p+1} \,.
\eea
The leading monomial in~\eqref{triple-cross} is the product of the three terms in $f$, 
\beq
A_{i_l} \prod_{q=0}^P e_q^{\mu_{q, i_l}} s^{a+i_l} t^{8-a-i_l}\,, \quad \text{for}\quad l=1,2,3\,,
\eeq
of the blown-up configuration $\mathcal Y$. They originate from
\beq
A_{i_l} u^{\mu_{0, i_l}} s^{a+i_l} t^{8-a-i_l} \,,
\eeq
in the initial configuration $\widehat{\mathcal Y}$. 
If we now start off with the base-changed configuration, they lead in $\bar{\mathcal Y}$ to the terms
\beq
A_{i_l} \prod_{\bar q =0}^{\bar P} e_{\bar q}^{\bar \mu_{\bar q, i_l}} s^{a+i_l} t^{8-a-i_l} \,,
\eeq
where the barred exponents $\bar \mu_{\bar q, i_l}$ are related to the unbarred ones as
\bea\label{relations}
\bar \mu_{ k p + r, i_l}= (k-r) \mu_{p, i_l} + r \mu_{p+1, i_l} \,, \quad \text{for}\quad r=0, \ldots, k\,,
\eea
which results simply from the linearity of the recursion relations~\eqref{recmu} for the configurations $\mathcal Y$ and $\bar{\mathcal Y}$. The product of this new triple gives the monomial in $\bar \Delta$ of the form
\beq\label{triple-cross-bar}
A_{i_1} A_{i_2} A_{i_3} \left(\prod_{\bar q=0}^{\bar P} \bar e_{\bar q}^{\bar \mu_{\bar q,i_1} + \bar \mu_{\bar q,i_2} + \bar \mu_{\bar q,i_3}} + \cdots\right) s^{3a+i_1+i_2+i_3} t^{24-3a-i_1-i_2-i_3} \,,
\eeq
whose exponents in the coordinates,
\beq
\bar e_{\bar p} \quad \text{for}\quad \bar p = k p + r\,,~ r=0, \ldots, k\,,
\eeq
are given as
\bea
\sum_{l=1}^3  \bar \mu_{kp+r, i_l}  &=& (k-r) \sum_{l=1}^3 \mu_{p,i_l} + r \sum_{l=1}^3 \mu_{p+1, i_i} \\ \label{upperbound}
&=& (k-r) n_p + r n_{p+1} \,.
\eea  
Here the relation~\eqref{relations} has been used in the first line and the constraints~\eqref{np-const} and~\eqref{np+1-const} in the second. This thus sets the upper bounds for $\bar n_{kp+r}$ as~\eqref{upperbound}. Furthermore, it is clear that the most general triple products of the form~\eqref{triple-cross} in the configuration $\mathcal Y$ are subject to the inequality version of the constraints~\eqref{np-const} and~\eqref{np+1-const}, 
\bea 
\mu_{p, i_1} + \mu_{p,i_2} +\mu_{p,i_3} &\geq& n_p \,,\\ \label{np+1-const}
\mu_{p+1, i_1} + \mu_{p+1,i_2} +\mu_{p+1,i_3} &\geq& n_{p+1} \,,
\eea
and hence that they correspond in the base-changed configuration $\bar{\mathcal Y}$ to~\eqref{triple-cross-bar}, for which the exponents in $\bar e_{\bar p=kp+r}$ are not less than~\eqref{upperbound}. It thus follows that the upper bounds ~\eqref{upperbound} for the exponents are saturated, and hence that the equation~\eqref{nkp+r_abs} holds.  \hfill\(\Box\) \\

\item 
We have seen that $\bar \Delta$ contains a monomial of the form~\eqref{triple-cross-bar} with the exponents in $\bar e_{\bar p}$ given as~\eqref{upperbound}. Furthermore, this monomial saturates the overall factors $\bar e_{\bar p}^{\bar n_{\bar p}}$ of the factorization~\eqref{DYbar} for all $\bar p = kp, \ldots, k(p+1)$, so that $\Delta'$ contains a non-trivial monomial of the form,
\beq
\cdots \prod_{\bar p=kp}^{k(p+1)}\bar e_{\bar p}^0  \cdots \,,
\eeq
independent of all $\bar p = kp, \ldots, k(p+1)$. We thus learn that there arise no special fibers at $\bar B^{\bar p, \bar p+1}$ for all $\bar p = kp, \ldots, k(p+1)-1$, as claimed.  \hfill\(\Box\) \\
\end{enumerate}

\vspace{.2\baselineskip}
{\Lemma\label{lemma2} Suppose that there are no special fibers at $B^{p,p+1}$ for a given $p \in [0, P-1]$. Let the factored-out discriminant $\Delta' (s, t; e_0, \ldots, e_P)$ be given by a $\IC^*$-linear combination of the monomials $M_m (s,t; e_0, \ldots, e_P)$ as
\beq
\Delta' = \sum_{m=1}^N c_m M_m \,, \quad c_m  \in \mathbb \IC^*\,,  
\eeq
where $M_m$ are written as
\beq\label{Mm}
M_m =  t^{\tau_m}\left(\prod_{q=0}^P e_q^{\epsilon_{q,m}}\right) s^{\sigma_m}  \,,
\eeq
and $N$ denotes the number of non-trivial terms in $\Delta'$. 
Then, the following holds: 
\begin{enumerate}
\item \label{L2-1} The quantities $\epsilon_p^{(0)}$ and $\epsilon_{p}^{(1)}$ defined as
\bea\label{ep0}
\epsilon_p^{(0)} &:=& {\rm min} \{\,\epsilon_{p,m} \;|\; \epsilon_{p+1, m}=0\,,\;m=1, \ldots, N\,\} \,,\\ \label{ep1}
\epsilon_p^{(1)} &:=& {\rm min} \{\,\epsilon_{p+1,m} \;|\; \epsilon_{p, m}=0\,,\; m=1, \ldots, N\,\} \,.
\eea
are both strictly positive, i.e., there exists a non-trivial brane stack both at $e_{p+1}=0$ in $B^p$ and at $e_p=0$ in $B^{p+1}$. 
\item \label{L2-2}The exponents $\bar n_{\bar p}$ in~\eqref{DYbar} for $kp \leq \bar p \leq k(p+1)$ are determined by
\beq\label{nkp+r}
\bar n_{kp+r} = (k-r) n_p + r n_{p+1} + \bar {\frak M}(r) \,, \quad r=0, \ldots, k\,,
\eeq
where $\bar{\frak M}: [0,k]  \to \IR_{\geq0}$ is a function defined as the minimum 
\beq\label{min-fn}
\bar{\frak M}(r) := {\rm min} \{\,\bar{\frak L}_m(r) \;|\; m=1, \ldots, N\,\} \,, 
\eeq
of the $N$ linear functions $\bar{\frak L}_m: [0,k] \to \IR_{\geq 0}$,
\beq\label{linearMbar}
\bar{\frak L}_m(r) := (k-r) \epsilon_{p,m} + r \epsilon_{p+1, m} \,, \quad m=1, \ldots, N \,.
\eeq
\item \label{L2-3} 
The function $\bar{\frak M}(r)$ defined as~\eqref{min-fn} is piece-wise-linear and concave. If we denote by $\frak M$ the $\bar{\frak M}$ with $k=1$, i.e., for the original configuration, the two functions are related as
\beq\label{Mscale}
\bar{\frak M}(kr) = k\, \frak M(r) \,,  
\eeq
for $r$ in the unit interval $I:=[0,1]$. 
The piece-wise linearity of $\frak M$ prompts us to naturally split its domain $I$ into a finite number of sub-intervals, 
\beq
I_\iota=[\cR_\iota, \cR_{\iota+1}] \,,\quad \iota=0, \ldots, \cN-1 \,,
\eeq 
in each of which $\frak M$ is linear, with rational boundary values $\cR_{\iota}$. Furthermore, $\frak M$ vanishes at the boundary of the domain, i.e.,
\beq\label{Mbdry}
\frak M(0)=0=\frak M(1) \,.
\eeq

\item \label{L2-4}There appear no special fibers at $\bar B^{\bar p,\bar p+1}$ for all $\bar p=kp, \ldots, k(p+1)-1$ if and only if the base-change order $k$ is a multiple of $k_0^{(p)}$, where the latter is defined as the smallest positive integer for which 
\beq\label{k0pdef}
k \cR_\iota \in \IZ \,,\quad \forall \iota = 1, \ldots, \cN-1\,.
\eeq
  
\end{enumerate}
}

\begin{figure}[h]
\centering
      \includegraphics[width=\textwidth]{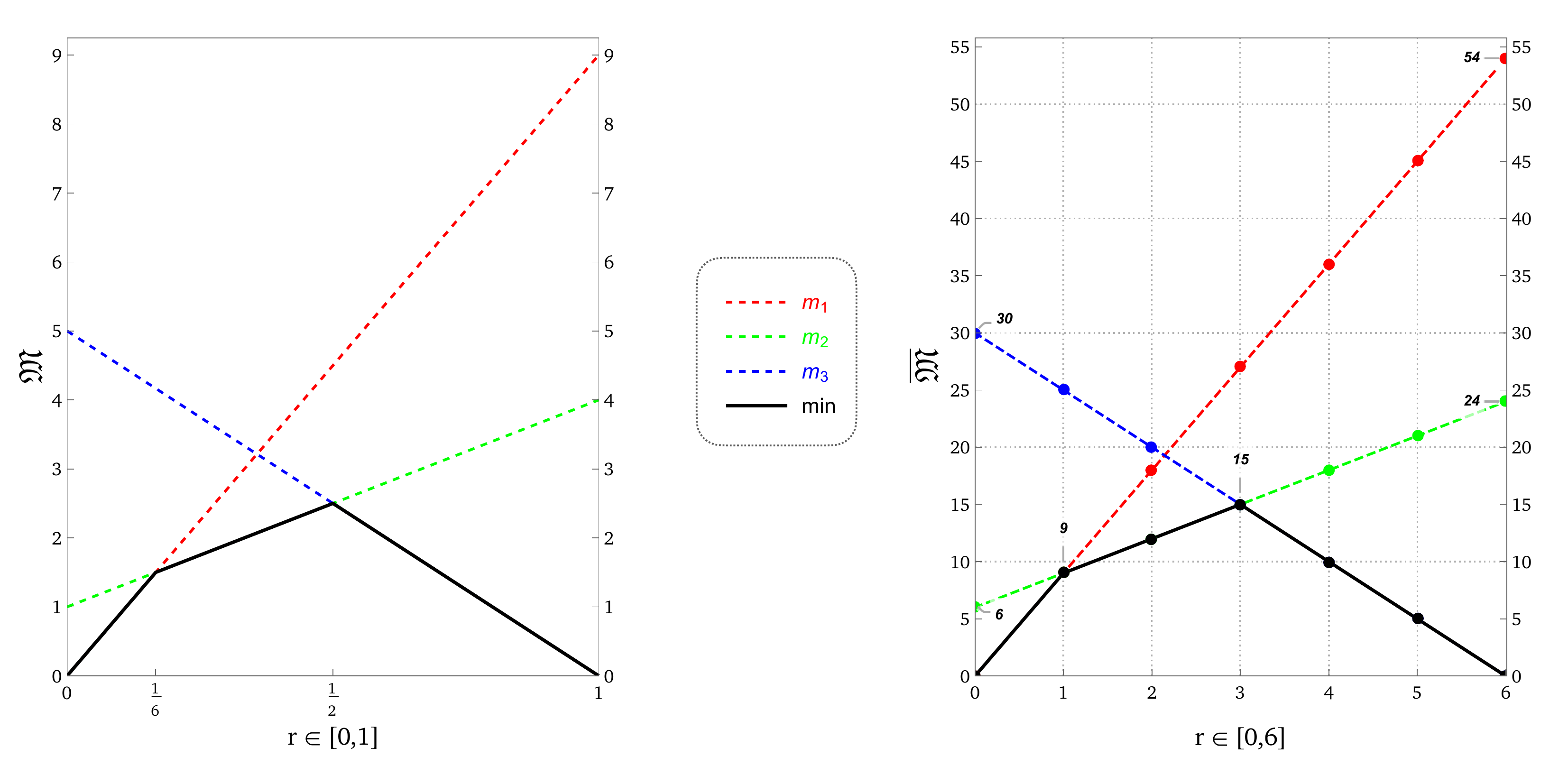}
    \caption{The minimum functions $\frak M$ and $\bar{\frak M}$ defined by~\eqref{min-fn} for $k=1$ and $6$, respectively, in their domains $I:[0,1]$ and $\bar I: [0, k]$. Sketched in the left figure is for an example configuration where a specific triple of linear functions $\frak L_m$, $m=m_{1,2,3}$, determine the minimum function $\frak M$ globally, and hence, $\cN=3$, with the corresponding two-vectors $(\epsilon_{p,m}, \epsilon_{p+1,m}) =(0,9), (1,4)$ and $(5,0)$, which indicates that $\epsilon_p^{(0)}=5$ and $\epsilon_p^{(1)}=9$; the dashed lines in red, green and blue are for the triple $\frak L_{m_{1,2,3}}$ and the solid black line is for their minimum $\frak M$. Note that $\cR_{1}=\frac16$ and $\cR_2=\frac12$ so that their denominators $\cQ_1$ and $\cQ_2$ determine $k_0^{(p)}$ as ${\rm lcm}(\cQ_1, \cQ_2)= {\rm lcm}(6,2) = 6$, which is taken as the value of $k$ for the right figure; the analogous objects defined for the base-changed configuration with $k=6$ are depicted on the right.}
    \label{fig:MMbar}
  \end{figure}

\vspace{.2\baselineskip}{\proof.}  
Before we start giving our proof, let us point out that Fig.~\ref{fig:MMbar} exemplifies the variety of objects defined in stating the Lemma for a concrete geometry, to which the readers are referred along the way for illustrative purposes. 

\begin{enumerate}
\item Note first that the vanishing of $\epsilon_p^{(0)}$ implies that of $\epsilon_p^{(1)}$ and vice versa. Let us thus suppose that they both vanish,
\beq
\epsilon_p^{(0)} = 0 = \epsilon_p^{(1)} \,.
\eeq
This means that $\Delta'$ contains a monomial of the form
\beq 
\cdots e_p^0 e_{p+1}^{0}\cdots
\eeq
independent of $e_p$ and $e_{p+1}$, which contradicts the starting assumption that branes are present at $B^{p,p+1}$.  \hfill\(\Box\) \\

\item 

We can use essentially the same reasoning as in the proof of Lemma 1-\ref{L1-1} to show the monomials
\beq
\prod_{p=0}^P e_p^{n_p} M_m 
\eeq 
appear in $\Delta$ if and only if the associated monomials
\beq\label{barDeltaMon}
\prod_{\bar p=0}^{\bar P} \bar e_{\bar p}^{\bar n_{\bar p}} \bar M_m 
\eeq
appear in $\bar \Delta$.
Here
\beq\label{Mbarm}
\bar M_m = t^{  \tau_m} \left(\prod_{\bar q=0}^{\bar P} \bar e_{\bar q}^{\bar \epsilon_{\bar q,m}}\right)s^{  \sigma_m}  
\eeq
has the same single-component-configuration origin as $M_m$ in that 
\beq
\Big[ \prod_{\bar p=0}^{\bar P} \bar e_{\bar p}^{\bar n_{\bar p}} \bar M_m\Big]_{\bar e_1=\ldots = \bar e_{\bar P} = 1} =\Big[ \prod_{p=0}^P e_p^{n_p} M_m\Big]_{e_0=\bar e_0^k ;\;e_1=\ldots = e_P = 1} \,.
\eeq
Here, once again, essentially due to the linearity of the recursion relations~\eqref{recmu} and~\eqref{recnu} for $\mathcal Y$ and $\bar{\mathcal Y}$, the exponents in the coordinates,
\beq
\bar e_{\bar p} \quad \text{for}\quad \bar p = k p + r\,,~ r=0, \ldots, k\,,
\eeq
for the monomials~\eqref{barDeltaMon} are given by
\beq\label{expmon}
(k-r) (n_p + \epsilon_{p,m}) + r (n_{p+1} + \epsilon_{p+1, m}) \,. 
\eeq
This in turn indicates that $\bar n_{kp+r}$ are determined by
\beq
\bar n_{kp+r} = (k-r) n_p + r n_{p+1} + \bar{\frak M}(r) \quad \text{for}\quad r=0, \ldots, k\,,
\eeq
where $\bar{\frak M}: [0,k] \to \IR_{\geq 0}$ is a function defined as
\bea
\bar{\frak M}(r)&:=& {\rm min} \{\,(k-r) \epsilon_{p, m} + r \epsilon_{p+1, m}\;|\; m=1, \ldots, N\,\} \\
&=& {\rm min} \{\,\bar{\frak L}_m(r) \;|\; m=1, \ldots, N\,\} \,, 
\eea
to extract the minimum of the $m$-dependent contributions to the exponents~\eqref{expmon}. \hfill\(\Box\) \\

\item  
While the symbol $\bar{\frak M}$ for the function does not make it explicit, we emphasize that the following set of quantities are assumed to have been fixed in order for $\bar{\frak M}$ to be well-defined via~\eqref{min-fn}: 
\bi
\item $k$ for the base-change order;
\item $p$ for the specific intersection $B^{p,p+1}$, the location of the special fiber in scrutiny;
\item $N$ for the number of monomials in $\Delta$; 
\item ${\vec \epsilon_m} := (\epsilon_{p,m}, \epsilon_{p+1, m}) \in \IZ_{\geq 0}^2$ where $m=1, \ldots, N$.
\ei
Since $\bar{\frak M}(r)$ is the minimum of the $N$ linear functions $\bar{\frak L}_m: [0,k] \to \IR_{\geq 0}$,
\beq\label{linearMbar}
\bar{\frak L}_m(r) = (k-r) \epsilon_{p,m} + r \epsilon_{p+1, m} \,, \quad m=1, \ldots, N \,,
\eeq
it is clearly a piece-wise linear function. 
Let us now consider the $k=1$ version, $\frak M(r)$, which is the minimum of ${\frak L}_m: [0,1] \to \IR_{\geq 0}$ defined analogously as
\beq\label{linearM}
{\frak L}_m(r) = (1-r) \epsilon_{p,m} + r \epsilon_{p+1, m} \,, \quad \quad m=1, \ldots, N \,.
\eeq
Given that the $N$ pairs of the individual linear functions $(\bar{\frak L}_m, \frak L_m)$ are identified through the scaling $r \to kr$ as
\beq\label{Lrescale}
\bar{\frak L}_m (kr) = k\,\frak L_m (r) \,,
\eeq
 their minimums should also be identified in the same manner,
\beq
\bar{\frak M}(kr) = k\, \frak M(r) \,. 
\eeq
Therefore, it suffices to analyze the structure of $\frak M$ as a reference function, on which we will focus now. Since $\frak M$ is the minimum of $N$ linear functions $\frak L_m$, it is natural to decompose the unit interval $I$ into a finite number of sub-intervals 
\beq
I_{\iota} := [\cR_\iota, \cR_{\iota+1}] \,,\quad \iota= 1, \ldots, \cN-1\,,
\eeq
in each of which $\frak M$ is given by a fixed linear function $\frak L_{m_\iota}$ and hence linear. Obviously, the number of such sub-intervals is not bigger than that of the individual linear functions, i.e.,
\beq
\cN \leq N \,,
\eeq
since a fixed $\frak L_m$ can only provide the minimal value of the total of $N$ functions in at most one $I_\iota$. Furthermore, each boundary value $r=\cR_\iota$ at the intersection $I_{\iota-1}\cap I_{\iota}$ for $1\leq \iota \leq \cN-1$ arises from the collision of the pair of linear functions $\frak L_{m_{\iota-1}}$ and $\frak L_{m_\iota}$, serving as the minimum  of the $N$ functions in the respective sub-intervals. Since the associated two-vectors $\vec\epsilon_m=(\epsilon_{p,m}, \epsilon_{p+1,m})$ for $m=m_{\iota-1}, m_\iota$ are both integral, it follows that the linear graphs of $\frak L_{m_{\iota-1}}$ and $\frak L_{m_\iota}$ meet at a rational value of $r$, i.e., that $\cR_\iota$ is rational. Furthermore, since there exists at least one two-vector $\vec\epsilon_m$ for which the first component vanishes, it follows that
\beq
\epsilon_{p, m}=0 \,,
\eeq
since, otherwise, the overall power of $e_p$ in $\Delta$ would be strictly bigger than $n_p$. Hence, $\frak L_m(0)=0$ for such $m$, which results in the vanishing of $\frak M$ at $r=0$, 
\beq
\frak M(0) = 0 \,.
\eeq
Similarly, the presence of a two-vector $\vec \epsilon_m$ for which $\epsilon_{p+1,m} =0$ results in 
\beq
\frak M(1) = 0\,.
\eeq

\hfill\(\Box\) \\

\item
First of all, since the end points $\cR_\iota$ are rational, there exist integral orders $k$ with the property~\eqref{k0pdef}. Specifically, if we write
\beq
\cR_\iota = \frac{\cP_\iota}{\cQ_\iota} \,, 
\eeq
where $\cP_\iota$ and $\cQ_\iota$ are relatively prime integers, the minimal such order, $k_0^{(p)}$, is obtained as
\beq\label{k0p-practical}
k_0^{(p)} = {\rm lcm} \{\cQ_\iota\}_{\iota=1}^{\cN} \,,
\eeq
and the condition~\eqref{k0pdef} is satisfied if and only if the base-change order $k$ is divisible by $k_0^{(p)}$. 
Let us now consider the fibers for a fixed $k$ and test if some special fibers sit at some of the $k$ intersections, 
\beq
\bar B^{\bar p, \bar p+1} \,, \quad \text{for} \quad \bar p = kp+ r\,,~ r=0, \ldots, k-1\,.
\eeq
Obviously, no special fibers are found at $\bar B^{kp+r, kp+r+1}$ for a fixed $r$ if and only if there exists a monomial $\bar M_{m_r}$ in $\Delta'$ with the  property
\beq
\bar\epsilon_{kp+r, m_r} =0 = \bar \epsilon_{kp+r+1, m_r} \,. 
\eeq
Such a monomial $\bar M_{m_r}$ can in turn be seen to exist if and only if the associated linear function $\frak L_{m_r}$, defined for the configuration with $k=1$ as~\eqref{linearM}, serves as the minimum in the sub-interval $I_{\iota_r}$ that contains both $\frac{r}{k}$ and $\frac{r+1}{k}$, i.e., iff
\beq\label{demand}
\exists \iota_r \in [0, \cN-1] \quad \text{such that}\quad k \cR_{\iota_r} \leq  r <  r+1 \leq k\cR_{\iota_r+1} \,.
\eeq
Clearly, demanding~\eqref{demand} for all values of $r=0, \ldots, k-1$ is equivalent to requiring that $k \cR_{\iota}$ be integral for all $\iota=1, \ldots, \cN-1$. \hfill\(\Box\) \\

\end{enumerate}

Note that Lemma~\ref{lemma1} is in fact follow as a corollary of Lemma~\ref{lemma2}. In any case, upon combining the explicit statements of Lemma 1-\ref{L1-2} and Lemma 2-\ref{L2-4}, we immediately learn that

\vspace{.2\baselineskip}
{\proposition\label{prop8} There are no special fibers at any $\bar B^{\bar p,\bar p+1}$ for $\bar p=0, \ldots, \bar P-1$ if and only if the base-change order $k$ is a multiple of $k_0$ defined as 
\beq\label{k0def}
k_0 := \prod_{p=0}^{P-1} k_0^{(p)} \,,
\eeq
where $k_0^{(p)}$ are the smallest positive integers with the defining property~\eqref{k0pdef}, given explicitly as~\eqref{k0p-practical}.} \\

While Proposition~\ref{prop8} teaches us that the ambiguities in interpreting the Kodaira type of the special fibers at intersections can be resolved by an appropriate base change, one may still question if base changes do not change the Kodaira type of the special fibers away from the intersections, which we will now call ``interior fibers'' for brevity. We observe that essentially they do not. To be more precise, we claim the following

\vspace{.2\baselineskip}
{\proposition\label{prop9} 
Suppose there are no special fibers  at any of the intersections $B^{p,p+1}$ in the configuration $\mathcal Y$. Then, the elliptic fiber of the central surface $Y_0$ at a point $[e_{p-1}:e_{p+1}]$ in $B^p$  has the same Kodaira type as that of $\bar Y_0$ at the point $[\bar e_{kp-1}:\bar e_{kp+1}]=[e_{p-1}:e_{p+1}]$ in $\bar B^{kp}$. Here, $e_p$ for $p=-1$ and $p=P+1$ respectively denote $t$ and $s$. 
}

\vspace{.2\baselineskip}{\proof.} 
Recall that, from the perspective of the central surface $Y_0$, the fiber over a point in $B^p$ has its Kodaira type determined by the K3-vanishing orders, i.e., the vanishing orders of the restrictions $f_p$, $g_p$ and $\Delta'_p$ of the sections $f$, $g$ and $\Delta'$. We will thus consider those sections for the base components $B^p$ in the left end ($p=0$), in the middle ($0<p<P$) and in the right end ($p=P$) of the chain in turn. 
\bi
\item For the left-end component, let us analyze how the restricted sections behave around the point with homogeneous coordinates $[t:e_1]$ in $B^0$. Since we are interested in relating the behavior of those sections to that of their analogues in $\bar B^0$, we start by noting that terms in $f$ and $g$ divisible (resp., not divisible) by $e_0$ turn into those in $\bar f$ and $\bar g$ divisible (resp., not divisible) by $\bar e_0$. Therefore, once we set $e_0=0$, $e_p=1$ for $p>1$ and $\bar e_0=0$, $\bar e_{\bar p} =1$ for $\bar p>1$, it immediately follows that the functional behavior of $(f_0, g_0)$ around $[t:e_1]$ is precisely the same as that of $(\bar f_0, \bar g_0)$ around $[t:\bar e_1]$. 

While a similar statement can be made about all terms in $\Delta$, there may arise a slight complication due to the common factors $e_0^{n_0}$ and $e_1^{n_1}$ that are to be factored out in defining $\Delta'$. 
In order to address this issue, let us first write, as in~\eqref{Mm} and~\eqref{Mbarm}, the monomials appearing in $\Delta'$ and $\bar \Delta'$, respectively, as
\bea
M_m &=&  t^{\tau_m}\left(\prod_{q=0}^P e_q^{\epsilon_{q,m}}\right) s^{\sigma_m}  \,,\\
\bar M_m &=& t^{  \tau_m} \left(\prod_{\bar q=0}^{\bar P} \bar e_{\bar q}^{\bar \epsilon_{\bar q,m}}\right)s^{  \sigma_m}  \,,
\eea
upon factoring out the overall factors $\prod_{p=0}^P e_p^{n_p}$ and $\prod_{\bar p=0}^{\bar P} \bar e_{\bar p}^{\bar n_{\bar p}}$ from $\Delta$ and $\Delta'$. 
As addressed by~\eqref{nkp+r_abs}, the overall exponents transform linearly as 
\beq
\bar n_0 = k n_0 \,,\qquad \bar n_1 = (k-1) n_0 +   n_1 \,. 
\eeq
Furthermore, the way the corresponding overall factors are taken out from the discriminant sections is uniform in that, for any monomial $M_m$ with $\epsilon_{0,m}=0$, the corresponding monomial $\bar M_m$ has the properties
\bea
\bar \epsilon_{0,m} &=& 0 \,,\\
\bar \epsilon_{1,m}&=& \epsilon_{1,m} \,.
\eea
This guarantees that the functional behavior of $\Delta'_0$ around $[s:e_1]$ is identical to that of $\bar \Delta'_0$ around $[s:\bar e_1]=[s:e_1]$. 

All in all, the relevant sections, $f$, $g$ and $\Delta'$, restricted to $B^0$ in the configuration $\mathcal Y$ and their counterparts in $\bar{\mathcal Y}$ may be viewed as being completely identical, and hence, the same applies to their vanishing orders, as well as the associated fiber types.

\item Next, for the middle components $B^p$ with $0<p<P$, it suffices to analyze the behavior of the discriminants, since we have already established in Proposition~\ref{prop3} that only special fibers of Type I$_n$ may arise there. Then, as for the discriminant behaviors, the rationale used for the left-end component persists; the overall exponents transform linearly as
\beq
\bar n_{kp} = k n_p \,,\qquad \bar n_{kp \pm 1} = (k-1) n_0 +  n_{p \pm 1} \,, 
\eeq
and the way the overall factors are taken out is uniform so that, for $M_m$ with $\epsilon_{p, m} = 0$, the corresponding $\bar M_m$ has the properties
\bea
\bar \epsilon_{kp, m} &=& 0 \,,\\
\bar \epsilon_{kp\pm1, m} &=& \epsilon_{p\pm1,m} \,.
\eea
This guarantees once again that the functional behavior of $\Delta'_p$ around $[e_{p-1}:e_{p+1}]$ is identical to that of $\bar \Delta'_{kp}$ around $[\bar e_{kp-1}: \bar e_{kp+1}]=[e_{p-1}:e_{p+1}]$. Therefore, the corresponding vanishing orders of $\Delta'_p$ and $\bar \Delta'_{kp}$ are identical there, and hence, the fibers thereon have the same Kodaira type. 

\item Finally, for the right-end component $B^P$, it is by now obvious that the same rationale is valid once again for the discriminants, so that the functional behavior of $\Delta'_P$ around $[e_{P-1}:s]$ is identical to that of $\bar\Delta'_{\bar P}$ around $[\bar e_{\bar P-1}: s]=[e_{P-1}:s]$. 
However, we need to carefully compare the behavior of $(f_P, g_P)$ with that of $(\bar f_{\bar P}, \bar g_{\bar P})$. In order to also identify them, it suffices to make sure that terms in $f$ and $g$ are divisible by $e_P$ if and only if the corresponding terms in $\bar f$ and $\bar g$ are by $\bar e_{\bar P}$. While this is not true in general, recall that we are already presuming the base change~\eqref{basec} in order to avoid physical non-minimality at $s=0$ in $B^P$, as discussed in Proposition~\ref{prop6}-1. As already argued in the proof of this proposition, either of the two index subsets, 
\bea\label{cI-0}
\cI_-^{(0)} = \{\,i \in \cI_-\,|\, \mu_{P,i}=0\,\} \,,\\ \label{cJ-0}
\cJ_-^{(0)} = \{\,j \in \cJ_-\,|\, \nu_{P,j}=0\,\} \,,
\eea
should thus be non-trivial (see Proof~\ref{proof6}). Then, the invariance of the definitions~\eqref{cI-0} and~\eqref{cJ-0} is guaranteed for any further base changes, so that 
\bea
\mu_{P,i} = 0 \quad &\Longleftrightarrow& \quad \bar\mu_{\bar P, i} =0 \,, \\
\nu_{P,j}=0 \quad &\Longleftrightarrow& \quad \bar \nu_{\bar P, j} = 0 \,.
\eea
Therefore, once we set $e_P=0$, $e_p = 1$ for $p<P$ and $\bar e_{\bar P} = 0$, $\bar e_{\bar p} = 1$ for $\bar p<\bar P$, it immediately follows that the functional behavior of $(f_P, g_P)$ around $[e_{P-1}:s]$ coincides with that of $(\bar f_{\bar P}, \bar g_{\bar P})$ around $[\bar e_{\bar P-1}: s]$. 
\ei
With all of this combined, we thus conclude that the types of the fibers of $Y_0$ are invariant under the base change in the sense of Proposition~\ref{prop9}. Hence, the same is true of the gauge algebras carried by the corresponding branes, as desired.  \hfill\(\Box\) \\

To summarize, according to Propositions~\ref{prop8} and~\ref{prop9}, the base change of an appropriately chosen order $k$ (so that it may be divisible by $k_0$ defined in~\eqref{k0def}) reconfigures the brane arrangement in $\mathcal Y$ into another in $\bar{\mathcal Y}$ in such a way that 
\bi
\item [1.] branes at intersections, if present in $\mathcal Y$, are ``scattered'' not to sit at any base-component intersections in $\bar{\mathcal Y}$;
\item [2.] interior brane stacks in $\bar B^{\bar p}$, for the configuration $\bar{\mathcal Y}$ free of branes at intersections, are only ``rescaled'' to interior stacks carrying the same gauge algebra, if a further base change is applied.{\footnote{In fact, we do not need to assume, in Proposition~\ref{prop9}, that the configuration $\mathcal Y$ does not have special fibers at component intersections. The content of  Proposition~\ref{prop9} generalizes in a straightforward manner even without this assumption. In case there were branes at $B^{p-1,p}$ and/or $B^{p,p+1}$, the restriction $\Delta_p'$ may admit non-trivial overall factors of $e_{p\pm 1}$. Similarly, the restriction $\bar \Delta_{kp}'$ may also admit factors of $\bar e_{kp \pm 1}$ unless $k$ is divisible by $k_0^{(p)}$. However, we may safely ignore such overall factors, even if present, away from the component intersections. Then, modulo this, the two restrictions $\Delta_p'$ and $\bar \Delta_{kp}'$ turn out once again to have exactly the same functional behavior. }}
\ei
To complete our understanding of the brane arrangement, we will end this appendix with a precise account of how exactly the branes at intersections $B^{p,p+1}$ transform into interior branes in the base-changed configuration $\bar{\mathcal Y}$. As is intuitively clear, it turns out that the configuration $\bar{\mathcal Y}$ has the following property on top of the above two.
\bi
\item [3.] the branes at $B^{p,p+1}$ turn into interior branes in $\bar B^{kp+r}$ with $r=1, \ldots, k-1$. 
\ei
To be precise, we have the following refinement of Lemma~\ref{lemma2}-\ref{L2-4}: 

\vspace{.2\baselineskip}
{\proposition\label{prop10} 
Let us consider the configuration $\mathcal Y$ containing $\epsilon_p^{(0)} + \epsilon_p^{(1)}$ branes at the intersection $B^{p,p+1}$ for a fixed $p \in [0, P-1]$, where $\epsilon_p^{(0)}$ and $\epsilon_p^{(1)}$ are defined by~\eqref{ep0} and~\eqref{ep1} as the number of branes at $e_{p+1}=0$ in $B^p$ and at $e_p=0$ in $B^{p+1}$, respectively. Let $\bar{\mathcal Y}$ be the configuration obtained for $k$ divisible by $k_0^{(p)}$ defined by~\eqref{k0pdef}, or practically by~\eqref{k0p-practical}. Then, the $\epsilon_p^{(0)} + \epsilon_p^{(1)}$ branes at $B^{p,p+1}$ split into $\bar \epsilon^{(r)}$ interior branes in $\bar B_{kp+r}$ for $r=1, \ldots, k-1$, so that
\beq\label{sum-inv}
\sum_{r=1}^{k-1} \bar \epsilon^{(r)} = \epsilon_p^{(0)} + \epsilon_p^{(1)}\,.
\eeq
In fact, the individual $\bar \epsilon^{(r)}$are given as
\beq\label{indiv}
\bar \epsilon^{(r)} = 2\, \bar{\frak M}(r) - \bar{\frak M}(r-1) - \bar{\frak M}(r+1)\,, \quad r=1, \ldots, k-1\,,  
\eeq
from which the relation~\eqref{sum-inv} follows. Here, $\bar{\frak M}$ is the minimum function defined by~\eqref{min-fn}.
}

\vspace{.2\baselineskip}{\proof.} 
We will first confirm that the formulas~\eqref{indiv} do lead to the sum rule~\eqref{sum-inv}. According to~{\eqref{indiv}}, the sum of the individual countings $\bar \epsilon^{(r)}$ is computed as
\bea\label{sum-barred}
\sum_{r=1}^{k-1} \bar \epsilon^{(r)} &=& \bar{\frak M}(0) + \bar{\frak M}(1) + \bar{\frak M}(k-1)+\bar{\frak M}(k) \\
&=& \bar{\frak M}(1)+ \bar{\frak M}(k-1) \,,
\eea
where in the second step the vanishing~\eqref{Mbdry} has been used (together with the scaling relation~\eqref{Mscale}). Let us now note that the linear functions $\frak L_{m_1}$ and $\frak L_{m_\cN}$ that govern the minimum function $\frak M$ in the first and the last sub-intervals $I_0$ and $I_{\cN-1}$, respectively, are related to their counterparts $\bar{\frak L}_{m_1}$ and $\bar{\frak L}_{m_{\cN}}$ for $\bar{\frak M}$ in the base-changed configuration by the rescaling~\eqref{Lrescale} as
\beq\label{Lrescale-again}
\bar{\frak L}_{m_1} (kr) = k\, {\frak L}_{m_1}(r) \,,\qquad \bar{\frak L}_{m_\cN} (kr) = k\,{\frak L}_{m_\cN}(r) \,. 
\eeq
Furthermore, according to the definitions~\eqref{ep0} and~\eqref{ep1}, we have
\beq\label{def-prop}
\frak L_{m_1} (1) = \epsilon_p^{(1)} \,,\qquad \frak L_{m_\cN} (0) = \epsilon_p^{(0)} \,.
\eeq
It thus follows that the RHS of the sum~\eqref{sum-barred} further simplifies as
\bea
 \bar{\frak L}_{m_1} (1) + \bar{\frak L}_{m_\cN} (k-1) 
 &=&  k (\frak L_{m_1}(1/k) + \frak L_{m_\cN} ((k-1)/k) ) \\
 &=& \frak L_{m_1}(1) + \frak L_{m_\cN} (0) \\
 &=& \epsilon_p^{(1)} + \epsilon_p^{(0)} \,,
\eea
where we have used the rescaling relation~\eqref{Lrescale-again}, the linearity of $\frak L_m$'s, and the defining property~\eqref{def-prop}, respectively, in the three steps. Therefore, the sum rule~\eqref{sum-inv} does follow from the formulas~\eqref{indiv}, which we will prove in the following. 

For a given $r \in [1, k-1]$, the branes in $\bar B^{kp+r}$ are located at the zeroes of $\bar\Delta'_{kp+r}$. Therefore, they are counted by the degree of the section $\bar\Delta'_{kp+r}$ in the homogeneous coordinates $\bar e_{kp+r\pm1}$ of $\bar B^{kp+r}$, and hence, $\bar \epsilon^{(r)}$ can be determined, e.g., from the two monomials $\bar M_{m_r}$ and $\bar M_{m_{r+1}}$ present in $\bar\Delta'_{kp+r}$. The former, for instance, has exponents $0$ and $\bar{\frak L}_{m_{\iota_-}}(r+1)-\bar{\frak L}_{m_{\iota_+}}(r+1)$ in $\bar e_{kp-1}$ and in $\bar e_{kp+1}$, respectively, where $\iota_\pm$ are the sub-interval labels such that 
\beq
r\pm 1 \in I_{\iota_\pm} \,.
\eeq
We therefore have 
\bea
\bar \epsilon^{(r)} &=& \bar{\frak L}_{m_{\iota_-}}(r+1)-\bar{\frak L}_{m_{\iota_+}}(r+1) \\
&=& ( \bar{\frak L}_{m_{\iota_-}}(r+1)- \bar{\frak L}_{m_{\iota_-}}(r))+(\bar{\frak L}_{m_{\iota_-}}(r)-\bar{\frak L}_{m_{\iota_+}}(r+1)) \\
&=& ( \bar{\frak L}_{m_{\iota_-}}(r)- \bar{\frak L}_{m_{\iota_-}}(r-1))+(\bar{\frak L}_{m_{\iota_+}}(r)-\bar{\frak L}_{m_{\iota_+}}(r+1)) \\
&=& (\bar{\frak M}(r) - \bar{\frak M}(r-1)) + (\bar{\frak M}(r)-\bar{\frak M}(r+1)) \\
&=& 2 \,\bar{\frak M}(r) - \bar{\frak M}(r-1)-\bar{\frak M}(r+1)\,. 
\eea
Here, in the second step $\bar{\frak L}_{m_r}(r)$ has been subtracted and then added back in and in the third step $\bar{\frak L}_{m_{\iota_-}}(r) = \bar{\frak L}_{m_{\iota_+}}(r)$ has been used (as the two linear functions, if $m_{\iota_-}$ and $m_{\iota_+}$ are distinct, should meet at $r$). Furthermore, we have used in the fourth step that $\bar{\frak L}_{m_{\iota_\pm}}$ govern the minimum function $\bar{\frak M}$ at $r \pm 1$ while both do at $r$. 
 \hfill\(\Box\) \\

Let us end this appendix by a final remark on the validity of our results. For simplicity of the presentation, we have so far been working under the assumption that the configuration $\mathcal Y$ is obtained by blowing up $\widehat{\mathcal Y}$ repeatedly at the ``same'' point, i.e., at $e_p=s=0$ for $p=0, \ldots, P-1$. As already addressed, however, in certain non-generic situations, several subchains of blowups, each at a different point, are to be performed in order to arrive at a configuration free of physical non-minimality; recall that Proposition~\ref{prop6} describes how one can deal with this potential complication. However, we can easily see how our results are also applicable to those non-generic situations, in which case $P$ blowups at a point should be followed by additional $Q$ blowups at another point, as suggested by Proposition~\ref{prop6} and its proof. All we need to do then is to blow down towards $B^P$ the $P$ base components $B^p$ for $0\leq p<P$, as well as the $Q$ components $B^{P+q}$ with $0<q\leq Q$. Viewing the geometry obtained by such blown-downs as the starting single-component configuration, we can for instance scatter all the special fibers at intersections $B^{p,p+1}$ for $p=0, \ldots, P-1$ and $B^{P+q, P+q+1}$ for $q=0, \ldots, Q-1$ by an appropriate base change; the criterion for this to work is, once again, that the base-change order $k$ is divisible by all of the minimal orders with the property~\eqref{k0pdef}, each defined for a given base-point intersection, regardless of the fact that two subchains are emanating from the two distinct points in $B^P$.  

\bibliography{papers}
\bibliographystyle{JHEP}

\end{document}